\documentclass[twocolumn]{aastex63}

\usepackage{xspace}
\usepackage{xfrac}

\received{x x, 2021}
\revised{x x, 2021}
\accepted{x x, 2021}
\submitjournal{AAS Journals}

\shorttitle{Multi-parameter degeneracy for extreme finite source effects }
\shortauthors{Johnson, Penny, \& Gaudi 2021}
\graphicspath{{./}{figures/}}
\usepackage{amsmath}
\begin{document}

\title{A Multi-Parameter Degeneracy in Microlensing Events with Extreme Finite Source Effects}

\newcommand\wfirst{{\textit{Roman}}\xspace}
\newcommand\romanst{{\textit{Roman}}\xspace}
\newcommand\tE{{$t_{\textrm{E}}$}}
\newcommand\thE{{$\theta_{\textrm {E}}$}}
\newcommand{\tsub}[2]{{#1}_\textrm{\tiny{#2}}}
\newcommand{\newCommandName}{text to insert}
\defcitealias{penny2019}{P19}
\newcommand{\epsScaleFactorOne}{1.1}
\newcommand{\epsScaleFactorTwo}{1.}
\newcommand{\bsgins}[1]{\textcolor{blue}{#1}}
\newcommand{\clarify}[1]{Clarify: \textcolor{red}{#1}}
\newcommand{\todo}[1]{\textcolor{violet}{TODO: #1}}

\correspondingauthor{Samson A. Johnson}
\email{johnson.7080@osu.edu}

\author[0000-0001-9397-4768]{Samson A. Johnson}
\affiliation{Department of Astronomy, The Ohio State University, 140 West 18th Avenue, Columbus OH 43210, USA}

\author[0000-0001-7506-5640]{Matthew T. Penny}
\affiliation{Department of Physics and Astronomy, Louisiana State University, Baton Rouge, LA 70803, USA}

\author[0000-0003-0395-9869]{B. Scott Gaudi}
\affiliation{Department of Astronomy, The Ohio State University, 140 West 18th Avenue, Columbus OH 43210, USA}



\begin{abstract}
For microlenses with sufficiently low mass, the angular radius of the source star can be much larger than the angular Einstein ring radius of the lens.  
For such extreme finite source effect (EFSE) events, finite source effects dominate throughout the duration of the event.
Here, we demonstrate and explore a continuous degeneracy between multiple parameters of such EFSE events.
The first component in the degeneracy arises from the fact that the directly-observable peak change of the flux depends on both the ratio of the angular source radius to the angular Einstein ring radius and the fraction of the baseline flux that is attributable to the lensed source star. 
The second component arises because the directly-observable duration of the event depends on both the impact parameter of the event and the relative lens-source proper motion.
These two pairwise degeneracies become coupled when the detailed morphology of the light curve is considered, especially when including a limb-darkening profile of the source star.
We derive these degeneracies mathematically through analytic approximations and investigate them further numerically with no approximations.
We explore the likely physical situations in which these mathematical degeneracies may be realized and potentially broken. 
As more and more low-mass lensing events (with ever decreasing Einstein ring radii) are detected with improving precision and increasing cadence from microlensing surveys, one can expect that more of these EFSE events will be discovered.
In particular, the detection of EFSE microlensing events could increase dramatically with the {\it Roman Space Telescope} Galactic Bulge Time Domain Survey. 
\end{abstract}

\keywords{Gravitational microlensing(672) - Finite-source photometric effect(2142) - Free floating planets(549)}


\section{Introduction} 
\label{sec:intro}

We only have weak constraints on the occurrence rate of isolated, planetary mass objects in the Milky Way, particularly for sub-Jovian mass objects.
As planetary systems form and host stars evolve, planets may become gravitationally unbound from their hosts through any number of dynamical processes or through post main sequence evolution \citep[see the introduction of][]{mroz2018}. 
In such scenarios, it is generally expected that a larger number of low-mass (i.e., terrestrial) planets or planetismals will be ejected than giant planets \citep[e.g.,][]{barclay2017}.
Thus, a determination of the mass function of these free-floating planets (FFPs) could provide important constraints on models of planet formation.
Alternatively, planetary-mass primordial black holes could compose some fraction of the mass budget of the Galactic halo \citep{niikura2019a,niikura2019b,montero-camacho2019}.
The only manner in which very low-mass and effectively dark objects can be detected is through gravitational microlensing \citep{distefano1999}. 

Recently, \citet{mroz2017} used the ground-based Optical Gravitational Lensing Experiment (OGLE) microlensing survey \citep{udalski2015} to place an upper limit on the occurrence rate of roughly Jupiter-mass FFPs.  
However, these authors also cautiously report a signal in the timescale distribution consistent with that of a population of terrestrial mass FFPs based on the detection of several events with extremely short (but highly uncertain) timescales. 
Since the analysis of \citet{mroz2017},  a total of seven additional, robust free-floating planet (or perhaps wide-orbit, see below) candidates have been discovered by ground-based microlensing surveys \citep{mroz2018,mroz2019,mroz2020a,mroz2020b,kim2021,ryu2021} primarily using data from the OGLE and Korea Microlensing Telescope Network (KMTNet) collaborations \citep{henderson2014,kim2016}.  
These events have denser photometric coverage than the tentative FFP events reported by \citet{mroz2017} and have estimated masses between that of Earth and Neptune, one of which may have a mass less than that of the Earth \citep{mroz2020b}.
A key factor in the discovery of these FFP candidates is the fact that they lens giant stars with angular radii larger than the angular Einstein ring radius, allowing for an estimate of the angular Einstein ring radius of the lens.
The masses of these candidate FFPs are only estimates, as additional measurements are required to break the mass-distance relationship of the Einstein ring radius and thus measure the true mass of the lens \citep[e.g.,][]{gould2021}.

Additionally, \citet{mcdonald2021} recently reported 4 candidate FFP events with extremely short effective timescales using data from \textit{Kepler K2} Campaign 9, the first blind space-based microlensing survey. 
Furthermore, the potential for the \textit{Nancy Grace Roman Space Telescope} to detect and characterize FFP events through its Galactic Bulge Time Domain Survey will be unprecedented and open new regions of parameters space not currently accessible by ground based microlensing surveys \citep{henderson2016,penny2019,johnson2020}.

It can be difficult to distinguish between a bound planet and a true FFP if the orbital separation between the bound planet and its host star is larger than $\gtrsim 10~{\rm au}$ \citep{han2005}.  
In this case, the host typically does not contribute any significant magnification to the event except for rare instances when the the source trajectory is parallel to the projected planet-host separation axis and thus is magnified by both objects.
While wide-orbit exoplanets are common \citep{poleski2021}, there are methods of rectifying this situation such as detecting flux from potential hosts or detailed modelling of the event, as summarized in \citet{han2005}.
The former method is typically used to place limits on the the presence of hosts in candidate FFP events \citep[e.g.,][]{mroz2020b} while the later was recently used by \citet{han2020} to identify deviations from a purely single-lens model and showed that an apparent FFP was actually bound to a star.
We will not explore this potential ambiguity in this paper, but refer to \citet{han2005, henderson2016} for a more thorough discussion.
However, we do note that this confusion between bound planets and true FFPs is an example of a degeneracy in the interpretation of candidate FFP microlensing events.

\subsection{Degeneracies in Microlensing Events}
\label{sec:degeneracies}

Microlensing events can be subject to degeneracies, i.e., when one event can be described by two or more models equally well within given photometric measurements.
There are two generic types of degeneracies: ``accidental'' degeneracies, in which the similarities between the models are not due to any fundamental underlying mathematical symmetry but rather just due to coincidence.  
Generally, such accidental degeneracies can be resolved with higher-precision photometric measurements or increased photometric monitoring. 
On the other hand, ``mathematical'' degeneracies exist because of some deeper underlying symmetries in the lens models.  
These symmetries typically appear in some extreme limits in one or more of the parameters that describe the model. 
For example, many of these degeneracies can be derived by expanding the lens equation\footnote{The lens equation describes the relation between the image positions created by the lens and the positions of the source and lenses.} in some small parameter.  
By keeping only the lowest order terms, the lens equation becomes degenerate with respect to that parameter or with the lens equation expanded in the same way with respect to another small parameter (see, e.g., \citealt{dominik99} for examples).
Mathematical degeneracies are more nefarious than accidental ones because the underlying models for the magnification can become nearly perfectly degenerate in some extreme limits, and thus such degeneracies cannot be resolved even with exquisite data. 

The nature of a degeneracy can be discrete or continuous.  
Discrete degeneracies occur when a finite number of models can be used to describe a given event. 
An infamous example is the $s \leftrightarrow s^{-1}$ degeneracy for low mass-ratio binary lenses, where $s$ is the instantaneous projected semi-major axis of the binary $a_\perp$ in units $\tsub{\theta}{E}$ \mbox{$s=(a_\perp/\tsub{D}{Lens})/\tsub{\theta}{E}$} \citep{griest1998,dominik99,yee2021}.
Here, $\tsub{\theta}{E}$ is the angular Einstein ring radius, given by
\begin{equation}
\label{eqn:thetae}
    \tsub{\theta}{E}=\sqrt{\kappa \tsub{M}{lens}\tsub{\pi}{rel}},
\end{equation}
where $\tsub{M}{lens}$ is the lens mass, \mbox{$\tsub{\pi}{rel}={\rm au}(1/\tsub{D}{Lens}-1/\tsub{D}{Source})$} is the lens-source relative parallax, and $\kappa=8.14$ mas $M_\odot^{-1}$ is a constant.

Continuous degeneracies occur for events in which a range of parameters can be used to model an event.
One example is the degeneracy between the impact parameter normalized to the Einstein ring radius $u_0$, the microlensing timescale $\tsub{t}{E}$, and the fraction of flux attributed to the source relative to the combined source and blend flux  $\tsub{f}{S}$.  
Here, the microlensing timescale $\tsub{t}{E}$ is given by 
\begin{equation}
    \label{eqn:te}
    \tsub{t}{E}=\frac{\tsub{\theta}{E}}{\tsub{\mu}{rel}}
\end{equation}
where $\tsub{\mu}{rel}$ is the relative lens-source proper motion, the impact parameter $u_0\equiv\theta_0/\tsub{\theta}{E}$ is the angular distance of closest approach between the lens and source on the sky $\theta_0$ in units of $\tsub{\theta}{E}$, and $\tsub{f}{S}=\tsub{F}{S}/(\tsub{F}{S}+\tsub{F}{B})$
where $\tsub{F}{S}$ and $\tsub{F}{B}$ are the source flux and any unresolved flux blended with the source flux, respectively. 
This degeneracy was first discussed in detail by \citet{wozniak1997}, and operates in two regimes: when $u_0\ll 1$ and when $u_0\gg 1$.
Another continuous degeneracy exists for bound planetary microlensing events between the lens mass ratio $q=\tsub{M}{planet}/\tsub{M}{host}$, the angular source size $\theta_*$ in units of $\tsub{\theta}{E}$
\begin{equation}
    \label{eqn:rho}
    \rho \equiv \frac{\theta_*}{\tsub{\theta}{E}},
\end{equation}
and $\tsub{\mu}{rel}$ for a subset of perturbations from a bound planet, described by \citet{gaudi1997}. 
Typically, the source star in a microlensing event can be well approximated as a point source. 
However, if there exists a significant second derivative of the magnification over the angular area covered by the source, the normalized angular size of the source star $\rho$ must be included in the model, i.e., the event exhibits finite source effects (FSEs). 
Both of these continuous degeneracies are similar in nature to the one we report here, but differ in detail.

\subsection{The Degeneracy for Extreme Finite Source Events}\label{sec:degefse}

Here we explore a continuous degeneracy between multiple parameters that emerges for isolated lenses with $\rho\gg1$,  which we refer to as extreme finite source effect (EFSE) events. 
Parts of this degeneracy have been identified in FFP candidate events (for which $\rho$ is typically $\gtrsim 1$) reported in \citet{mroz2018,mroz2019,mroz2020a,mroz2020b}.
Specifically in \citet{mroz2020a}, when fitting a lensing model to their event, they noted a strong correlation between four parameters $\rho$, $\tsub{t}{E}$, $u_0$, and $\tsub{f}{S}$.   
As we elucidate here, the correlations identified by the above authors are due to a continuous, mathematical degeneracy between the parameters that describe EFSE microlensing events.

As we will show, in the EFSE regime and assuming no limb-darkening, light curves can be characterized by three gross observables, but their complete models have four free parameters.
Namely, these observables are the peak flux above the baseline of the event $\Delta \tsub{F}{max}$ (which is approximately constant for events with $\rho\gg1$ when the center of the lens and source are separated by less than $\theta_*$), the full-width half-maximum (FWHM) duration of the event $\tsub{t}{FWHM}$, and the fraction of time spent in the wings/shoulders of an event $\tsub{f}{ws}$.
We note that there are two other observables, namely the baseline flux $\tsub{F}{base}=\tsub{F}{S}+\tsub{F}{B}$, which is only constraining when multi-band photometry is collected while the source is magnified, and $t_0$, which is time symmetric and is not a part of this degeneracy. 
Thus, in the absence of limb-darkening and for single-band photometry, four free parameters characterize the flux as a function of time $F(t)$ and there are only three observables, resulting in a degeneracy. 
These parameters can be analytically approximated to be pairwise degenerate.
The source size $\rho$ is degenerate with the source flux $\tsub{F}{S}$ in such a way that they can be varied in order to maintain the peak magnification of an event ($\Delta\tsub{F}{max}$). 
Similarly, the source star angular radius crossing time \begin{equation}
    t_*\equiv \frac{\theta_*}{\tsub{\mu}{rel}}
    \label{eqn:tstar}
\end{equation} and the impact parameter scaled to the angular size of the source star $b_{0}=\theta_0/\theta_*$ are degenerate in that they can be varied to maintain the observed duration ($\tsub{t}{FWHM}$) of an event. 

When limb-darkening is included however, the situation becomes more complex.
A fourth observable becomes measurable, which is the flux ratio of the event at its peak to when the lens is centered on the limb of the source $\tsub{f}{pl}$. 
For a linear limb-darkening profile with coefficient $\Gamma$, this introduces a fifth parameter for four observables, meaning a degeneracy remains. 
Furthermore, for non-zero $\Gamma$, $\Delta \tsub{F}{max}$ now also depends on $b_0$, which results in the coupling between the four previous parameters into a larger five-parameter degeneracy through the variation in the flux during the event due to limb-darkening. 

\subsection{Plan for this Paper}
\label{sec:plan}

In Section \ref{sec:psmodels} we review the mathematical models, parameters, and observables of single lens microlensing events in the $\rho\ll1$ regime.
In Section \ref{sec:efsenold}, we repeat this process and describe the morphology and observables of EFSE events without limb-darkening.
We reparameterize the canonical set of single lens parameters to a new set of parameters that are more closely tied to the observables.  
We then derive the degeneracy for a single lens event in the EFSE regime without considering limb-darkening.
We then add another observable and extend the degeneracy to include limb-darkening in Section \ref{sec:efseld} .
We explore these degeneracies qualitatively and quantitatively for fixed limb-darkening in Section \ref{sec:degenfixedld}, and explore the full degeneracy including limb-darkening as a free parameter in Section  \ref{sec:degenfull}.
We then discuss physical constraints on the severity of these mathematical degeneracies in Section \ref{sec:mathvsphysical}. Finally, we consider the implications of our findings and conclude in Section \ref{sec:discussconclude}. 

Throughout this paper, we calculate the magnifications using either the \citet{witt1994} or the \citet{lee2009} method as implemented in \texttt{MulensModel} \citep{poleski2019}.

\section{Single Lens Events without EFSE}
\label{sec:psmodels}

In this section we review the mathematical model, parameters, and observables for single lens events for which $\rho\ll1$ for context (Sections~\ref{subsec:singlelens} and \ref{subsec:paramstypical}).
We will then repeat this exercise for single lens events for which $\rho\gg1$ (i.,e., EFSE events), demonstrating that the canonical set of parameters for the former case are not optimal for the latter case.
We will therefore introduce a new set of parameters that are more appropriate for EFSE events.

\subsection{Single Lens Events with $\rho\ll 1$}
\label{subsec:singlelens}

For the majority of the single-lens microlensing events that have been observed to date, the size of the source star can be ignored, as its angular size is much smaller than the angular Einstein ring radius of the lens (i.e., $\rho\ll 1$) and the source does not pass within ${\sim}1$ source radius of the lens.
In this approximation, the point-source point-lens (PSPL) model magnification $\tsub{A}{ps}(t)$ is given by the standard formula \citep{paczynski1986}
\begin{equation}
    \tsub{A}{ps}(t) = \frac{u(t)^2+2}{u(t)\sqrt{u(t)^2+4}},
    \label{eqn:apspl}
\end{equation}
where $u(t)^2=u_0^2+\tsub{\tau}{E}^2(t)$, and $\tsub{\tau}{E}(t)\equiv (t-t_0)/\tsub{t}{E}$ is the time from the peak of the event in units of the microlensing timescale.
Note that, such events reach peak magnification at a time $t_0$ when $u(t)$ is at its minimum angular separation $u_0$.   

The magnification is not a direct observable; rather one measures the flux as a function of time
\begin{equation}
    F(t)=\tsub{F}{S} A(t) + \tsub{F}{B}.
    \label{eqn:fluxtime}
\end{equation}
Note that $\tsub{F}{B}$ can include flux from the lens, any flux from companions to either or both the lens and source, as well as flux from unrelated stars that are blended in the point spread function of the source.

The point-source magnification in Equation~\ref{eqn:apspl} diverges as $u\rightarrow{0}$ and in this regime can be approximated by $\tsub{A}{ps}(t)\simeq [u(t)]^{-1}$.  
When $u\gg1$, the magnification can be approximated as $\tsub{A}{ps}(t) \simeq 1+2[u(t)]^{-4}$.  
As mentioned previously, \citet{wozniak1997} demonstrated that in these two limits, there is a continuous mathematical degeneracy between $u_0$, $\tsub{t}{E}$, and $\tsub{f}{S}$.

\subsection{Summary of the Parameters for $\rho\ll1$}
\label{subsec:paramstypical}

A PSPL microlensing event can be described by five parameters when FSEs are negligible: ($u_0, t_0, \tsub{t}{E}, \tsub{F}{S}, \tsub{F}{B}$).  
To the extent that there are only four gross observables, namely $t_0$, the baseline flux $\tsub{F}{base}$, the difference between the peak flux and the baseline flux $\Delta \tsub{F}{max}$, and some measure of the characteristic timescale of the event (such as full-width at half-maximum), it is clear that $u_0,~\tsub{t}{E}$, and $\tsub{f}{S}$ cannot be uniquely determined and there is a continuous degeneracy in these parameters.  
This is the basic underlying cause of the degeneracy described by \cite{wozniak1997}.  
In reality, when one is not deep in the limits noted above by \cite{wozniak1997}, there are additional observables related to the detailed shape of the microlensing light curve that allow one to break this degeneracy with sufficiently good light curve coverage and photometric precision. 

The magnification of a finite source begins to deviate from Equation \ref{eqn:apspl} when $\rho \gtrsim 2 u_0$ \citep{liebes1964,gould1997}.
In this case, one must include the additional parameter $\rho$, as well as the limb-darkening profile of the source, which is commonly described as a linear limb-darkening profile with a coefficient $\Gamma$ \citep{yoo2004}. 
Thus, point-lens events can be described by seven parameters when FSEs are significant: $u_0, ~t_0, ~\tsub{t}{E}, ~\tsub{F}{S},~ \tsub{F}{B},~\rho$, and $\Gamma$.
For completeness, we note that for stars in the bulge, and typical lenses with masses in the brown dwarf, stellar, or remnant regimes, $\rho\ll1$.
Thus, FSEs only begin to manifest as deviations from the point source magnification (Equation~\ref{eqn:apspl}) in high-magnification events when the source approaches within a few stellar radii of the lens, i.e.\ events for which $u_0 \lesssim  \rho/2$.  
These events are relatively rare.  
In this case, the majority of the light curve is well-approximated by the point source assumption, except for a deviation within a few source crossing times $t_*$ of $t_0$.  
This deviation takes the form of a ``rounding'' of the peak of the microlensing event.  
Because the deviations due to FSEs are localized to a small time window near the peak, the parameters $\rho$ and $\Gamma$ do not participate in the single lens degeneracy identified in \citet{wozniak1997}. 

\section{EFSE Events without Limb-darkening}
\label{sec:efsenold}

We next consider the opposite extreme when $\rho\gg1$ (EFSE events), which are typically caused by low-mass lenses.  
In these cases, FSEs become important for trajectories in which the center of the lens passes within $\theta_*$ of the center of the source. 
Indeed, there is no significant magnification if this condition is not met.  
As a result, the basic event morphology for EFSE events changes dramatically~\citep[e.g.,][]{gould1997, agol2003}.  
This has implications for the kind of information that can be extracted from such events. 

\begin{figure*}[t]
\plottwo{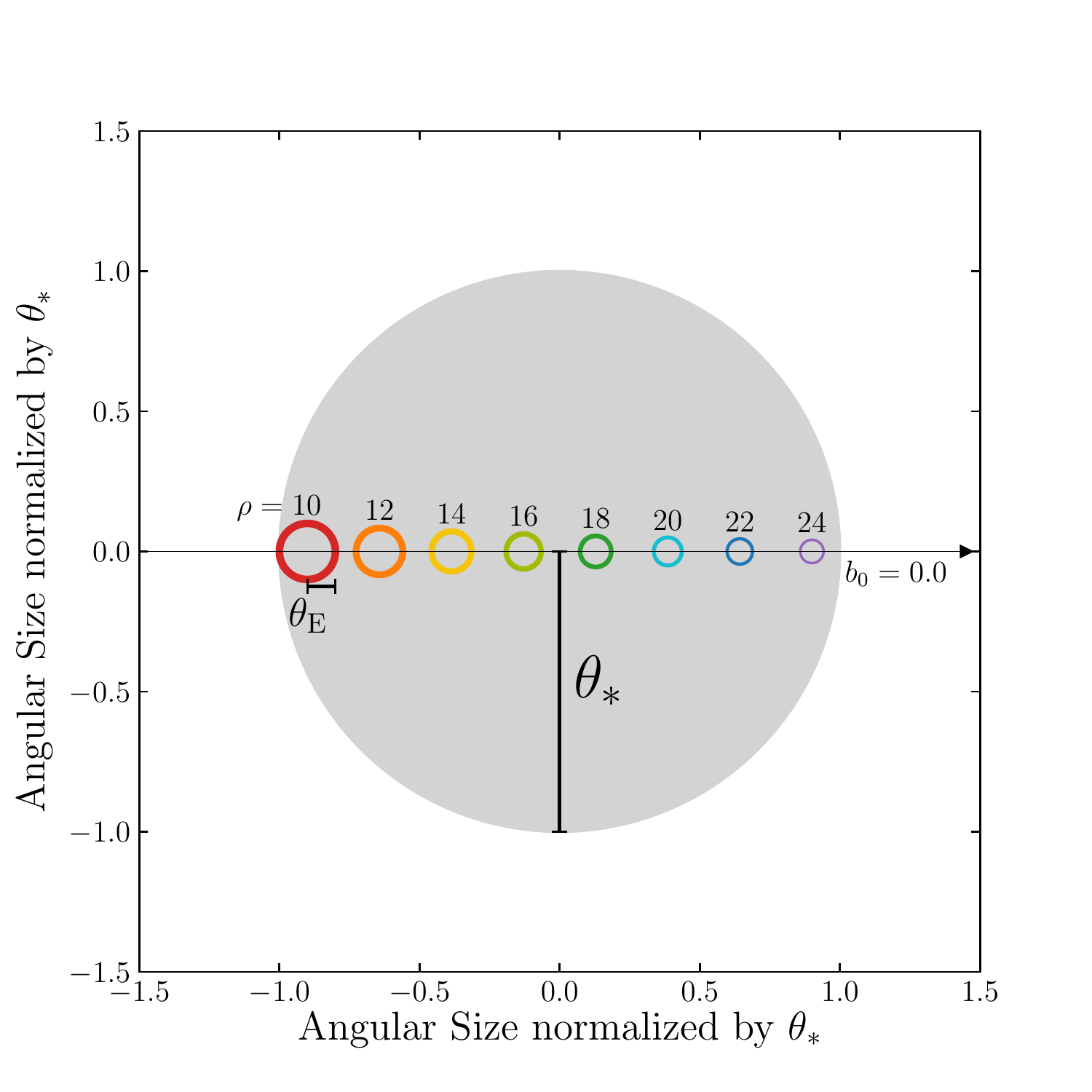}{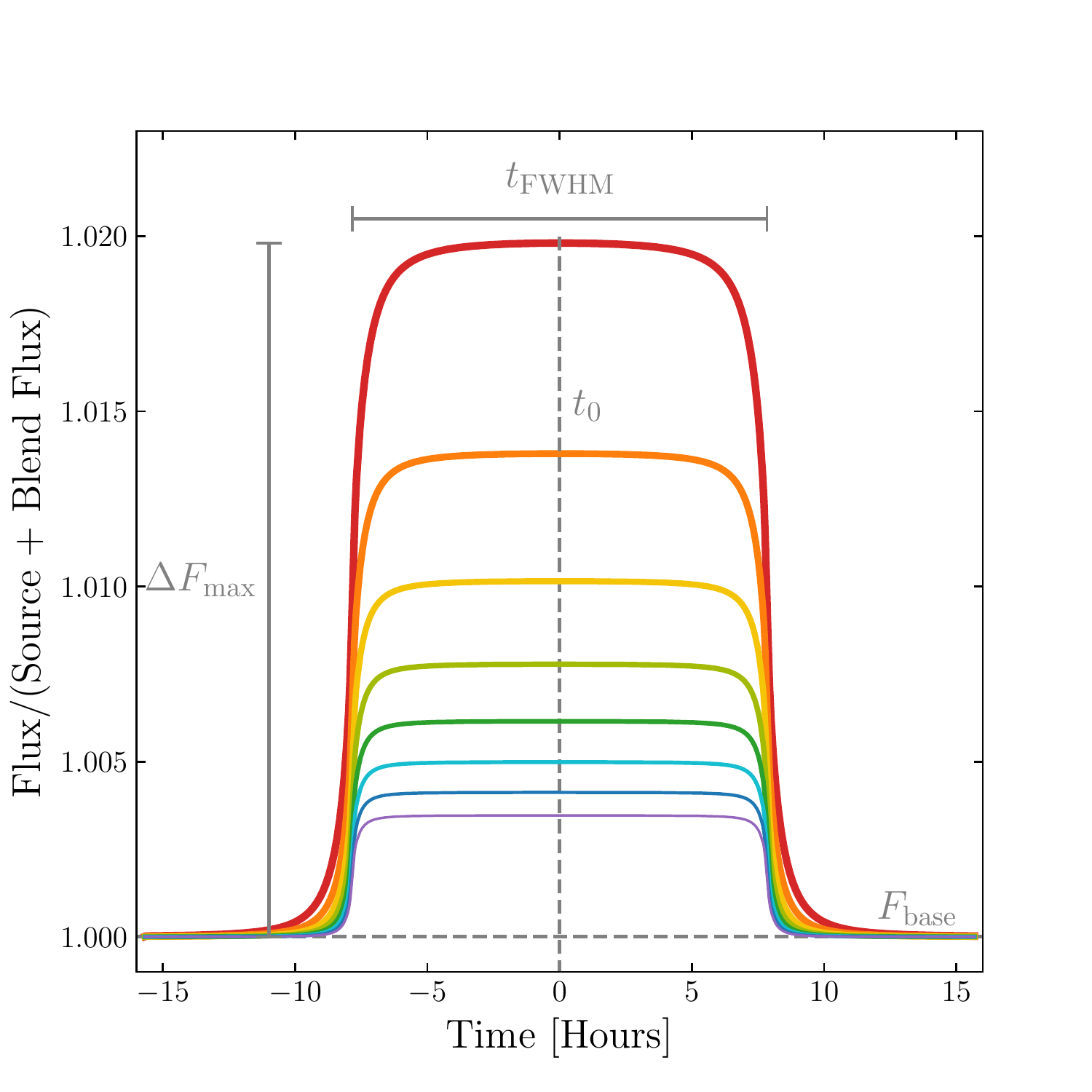}
\caption{A demonstration of changing $\rho$ for a fixed angular source size $\theta_*$, impact parameter $b_0$, and relative proper motion $\tsub{\mu}{rel}$.
\textit{Left:} The geometry of eight events with $\rho=(10,12,...,24)$, each represented by a separate circle the size of the Einstein ring radius $\tsub{\theta}{E}$. 
Each event has a trajectory that crosses the center of the source $b_0=0.0$, where the to-scale size of the source is the filled gray circle.
\textit{Right:} The resulting magnification curves for the eight Einstein rings (with corresponding line color and line thickness) in the Left panel. 
We label four observables for these events: $t_0$, $\Delta\tsub{F}{max}$, $\tsub{t}{FWHM}$, and $\tsub{F}{base}$.
As $\rho$ increases (and $\tsub{\theta}{E}$ decreases), the peak change in flux $\Delta \tsub{F}{max}$ decreases but $\tsub{t}{FWHM}$ remains roughly constant. 
%$\Delta \tsub{F}{max}$ of the $\rho=10$ event could be decreased to become equal to any of the events with $\rho>10$ by increasing the amount of blending.
\label{fig:rho_demo}}
\end{figure*}

The magnification of a uniform, finite source can be found by integrating the point source magnification of a lens across the area of the source star $\tsub{\cal A}{source}$,
\begin{equation}
    \tsub{A}{fs}(u) =\frac{1}{\pi\rho^2} \int_{\tsub{\cal A}{source}} \tsub{A}{ps}d{\cal A}
\end{equation}
\citep[e.g.,][]{gould1994,lee2009}.
As $\rho\to\infty$ and assuming no limb-darkening of the source, the magnification curve will take on a ``top hat'' or boxcar shape. 
The maximum magnification is essentially constant regardless of the angular separation between the lens relative to the source, provided that the lens is not near the edge or outside of the source.
This becomes apparent when considering that $\tsub{A}{fs}$ can be approximated as
\begin{equation}
    \tsub{A}{fs}\approx1+\frac{2}{\rho^2}+\mathcal{O}\left(\frac{1}{\rho^4}\right)
    \label{eqn:fse_peak}
\end{equation}
\citep{liebes1964,gould1997,agol2003}.
In this case, the flux of the event as a function of time is 
\begin{equation}
F(t) = \tsub{F}{S} \tsub{A}{fs}[b(t)] +\tsub{F}{B},
\label{eqn:simpleflux}
\end{equation}
where (keeping terms to second order in $1/\rho$)
\begin{equation}
A[b(t)]\approx1+\frac{2}{\rho^2}H[1-b(t)],
\end{equation}
$H(x)$ is the Heaviside step function, and $b(t)=\theta(t)/\theta_*$ is the angular separation $\theta$ between the center of the source and lens in units of the angular size of the source star $\theta_*$. 
This is explicitly given by 
\begin{equation}
b(t;b_{0},t_0,t_*) = \sqrt{b_{0}^2 + \tau_*^2}.
\end{equation}
Here we define $\tau_*(t)\equiv (t-t_0)/t_*$ for the source radius crossing time $t_*$ (Equation~\ref{eqn:tstar}), and the minimum source-lens angular separation in units of $\theta_*$ is  $b_{0}\equiv\theta_{0}/\theta_*$. 
We note that this can also be written as $b_0=u_{0}/\rho=(\theta_{0}/\tsub{\theta}{E})(\tsub{\theta}{E}/\theta_*)$.

Note that in this paper we will adopt the convention
\begin{eqnarray}
H(x)=\begin{cases} 0, &x < 0 \\
1, &x \ge 0, \end{cases}
\label{eqn:Hofxconv1}
\end{eqnarray}
such that $A(b)$ is defined and equal to $1+2\rho^{-2}$ when $b=1$ (e.g., when the lens is centered on the limb of the source).  
This convention is appropriate for $\rho\rightarrow \infty$.
However, when $\rho$ is large but finite, the magnification at $b=1$ is approximately $A\simeq 1+\rho^{-2}$ because the disk of the source fills roughly half of Einstein ring of the lens.
This latter situation corresponds to the half-maximum convention for the Heaviside step function
\begin{eqnarray}
H(x)=\begin{cases} 0, &x < 0 \\
\frac{1}{2}, &x=1\\
1, &x > 1. \end{cases}
\label{eqn:Hofxconv2}
\end{eqnarray}
Although our mathematical formalism could be derived using the half-maximum convention, we found that this did not lead to any qualitatively new insights but did obfuscate some of the points we make below.

We demonstrate the impact of $\rho$ on EFSE events in Figure \ref{fig:rho_demo}.
Figure \ref{fig:rho_demo} shows the morphology of eight magnification curves with fixed $b_0=0$ and various $\rho$ values.
The left panel shows a to-scale depiction of the geometry of these events with different values of $\tsub{\theta}{E}$ such that $\rho=(10,12,14,...,24)$.
In each case, all other parameters are held constant, the flux is normalized to that of the baseline, and no blending is included. 
The resulting light curves in the right panel of Figure \ref{fig:rho_demo} shows that for smaller values of $\rho$, $\Delta\tsub{F}{max}$ is larger.
As $\rho$ increases, the shape of the light curve becomes boxier and is more akin to a true top hat shape. 

\begin{figure*}
\plottwo{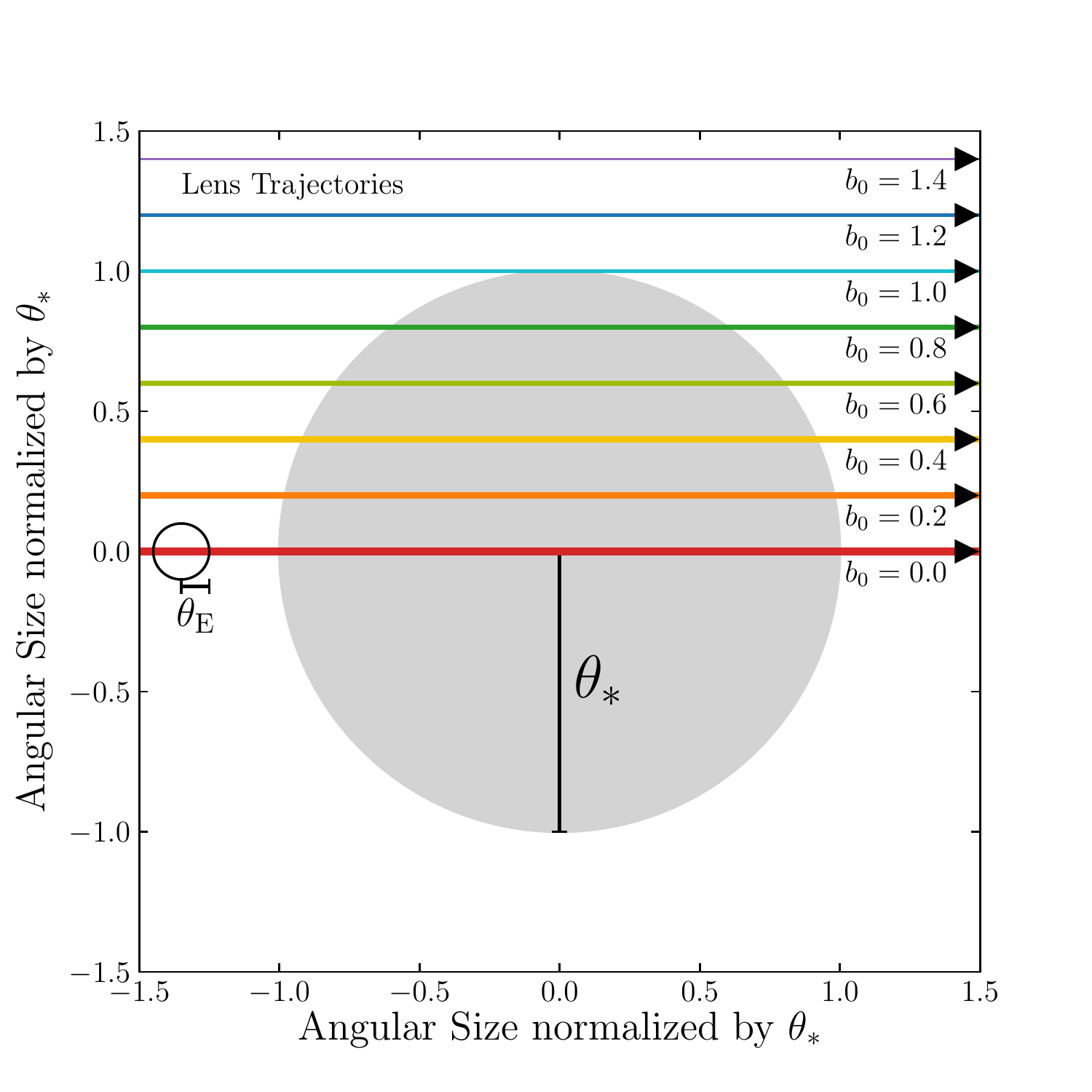}{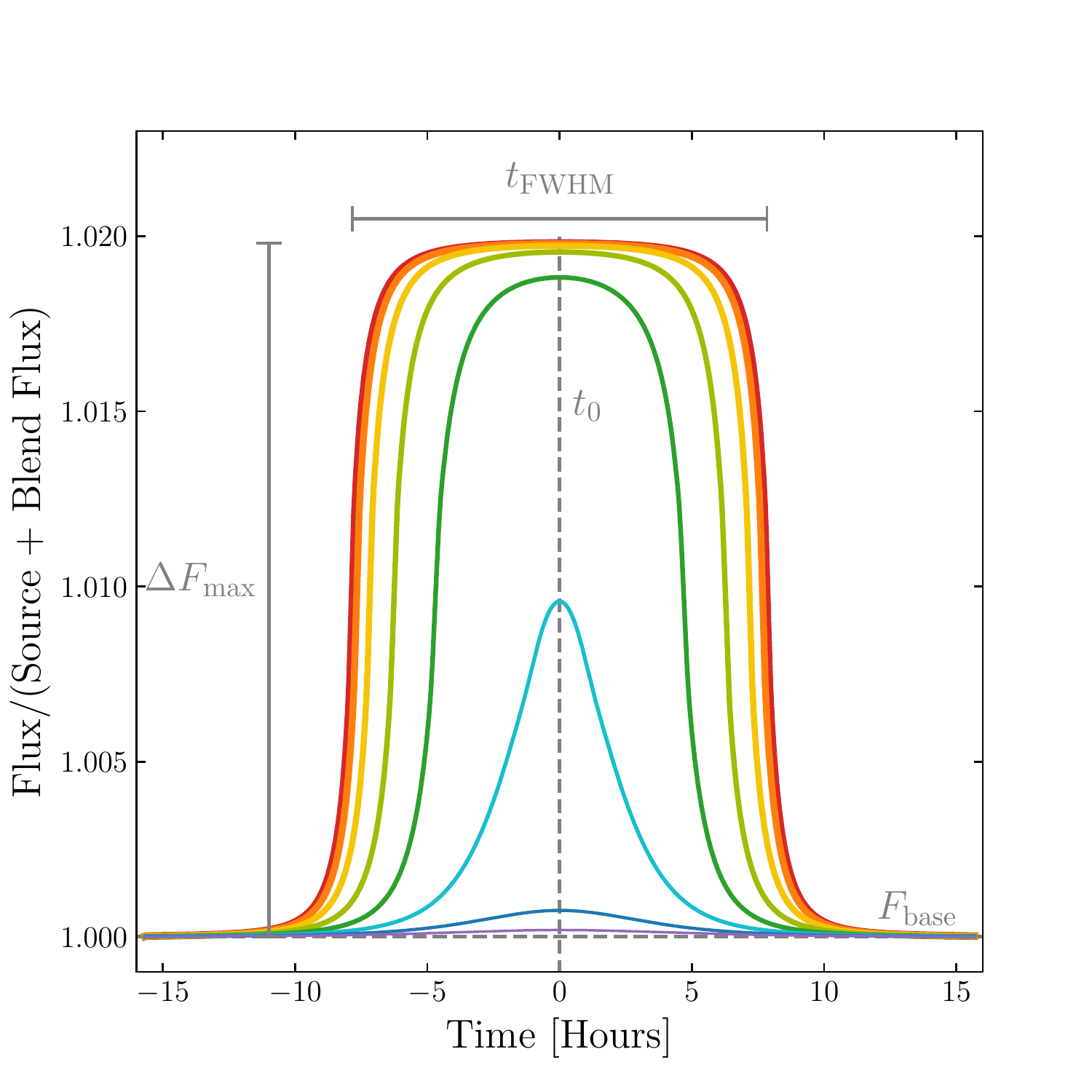}
\caption{A demonstration of changing the impact parameter $b_0$ for a fixed angular source size $\theta_*$, angular Einstein ring radius $\tsub{\theta}{E}$, and relative proper motion $\tsub{\mu}{rel}$.
\textit{Left:} The geometry of eight events with impact parameters $b_0=(0.0,0.2,...,1.4)$, each represented by a separate lens trajectory. 
Each event has the same angular Einstein ring radius (black circle) with $\rho=10$, and the to-scale size of the source is the filled gray circle.
\textit{Right:} The resulting magnification curves from the varying impact parameters (with corresponding line color and line thickness) in the left panel. 
The highest magnification/longest event corresponds to the $b_0=0.0$ event.
As $b_0$ increases, events become shorter and have progressively lower peak magnifications.
Note that for sufficiently large impact parameters ($b_0\gtrsim0.8$), the top-hat shape disappears and eventually no magnification occurs. 
\label{fig:b0_demo}
}
\end{figure*}

Generally, the baseline flux $\tsub{F}{base}$ is a direct (and typically precisely-measured) observable based on observations well before and after the microlensing event. 
Thus, a parameterization of $F(t)$ that is more directly related to the observables is $\tsub{F}{base}$ plus the difference flux $\Delta F$, which is given by
\begin{eqnarray}
\label{eqn:dfexpannold}
    \Delta F (t)
    &\equiv& F(t)-F_{\rm base}\nonumber\\
    &=& \tsub{F}{S} [A(t)-1]\nonumber\\
    &\simeq&\frac{2\tsub{F}{S}}{\rho^2}H(1-b(t)).
\end{eqnarray}  

Next we consider the duration of EFSE microlensing events.
The source crossing time $t_*$ is generally not the actual duration of the event.  
The observed duration of an EFSE event is rather $\tsub{t}{FWHM}$ which is well-approximated by twice the source half-chord crossing time
\begin{equation}
    \tsub{t}{c} =\frac{\theta_*}{\tsub{\mu}{rel}}\sqrt{1-\left(\frac{u_{0}}{\rho}\right)^2}=t_*\sqrt{1-b_{0}^2}=\beta t_*,
    \label{eqn:tchord}
\end{equation}
where we have defined
\begin{equation}
\beta\equiv\sqrt{1-b_0^2}
\end{equation}
for convenience \citep[see, e.g.,][]{agol2003,mroz2017}.
In this approximation for the event duration, there is no dependence on $\tsub{\theta}{E}$ and thus the mass of the lens.
With this and the approximation for $\Delta F$ in Equation \ref{eqn:dfexpannold}, one would naively believe the duration and flux of an EFSE event to be completely decoupled.
However, this will turn out to not be strictly true.

To demonstrate the result of changing impact parameter on $\tsub{t}{FWHM}$ and the event morphology, Figure \ref{fig:b0_demo} shows a set of eight light curves with fixed $\rho=10$ and increasing values of $b_0$.
In the left panel of Figure \ref{fig:b0_demo}, we include a to-scale depiction of the geometry of these events with impact parameters from $b_0=(0.0,0.2,0.4,...,1.4)$.
In each case, all other parameters are held constant, the flux is normalized to that of the baseline, and no blending is included. 
The resulting light curves in the right panel of Figure \ref{fig:b0_demo} show events with decreasing $\tsub{t}{FWHM}$ as $b_0$ increases, with an eventual departure from the top-hat shape for $b_0\gtrsim0.8$.
Also note that as the impact parameter approaches the limb of the source, $\Delta\tsub{F}{max}$ progressively decreases.

There are several points to note.  
First, in the EFSE regime the magnification of the event is approximated as constant and depends only on $\rho$, making the change in flux during the event constant and proportional to $2\tsub{F}{S}/\rho^2$. 
Thus, when $\rho\to\infty$, the characteristic timescale of the event becomes independent of $\tsub{\theta}{E}$, and the magnification is independent of the duration of the deviation and depends solely on $\rho$. 
Furthermore, because the duration of the deviation is independent of $\tsub{\theta}{E}$, it is also independent of the lens mass. 
Rather, the only observable parameter that depends on the lens mass is the amplitude of the deviation, as it depends on $\rho$, which in turn depends on $\tsub{\theta}{E}$.
However, as we discuss next, when $\rho$ is large but finite, this independence is only approximate. 

We now consider the deviations from the top-hat morphology for finite $\rho$.  
For a uniform source but finite $\rho$, the finite size of the angular Einstein ring radius compared to the angular radius of the source is not negligible. 
As a result, the morphology of the event deviates from the strict top-hat shape; in particular the event exhibits ``wings'' of magnification just before the lens enters the source and just after the lens exits the source, and ``shoulders'' just after the lens enters the source and just before it exits the source (see the $\rho=10$ lightcurve in Figure \ref{fig:rho_demo} as an example).  
The characteristic time $\tsub{t}{ws}$ of each tail or shoulder deviation from the top-hat form is simply half the time between first and second contact or third and fourth contact of the angular Einstein ring and the source limb,
\begin{equation}
\tsub{t}{ws}\equiv \frac{\tsub{t}{E}}{\beta}.
\label{eqn:tts}
\end{equation}
This corresponds to a fraction of the primary event duration of
\begin{equation}
\tsub{f}{ws}\equiv\frac{\tsub{t}{ws}}{\tsub{t}{c}} = (\beta^2\rho)^{-1}
\label{eqn:fws}
\end{equation}
Thus these wings and shoulders increase in duration relative to the total event duration with decreasing $\beta$ (i.e., events with larger impact parameters $b_0$) and decrease in duration relative to the total event duration with increasing $\rho$ \citep{agol2003}. 

The impacts of $\rho$ and $\beta$ on $\tsub{f}{ws}$ are apparent in Figures \ref{fig:rho_demo} and \ref{fig:b0_demo}.
The right panel of Figure \ref{fig:rho_demo} shows that events with smaller values of $\rho$ have more prominent wings and shoulders for those events (Equation \ref{eqn:fws}).
As $\rho$ increases, $\Delta\tsub{F}{max}$ decreases and the wing and shoulder features become less prominent as $\tsub{f}{ws}$ decreases.
Thus, as $\rho$ increases the light curves are better approximated by a top-hat shape.
In Figure \ref{fig:b0_demo} as $b_0$ increases (and $\beta$ decreases), the wing and shoulder features become more prominent as $\tsub{f}{ws}$ increases following Equation \ref{eqn:fws}.
The shared dependence of the observable $\tsub{f}{ws}$ on $\rho$ and $\beta$ links the two sets of parameters that control the amplitude and duration of EFSE events in the derivation of this EFSE degeneracy.

\subsection{The Degeneracy with No Limb-Darkening}
\label{subsec:degenmathnold}

We now derive the degeneracy for EFSE events for a source without limb-darkening.
As discussed previously, in the case of no limb-darkening and $\rho\to\infty$ there are two gross observables, namely $\Delta \tsub{F}{max}$ (which is roughly constant during the event), and $\tsub{t}{FWHM}$. 
Furthermore, these two observables are decoupled under these assumptions.  
We therefore first consider the flux degeneracy and duration degeneracy separately. 

The maximum difference flux when $\Gamma=0$ is simply
\begin{equation}
    \Delta \tsub{F}{max} = \frac{2\tsub{F}{S}}{\rho^2},\qquad (\Gamma=0).
     \label{eqn:dfmaxnold}
\end{equation}
It is straightforward to verify that substituting the parameters
\begin{equation}
    \tsub{F}{S}'=\zeta \tsub{F}{S},\qquad \rho'=\zeta^{1/2}\rho
    \label{eqn:fsrhotransform}
\end{equation}
into Equation \ref{eqn:dfmaxnold}, we recover the same observable difference flux (e.g., $\Delta \tsub{F}{max}'=\Delta \tsub{F}{max}$).
Therefore, the difference flux is constant under the transformations $\tsub{F}{S}' \rightarrow \zeta \tsub{F}{S}$ and $\rho' \rightarrow \zeta^{1/2}\rho$ for any arbitrary positive constant $\zeta$. 
Thus in the limit of $\rho\to\infty$, there is a perfect degeneracy between $\tsub{F}{S}$ and $\rho$ such that
\begin{equation}
    \tsub{F}{S}' = \tsub{F}{S}\left(\frac{\rho'}{\rho}\right)^2.
    \label{eqn:flux_degen}
\end{equation}
It is also trivial to show that by dividing both sides of Equation \ref{eqn:flux_degen} by $\tsub{F}{base}$ that the relationship holds for two blending parameters $\tsub{f}{S}$ and $\tsub{f}{S}'$.

Now consider the duration of the event as parameterized by the source half-chord crossing time, which is related to the model parameters by $t_c=\beta t_*$ (Equation~\ref{eqn:tchord}). 
It is trivial to verify that substituting the following parameters 
\begin{equation}
    \beta'=\xi \beta,\qquad t_*'=\xi^{-1}t_*
\end{equation}
into Equation \ref{eqn:tchord} will result in an equal source chord crossing time ($\tsub{t}{c}'=\tsub{t}{c}$) for any arbitrary positive constant $\xi$ satisfying $0\le\xi \beta\le1$.
Thus, in the limit of $\rho\to\infty$, there is a perfect degeneracy between $b_0$ and $t_*$
such that 
\begin{equation}
    t_*' = t_*\frac{\beta}{\beta'}=t_*\frac{\sqrt{1-b_0^2}}{\sqrt{1-b_0'^2}}.
    \label{eqn:time_degen}
\end{equation}
Because the duration of the event is decoupled from the flux during the event, $\xi$ does not need to be equal to $\zeta$ (and, in general, will not be). 

The above mathematical degeneracies are only strictly valid in the limit $\rho \rightarrow \infty$, or equivalently in the limit that $\tsub{t}{E}/t_* \rightarrow 0$.  
As discussed previously, for values of $\rho$ that are large but finite, EFSE events deviate from the strict top-hat morphology.  
In particular and as illustrated in Figures~\ref{fig:rho_demo} and \ref{fig:b0_demo}, events with finite $\rho$ exhibit wings and shoulders near the source limb crossing point of the event.  
The duration of these features relative to the half-chord crossing time is equal to $\tsub{f}{ws}=(\beta^2\rho)^{-1}$ (see Equations~\ref{eqn:tts}, \ref{eqn:fpl_simp}, and the surrounding discussion). 
Effectively, $\tsub{f}{ws}$ provides another observable parameter when $\rho$ is large but finite.  

The fact that the duration of the wings and shoulders relative to $\tsub{t}{c}$ depends on both $\beta$ and $\rho$ has two important implications.  
First, the conclusion that $\tsub{F}{S}$ and $\rho$ have mathematically equivalent effects on the peak flux and morphology of EFSE events provided that they satisfy Equation \ref{eqn:flux_degen} is not strictly true. 
While varying $\tsub{F}{S}$ does not change the morphology of EFSE events, varying $\rho$ does as $\tsub{f}{ws}$ is a function of $\rho$, but not a function of $\tsub{F}{S}$.  
This is illustrated in Figure~\ref{fig:rho_demo}, which shows how as $\rho$ increases, the duration of the wings and shoulders relative to the total duration of EFSE events decreases, and the morphology of the light curves become increasingly well-approximated by a strict top hat.  

Second, because $\tsub{f}{ws}$ depends on $\beta$ and thus the impact parameter $b_0$, the morphology of events with large but finite $\rho$ is not strictly independent of $b_0$.  
In particular, events with larger impact parameter (smaller $\beta$) exhibit more pronounced wings and shoulders.  
Thus, for EFSE events with large but finite values of $\rho$, the morphology of the event depends on $b_0$, and thus the flux during the event is coupled to the duration of the event. 
In the specific case when $\Gamma=0$ and $\rho$ is large but finite, there are four parameters ($\tsub{F}{S}, \rho, \beta, t_*$) and three observables ($\Delta\tsub{F}{max}, \tsub{t}{c}, \tsub{f}{ws}$). 
Given the definitions of the observables, and assuming that the morphology of the wings and shoulders is directly proportional to their fractional duration $\tsub{f}{ws}$, it is straightforward to show that there is a continuous mathematical one-parameter degeneracy.  
Specifically, by substituting the following parameters
\begin{equation}
    \tsub{F}{S}'=\eta \tsub{F}{S},\,\,\, \rho'=\eta^{1/2}\rho,\,\,\, \beta'=\eta^{-1/4}\beta,\,\,\, t_*'=\eta^{1/4}t_*
    \label{eqn:degennoldfiniterho}
\end{equation}
into Equations~\ref{eqn:tchord}, \ref{eqn:dfmaxnold}, and \ref{eqn:fws} we recover the same values for the observables (i.e., $\Delta\tsub{F}{max}'=\Delta\tsub{F}{max}, \tsub{t}{c}'=\tsub{t}{c}, \tsub{f}{ws}'=\tsub{f}{ws}$).
Here $\eta$ is any arbitrary positive constant satisfying $0\le\eta^{-1/4}\beta\le 1$.

\begin{figure}[t]
\plotone{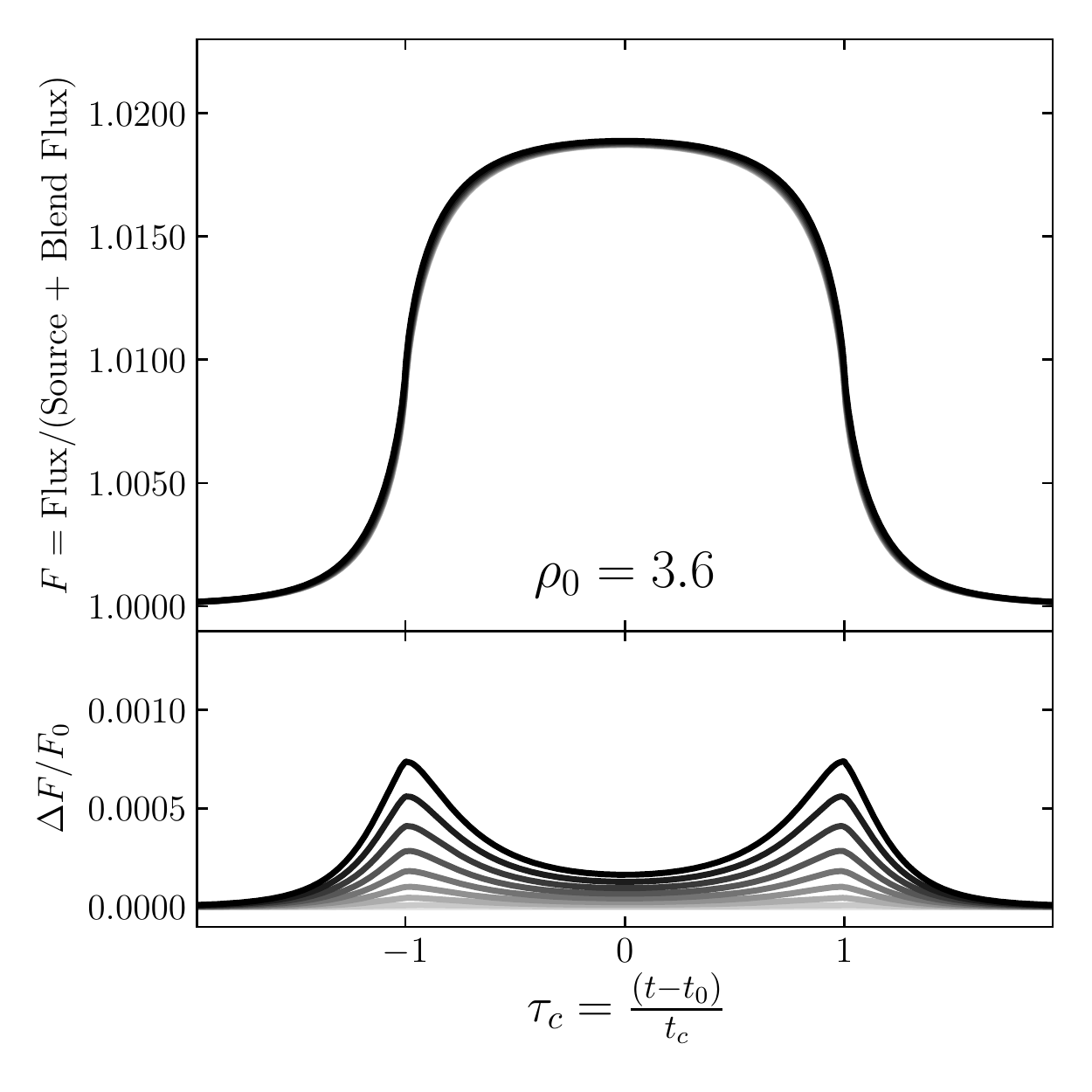}
\plotone{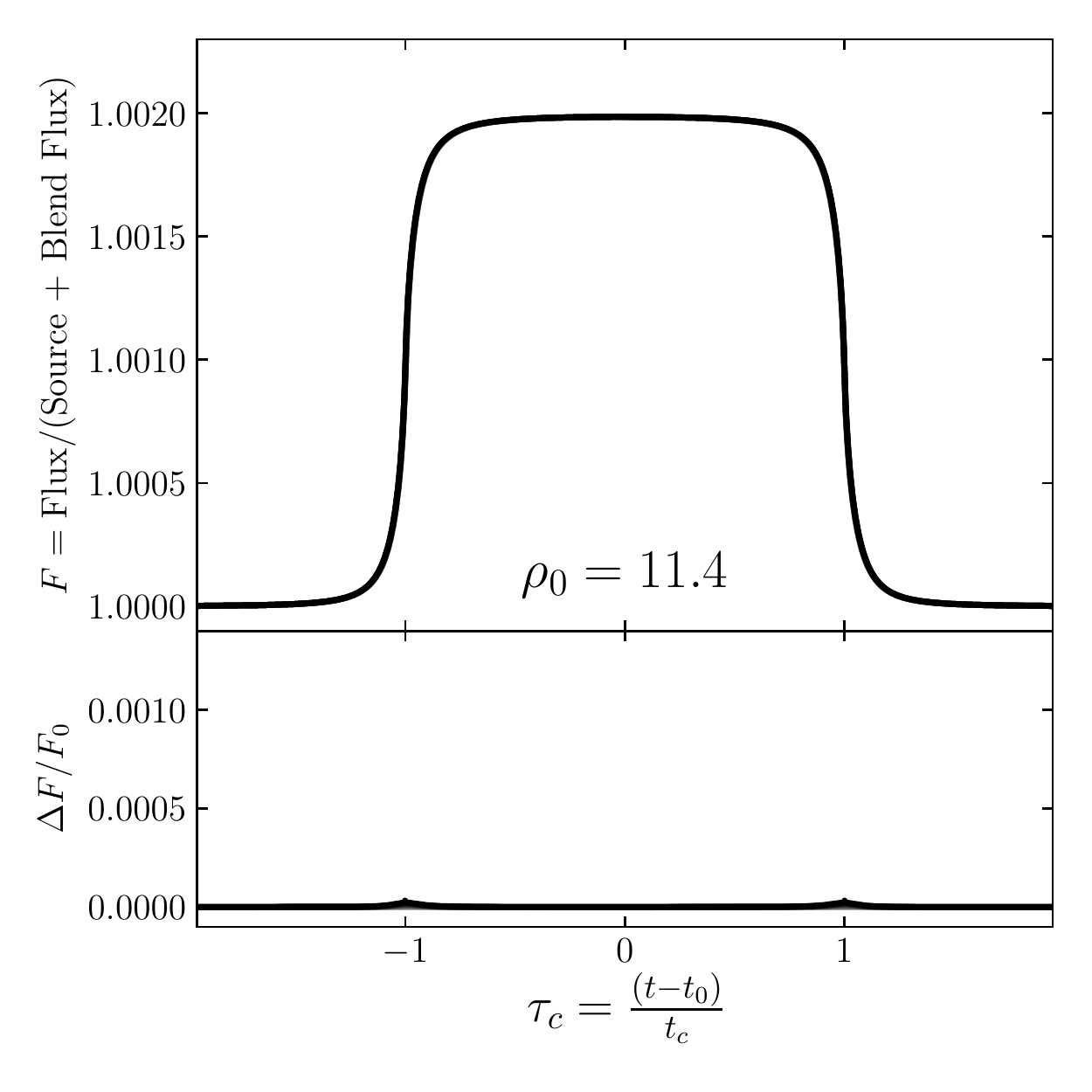}
\caption{Demonstration of the degeneracy for no limb-darkening with a fiducial value of $\rho=3.6$ (top panel) and $\rho=11.4$ (bottom panel). 
Both panels show nine different light curves that serve to demonstrate the degeneracy between the parameters $\tsub{F}{S}, \rho, \beta$ and $t_*$ for $\Gamma=0$ and $\rho\gg 1$ but finite.  
We choose fiducial values for the parameters, and for each of nine values of $b_0$ uniformly spaced between 0.0 and 0.8 we use the transformation in Equation~\ref{eqn:degennoldfiniterho} to scale the event parameters and plot the resulting light curves in the top subpanels.
Here, the line shadings go from lightest to darkest for increasing values of $b_0$.
The bottom subpanels show the fractional residuals, where $F_0$ is the flux from the event with $b_0=0.0$ and $\Delta F=F-F_0$.}
\label{fig:degennoldfiniterho}
\end{figure}

We illustrate the mathematical severity of the degeneracy between these four variables in Figure~\ref{fig:degennoldfiniterho}. 
We use a fiducial values of $b_0=0.0$, $\tsub{\mu}{rel}=6.5$ mas yr$^{-1}$, and $\tsub{f}{S}\simeq0.13$ but use $\rho=3.6$ in the top panel and $\rho=11.4$ in the bottom panel.
We then scale the other parameters by using nine uniformly spaced values of $b_0$ from 0.0 to 0.8, transforming these to values of $\beta$, and calculating $\eta$ using Equation~\ref{eqn:degennoldfiniterho}.
In the upper subpanels we show the resulting light curves and in the lower subpanels their relative residuals compared to the $b_0=0.0$ event which is the lightest gray and backmost light curve.
In both the top and bottom panels, the shading of the light curves and their residuals go from lightest to darkest for increasing values of $b_0$.
We see that the light curves are nearly perfectly degenerate, and in particular the $\rho=11.4$ case are noticeably more degenerate than those for the $\rho=3.6$ case.
The prominent departures in the $\rho=3.6$ case are due to the difference in the wing and shoulder shapes of the events due to detailed differences in the light curve caused by increasing $\rho$ or $\beta$ (centered on $\tau_c=\pm1$).
Furthermore, there is a ``trough'' between the two peaks in the residuals that results from the slight decrease in flux as the impact parameter approaches the limb of the source star.
However, these residuals decrease significantly in the $\rho=11.6$ case. 
From this we conclude that for $\rho\ga 10$, the morphology of these wings and shoulders do scale approximately with $\tsub{f}{ws}$ and the mathematical degeneracy in Equation~\ref{eqn:degennoldfiniterho} is perfectly realized in the limit $\rho\rightarrow \infty$.

Although the wings and shoulders of EFSE events due to finite values of $\rho$ indeed provide a formal constraint on $\rho$ and $\beta$, the magnitude of the differences in the light curves for impact parameters in the range $b_0=0-0.8$ is extremely small, as can be seen in the lower panels of Figure~\ref{fig:degennoldfiniterho}.  
For the $\rho=3.6$ case, the full range of the deviation of the light curves relative to the fiducial ($b_0=0$) case is $\la 1\times 10^{-3}$, whereas for the $\rho=11.4$ case it is $\la 5\times 10^{-5}$.
In both cases, these deviations are a factor of $\sim 40$ times smaller than the magnitude of the event itself, and thus if the event is detected with a signal-to-noise ratio of ${\rm S/N}=X$, the deviations will only be detectable at ${\rm S/N}\la X/40$. 
Thus, even for events that are detected at very high ${\rm S/N}$, the deviations will be essentially undetectable, and the mathematical degeneracy in Equation~\ref{eqn:degennoldfiniterho} will hold.

However, we note that Figure \ref{fig:degennoldfiniterho} may make the degeneracy appear more pernicious than is likely to be realized in practice.  
In particular, and as we discuss in more detail in Section \ref{sec:mathvsphysical}, when determining how deleterious these mathematical degeneracies will be for actual detected events one must also consider the expected prior distributions for the underlying parameters.  
For example, for microlensing events with giant sources the fraction of the baseline flux due to the source is known to be bimodal \citep{mroz2020a} such that the source flux is either very close to the baseline flux (i.e., the blend flux is very small) or the source flux is much smaller than the baseline flux (i.e., the majority of the baseline flux is due to the blend).  
This is simply because the luminosity function of stars in the bulge has a local minimum between giants and the main sequence turn-off. 
As a result, EFSE events with giant sources will generally have $\tsub{F}{S}\simeq \tsub{F}{base}$ as strongly blended events will be difficult or impossible to detect.  
This is important because events with $\tsub{F}{S}\simeq \tsub{F}{base}$ and $\beta\simeq 1$ ($b_0\simeq 0$) are not strongly subject to the degeneracy in Equation~\ref{eqn:degennoldfiniterho}, since the source flux is bounded such that $\tsub{F}{S}/\tsub{F}{base} \la 1$ and the impact parameter is bounded such that $b_0\ge 0$ ($\beta\le 1$), implying that there is a narrow range of $\eta$ that can satisfy Equation~\ref{eqn:degennoldfiniterho} which shrinks to nearly zero as $\beta\rightarrow 1$ and $\tsub{F}{S}/\tsub{F}{base} \rightarrow 1$.  
Furthermore, events with larger values of $b_0$ (smaller values of $\beta$) are less likely to be detected because they have a shorter duration and smaller peak difference flux (all else being equal).  
Finally, even when the degeneracy is realized, the posterior distribution of, e.g., $\tsub{F}{S}$, will be narrower than the full allowed range.  
This is because events occur (although are not necessarily detected) with uniform values of $b_0$, which leads to a distribution of $\eta$ that is not uniform, and in particular is weighted toward smaller values of $\eta$.  
This also implies that the distributions of $\tsub{F}{S}$, $\rho$, $\beta$, and $t_*$ are not uniform, and in particular are weighted toward smaller, smaller, larger, and smaller values, respectively.  

Another example of prior information that is important to note is that we placed no upper limit on the value of $\tsub{f}{S}$, therefore allowing $\tsub{F}{S}>\tsub{F}{base}$. 
This is commonly known as negative blending, as $\tsub{F}{B}$ must be less than zero for $\tsub{F}{S}=\tsub{F}{base}-\tsub{F}{B}>\tsub{F}{base}$ \citep[e.g.,][]{smith2007}.  
While seemingly unphysical, in crowded fields such as those toward that Galactic bulge that are typically monitored by microlensing surveys, it is possible to have some negative blending if the source happens to be located in a local minimum in the background which is typically dominated by an inhomogeneous `sea' of partially resolved faint stars. 
However, very large negative blending, i.e., $\tsub{F}{S}\gg\tsub{F}{base}$ or $\tsub{f}{s}$ significantly greater than unity, is essentially never realized in nature. 
Therefore, we will place an upper limit on the value of $\tsub{f}{S}$ in Section 5 and beyond.

\section{EFSE Events with Limb-Darkening}
\label{sec:efseld}

Any surface brightness features on the source can affect the magnification and thus the morphology of the light curve \citep[e.g.][]{witt1994,gould1996,agol2003,herovsky2003}.  
Here we only consider inhomogeneities in the surface brightness distribution of the source due to limb-darkening.
In this case, the shape of the light curve and the peak magnification depend not only on $\rho$, but also on the amount and form of the limb-darkening and on the impact parameter of the source center with respect to the lens, $b_{0}$.  
However, as we will show, the two pairwise degeneracies that appear for uniform sources remain in the presence of limb-darkening, and indeed become even more linked resulting in a larger, five-parameter degeneracy when the limb-darkening parameter is unknown.   

For a limb-darkening profile, we adopt a linear limb-darkening profile of the form
\begin{equation}
{\cal S}[b(t)]=\left[1-\Gamma\left(1-\frac{3}{2}\sqrt{1-b^2}\right)\right],
\label{eqn:LD}
\end{equation}
where ${\cal S}(b)$ is the surface brightness of the source normalized to the average surface brightness ${\tilde S}=\tsub{F}{S}/\pi \theta_*^2$ as a function of $b$ \citep{yoo2004}.  
Note that ${\cal S}, \tsub{F}{S}$, and $\Gamma$ are all formally functions of wavelength (or bandpass) and we have not explicitly noted this for simplicity. 
Also note that ${\cal S}(0)=(1+\Gamma/2)$ (the center of the star) and ${\cal S}(1)=(1-\Gamma)$ (the limb of the star). 

The flux of the event as a function of time including the limb-darkening profile is
\begin{equation}
    F(t;b(t))=\tsub{F}{S} {\cal S}(b,\Gamma) A(t;\rho,b_{0},t_0,t_*) + \tsub{F}{B}.
\end{equation}
Note that as the position of the lens on the source is a function of time, so to is the coordinate of the limb-darkening profile $b$ being sampled. 

In analogy to $\tsub{\tau}{E}$ and $\tsub{\tau}{*}$, it is useful to define the time from the midpoint of the event in units of the half-chord crossing time
\begin{equation}
    \tau_c \equiv \frac{(t-t_0)}{t_c} = \frac{\tau_*}{\beta}.
\end{equation} 
In fact, by recognizing that $b^2 = b_0^2 + \tau_*^2=b_0^2+\beta^2\tau_c^2$, and $\sqrt{1-b^2}=\beta\sqrt{1-\tau_c^2}=\beta T_c(t)$, where we have further defined $T_c \equiv \sqrt{1-\tau_c^2}$, we can rewrite the limb-darkening profile including the time dependence as
\begin{equation}
    {\cal S}(t)=1-\Gamma\left[1-\frac{3}{2}\beta T_c(t)\right] = (1-\Gamma) + \frac{3\Gamma}{2} \beta T_c(t)
    \label{eqn:LDreparam},
\end{equation}
which will be advantageous for later use. A key insight is that the parameters of $T_c(t)$ are directly constrained by the observables $t_0$ and $t_c=\tsub{t}{FWHM}/2$.

The difference flux is given by
\begin{equation}
\label{eqn:dfexpan}
    \Delta F (t)\simeq\frac{2\tsub{F}{S}}{\rho^2}{\cal S}(t;b_{0},t_0,t_*,\Gamma)H(1-b(t)).
\end{equation}  
Using the reparameterized version of the surface brightness profile (Equation~\ref{eqn:LDreparam}), the difference flux can be written as
\begin{eqnarray}
    \Delta F(t) &=&\frac{2\tsub{F}{S}(1-\Gamma)}{\rho^2}\times\nonumber\\
    &&\left[1+\frac{3\Gamma\beta}{2(1-\Gamma)} T_c(t)\right]H(1-|\tau_c|).
    \label{eqn:dfreparam}
\end{eqnarray}
The form of this equation deserves careful study.  
Note that when $|\tsub{\tau}{c}|\le 1$ the difference flux is a linear function of $T_c$, which has a slope of $3\tsub{F}{S}\Gamma\beta/\rho^2$ and an intercept of $2\tsub{F}{S}(1-\Gamma)/\rho^2$.
This, combined with the fact that $T_c(t)$ is a well constrained function, means that the overall shape of the event is set by the coefficient of the $T_c$ term in the square brackets in Equation~\ref{eqn:dfreparam}, whereas the overall scale of the event is set by $2F_S(1-\Gamma)/\rho^2$.  
We can use this to define an observable shape parameter for the lightcurve 
\begin{equation}
    \tsub{f}{pl} \equiv \frac{\Delta F(t_0)-\Delta F(t_c)}{\Delta F(t_0)}
    \label{eqn:fpl}
\end{equation}
which is the difference between the flux at the peak of the event ($t=t_0$) and the flux when the lens is positioned on the limb of the source ($t=\pm t_c$), relative to the peak difference flux. 
Substituting Equation~\ref{eqn:dfreparam} into Equation \ref{eqn:fpl} and simplifying, we find
\begin{equation}
    \tsub{f}{pl}= \frac{\frac{3}{2}\Gamma\beta}{(1-\Gamma)+\frac{3}{2}\Gamma\beta}.
    \label{eqn:fpl_simp}
\end{equation}
Note that $\tsub{f}{pl}=0$ for $\Gamma=0$ and $\tsub{f}{pl}=1$ for $\Gamma=1$.  
Also, $\tsub{f}{pl}=0$ for $\beta=0$ ($b_0=1$) and $\tsub{f}{pl}=\frac{3}{2}\Gamma/[(1-\Gamma)+\frac{3}{2}\Gamma]$ for $\beta=1$ ($b_0=0$).

The definition of $\tsub{f}{pl}$ allows us to write the maximum flux difference as
\begin{equation}
    \Delta \tsub{F}{max} \equiv \frac{2\tsub{F}{S}(1-\Gamma)}{\rho^2}\left[1+\frac{\tsub{f}{pl}}{1-\tsub{f}{pl}}\right].
    \label{eqn:dfmax}
\end{equation}
Note then that for $\Gamma=0$, $\tsub{f}{pl}=0$, and we recover the previous forms for $\Delta \tsub{F}{max}=2\tsub{F}{S}/\rho^2$ and $\Delta F(t) = \Delta \tsub{F}{max}H(1-|\tau_c|)$, and that the magnitude of the event is decoupled from the duration of the event.
In the opposite limit of $\Gamma=1$,  $\tsub{f}{pl}=1$, so $\Delta \tsub{F}{max}=3\tsub{F}{S}\beta/\rho^2$, and $\Delta F(t) = \Delta \tsub{F}{max}T_c(t)H(1-|\tau_c|)$. 
Thus, in the limit that $\Gamma \rightarrow 1$, the magnitude and shape of the event is also decoupled from the duration of the event.

\subsection{The Degeneracy for Fixed Limb-Darkening}
\label{subsec:degenmathfixedld}

We now consider the degeneracies that exist for a limb-darkened source ($\Gamma \ne 0$), assuming the limb-darkening parameter $\Gamma$ is known {\it a priori} and thus is not a free parameter.
We preface this discussion by noting that for a limb-darkened source, the peak flux and shape of the EFSE event depends not only on $\tsub{F}{S}$ and $\rho$, but also on the impact parameter $b_0$. 
This is because the peak flux observed during the event now depends on the location of the center of the lens with respect to the center of the source, such that larger lens-source separations result in smaller peak fluxes (see Equation ~\ref{eqn:dfexpan}). 
On the other hand, as with the $\Gamma=0$ case, the observed duration $\tsub{t}{c}$ is related to $t_*$ and $b_0$.  
Thus the duration and magnitude of the deviation are no longer decoupled.  
Nevertheless, as we will show, there is an approximate degeneracy in the case of fixed limb-darkening that becomes a perfect mathematical degeneracy as $\Gamma \rightarrow 0$ and an another perfect mathematical degeneracy as $\Gamma \rightarrow 1$.
As we will show, this approximate degeneracy is actually quite severe.

We first recall that there are four primary observables\footnote{Again, we ignore the parameters $\tsub{F}{base}$ and $\tsub{t}{0}$ as they are well-constrained by the observations and thus do not participate in the degeneracy.} ($\Delta \tsub{F}{max}$, $\tsub{t}{c}$, $\tsub{f}{ws}$, and $\tsub{f}{pl}$), and for fixed limb-darkening with known $\Gamma$ there are four free parameters ($\tsub{b}{0}$, $\tsub{t}{*}$, $\tsub{F}{S}$, and $\rho$). 
Thus, given that there are an equal number of observables as free parameters, we might anticipate that there would not be a degeneracy.  
From Equation~\ref{eqn:fpl_simp} and assuming fixed $\Gamma \ne 0$, a measurement of $\tsub{f}{pl}$ yields a constraint on $\beta$ and thus $\tsub{b}{0}$.  
A measurement of $\tsub{t}{c}$ and a constraint on $\beta$ thus yields a constraint on $t_*$.  
A measurement of $\tsub{f}{ws}$ and a constraint on $\beta$ also yields a constraint on $\rho$.  
Finally, a measurement of $\Delta\tsub{F}{max}$, combined with a constraint on $\rho$, yields a constraint on $\tsub{F}{S}$.  
Thus there is no mathematical degeneracy. 

\begin{figure}[t]
\plotone{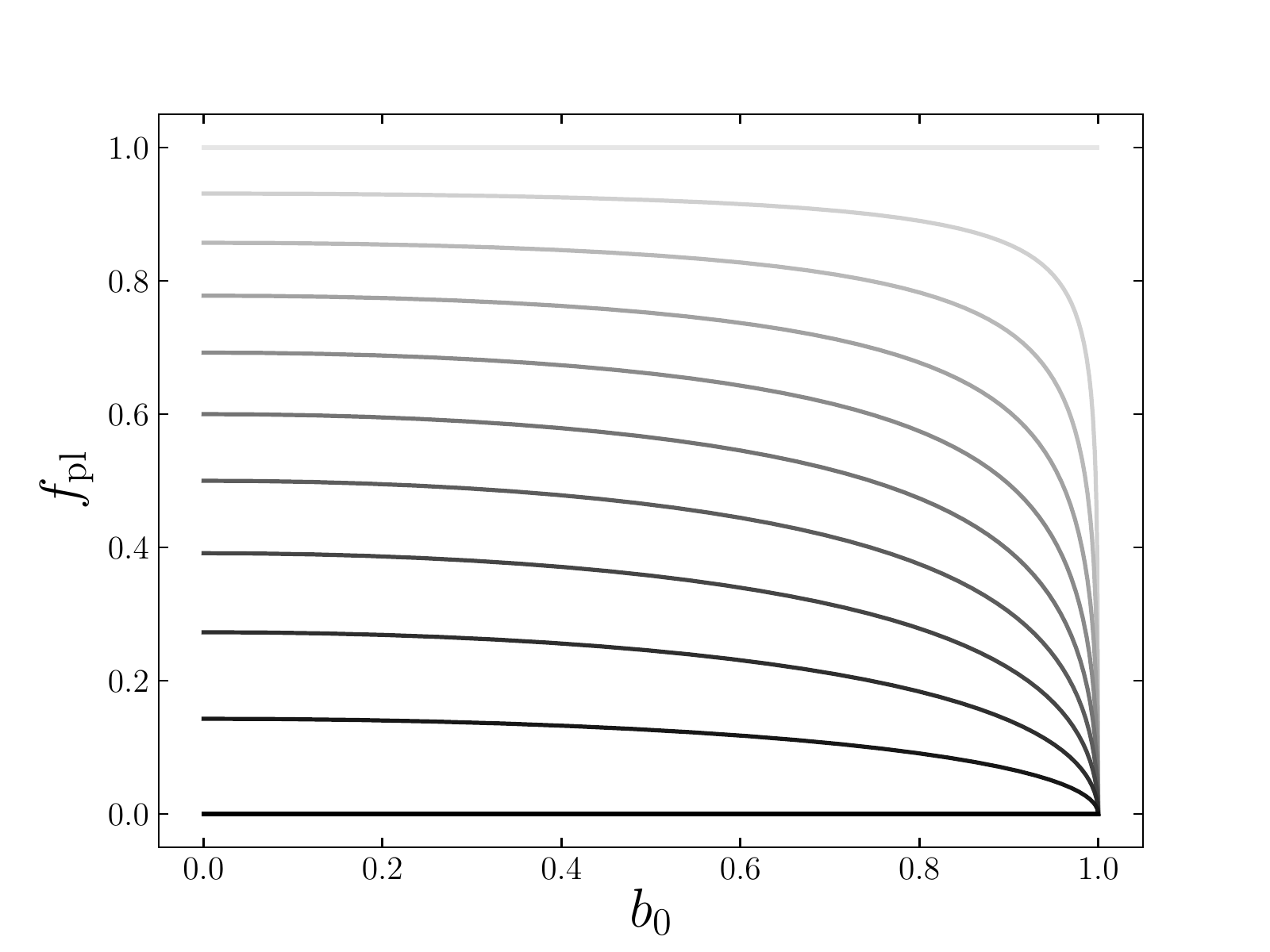}
\caption{The fractional peak-to-limb flux difference $\tsub{f}{pl}$ as a function of the dimensionless impact parameter $b_0$, for $\Gamma=0.0,0.1,0.2,...1.0$ from darkest to lightest, respectively.
As $\Gamma$ increases, so does the value of $\tsub{f}{pl}(b_0=0.0)$.
Also, the dependence of $\tsub{f}{pl}$ for $b_0\lesssim0.8$ becomes weaker for increasing $\Gamma$. 
By definition, $\tsub{f}{pl}$ for $\Gamma=0.0$ and $1.0$ are flat, as described in Section \ref{sec:efseld}.}
\label{fig:fplvaryb0}
\end{figure}

However, the lack of mathematical degeneracy rests on the fact that $\tsub{f}{pl}$ depends on $b_0$.  
Therefore, we next explore how the observable $\tsub{f}{pl}$ depends on $b_0$ for various values of $\Gamma$ in order to provide a qualitative understanding of how well $\beta$ can be constrained with a measurement of $\tsub{f}{pl}$ of a given precision.  
Figure \ref{fig:fplvaryb0} shows $\tsub{f}{pl}$ as a function of $b_0$ for various values of $\Gamma=0.0,0.1,0.2,...,1.0$, where $\Gamma=0.0$ is the darkest line and $\Gamma=1.0$ the lightest. 
There are several points to note.  
First, when $\Gamma=0$, $\tsub{f}{pl}=0$ and thus is independent of $\tsub{b}{0}$, as noted previously. 
Second, for non-zero $\Gamma$, $\tsub{f}{pl}$ is a weak function of $\tsub{b}{0}$ over a relatively broad range of $\tsub{b}{0}$.  
Furthermore, the range of $\tsub{b}{0}$ for which the difference in $\tsub{f}{pl}$ is smaller than some fixed value is larger for larger $\Gamma$.  
Thus the shape of the light curves become increasingly degenerate as $\Gamma$ increases.

For $\Gamma=1$, the light curves are completely self-similar.  
This is due to the fact that for $\Gamma=1$, $\tsub{f}{pl}=1$, and therefore $\Delta\tsub{F}{max}=3\tsub{F}{S}\beta/\rho^2$, and $\Delta F(t) = \Delta\tsub{F}{max}T_c(t)H(1-|\tau_c|)$ (see Section~\ref{sec:efseld}).
Thus for $\Gamma=1$, there are only three observables ($\Delta\tsub{F}{max}$, $\tsub{t}{c}$, $\tsub{f}{ws}$), and four parameters ($\tsub{F}{S}$, $\rho$, $\beta$, $t_*$). 
All four parameters are therefore degenerate with each other, such that
the observables are unchanged (i.e., $\Delta\tsub{F}{max}'=\Delta\tsub{F}{max}$,  $\tsub{t}{c}'=\tsub{t}{c}$, and $\tsub{f}{ws}'=\tsub{f}{ws}$) under the transformation
\begin{eqnarray}
    \tsub{F}{S}'&&=\eta \tsub{F}{S},~~\beta'=\eta^{-1/5} \beta,\nonumber\\
    \rho'&&=\eta^{2/5} \rho,\,\,\, t_*'=\eta^{1/5} t_*,
    \label{eqn:degengammaequalone1}
\end{eqnarray}
for any arbitrary positive value of $\eta$ that satisfies \mbox{$0\le\eta^{-1/5}\beta\le 1$}.  

Thus for fixed $\Gamma=0$, there is a continuous one-parameter mathematical degeneracy between $\tsub{b}{0}$, $\tsub{t}{*}$, $\tsub{F}{S}$, and $\rho$ (Equation \ref{eqn:degennoldfiniterho}), whereas for $\Gamma=1$, there is also a continuous one-parameter mathematical degeneracy between these parameters, albeit with a different scaling (Equation~\ref{eqn:degengammaequalone1}).  
Therefore, for intermediate values of $\Gamma$, we expect an approximate degeneracy, such that for smaller values of $\Gamma$, the scalings should more closely approximate Equations~\ref{eqn:degennoldfiniterho}, whereas for larger values of $\Gamma$, the scalings should more closely approximate those of Equations~\ref{eqn:degengammaequalone1}.
Indeed, this is what we find via numerical investigations (see Section \ref{sec:degenfixedld} and Appendix \ref{apdx:tables})

\subsection{The Degeneracy for Free Limb-Darkening}
\label{subsec:degenmathfreeld}

Finally, we consider the full degeneracy that exists for a limb-darkened source ($\Gamma \ne 0$) when the limb-darkening as parameterized by $\Gamma$ is either completely unconstrained, or has some finite uncertainty.  
In the case of $\rho\to\infty$ ( and thus $\tsub{f}{ws}=0$), we still have three primary observables ($\Delta \tsub{F}{max}$, $\tsub{t}{c}$, and $\tsub{f}{pl}$), but we now have five free parameters ($\tsub{F}{S}$, $\rho$, $\beta$, $t_*$, and $\Gamma$).  

We proceed in the same manner as the previous sections.  Consider an event with fiducial values of $\beta$, $t_*$, $\tsub{F}{S}$, $\rho$, and $\Gamma$. 
Given the definitions of the observables, it is straightforward to show that there is a continuous mathematical one-parameter degeneracy between $\beta$, $t_*$, $\Gamma$, and $\tsub{F}{S}\rho^{-2}$ such that
\begin{eqnarray}
    \Gamma'&&=\eta\Gamma,~~\beta'=\gamma\beta,~~t_*'=\gamma^{-1}t_*, \nonumber\\
    &&\left(\frac{\tsub{F}{S}}{\rho^2}\right)'=\frac{1}{\eta\gamma} \left(\frac{\tsub{F}{S}}{\rho^2}\right),
    \label{eqn:degenldfiniterho1}
\end{eqnarray}
where we have defined\footnote{The expression for $\gamma$ can be derived by solving for $\gamma$ if $\tsub{f}{pl}$ is to remain constant under the transformation of $\Gamma$ and $\beta$. 
Both constant and time variable terms of Equation~\ref{eqn:dfreparam} can then be shown to be invariant if $\tsub{F}{S}/\rho^2$ is so transformed.}
\begin{equation}
    \gamma\equiv \frac{1-\eta\Gamma}{\eta(1-\Gamma)},
\end{equation}
and $\eta$ is an arbitrary positive constant such that $\eta\Gamma\leq 1$ and $\gamma \beta \leq 1$.
Thus, $\tsub{F}{S}$ and $\rho$ only participate in the degeneracy through the combination of $\tsub{F}{S}/\rho^2$, and are degenerate with each other such that
$\Delta \tsub{F}{max}$ is constant under the transformations in Equation \ref{eqn:fsrhotransform} for fixed $(\eta\gamma)^{-1}$. 

We next consider the case when $\rho\gg 1$ but finite, in which $\tsub{f}{ws}$ is non-zero.   
In this case, we still have five free parameters ($\tsub{F}{S}$, $\rho$, $\beta$, $t_*$, and $\Gamma$), but there are now four primary observables ($\tsub{t}{c}$, $\tsub{f}{pl}$, $\tsub{f}{ws}$, and $\Delta \tsub{F}{max}$). 
We again consider an event with fiducial values of the parameters. 
Given the definitions of the observables, it is straightforward to show that there is a continuous mathematical degeneracy between $\tsub{F}{S}$, $\rho$, $\beta$, $t_*$, and $\Gamma$ such that
\begin{eqnarray}
    \Gamma'&=&\eta\Gamma,~~\beta'=\gamma\beta,~~t_*'=\gamma^{-1}t_*,\nonumber\\
    \rho'&=&\gamma^{-2}\rho,~~\tsub{F}{S}'=\eta^{-1}\gamma^{-5}\tsub{F}{S},
    \label{eqn:degenldfiniterho2}
\end{eqnarray}
and $\eta$ is an arbitrary positive constant such that $\eta\Gamma\le 1$ and $\gamma \beta \le 1$. 
In this case, $\tsub{F}{S}$ and $\rho$ become coupled to the larger degeneracy through the observables $\tsub{f}{ws}$ and $\Delta \tsub{F}{max}$. 

\begin{figure}[t]
\plotone{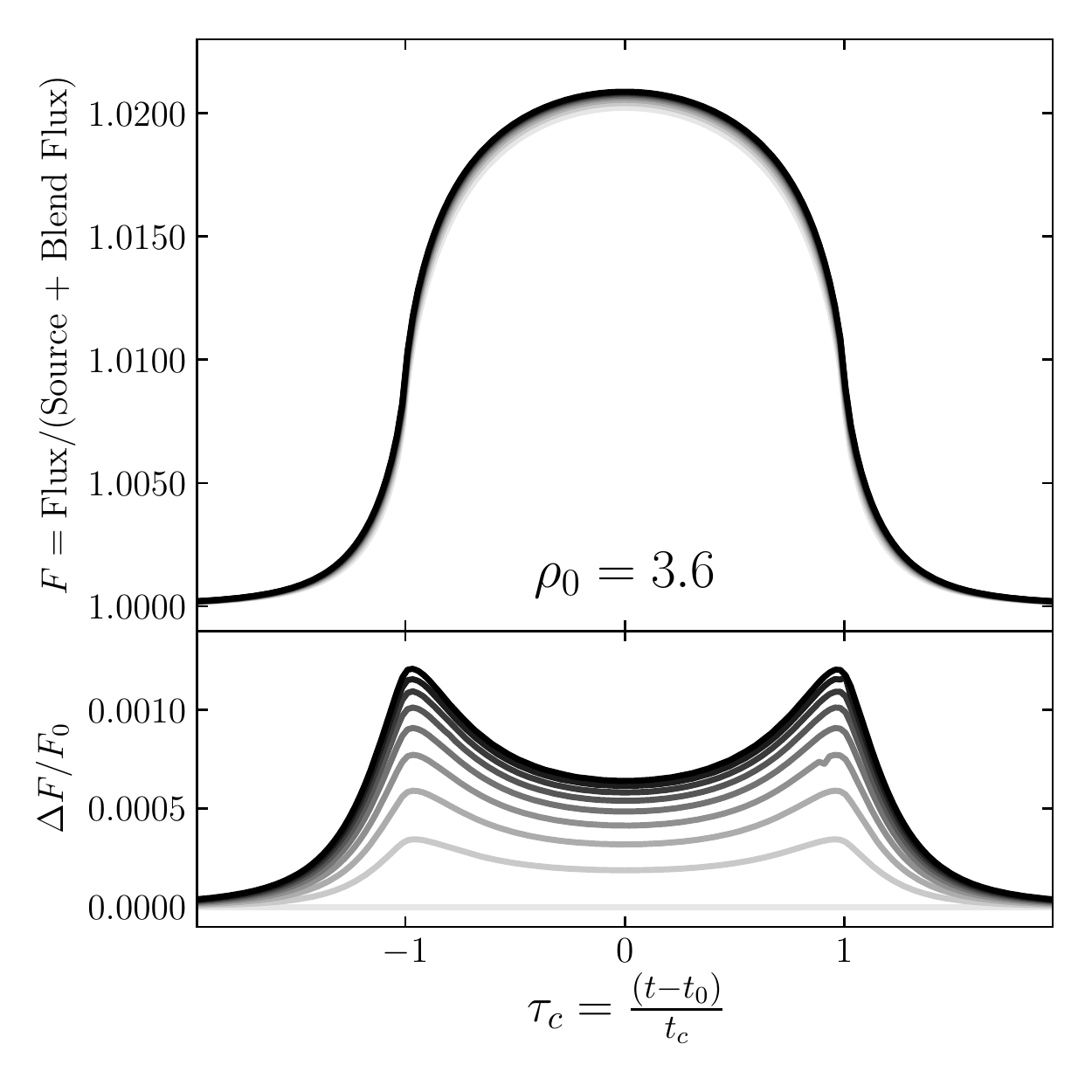}
\plotone{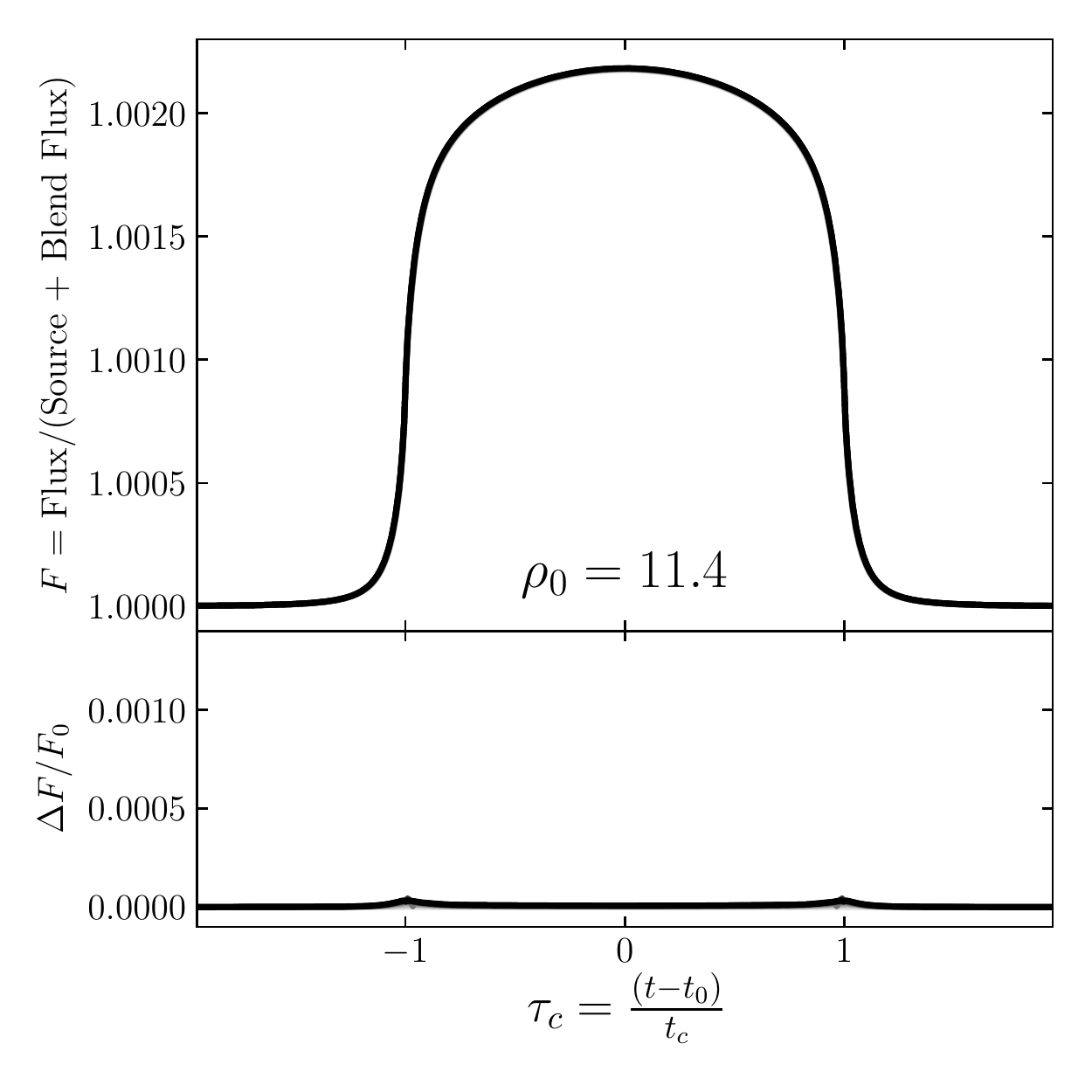}
\caption{Demonstration of the full degeneracy for free limb-darkening with a fiducial value of $\rho=3.6$ (top panel) and $\rho=11.4$ (bottom panel). 
Both panels show nine different light curves that serve to demonstrate the degeneracy between the parameters $\tsub{F}{S}, \rho, \beta$, $t_*$, and $\Gamma$ for $\rho\gg 1$ but finite.  
We choose fiducial values for the parameters, and for each of nine values equally spaced between $\Gamma=$ 0.2 to 0.4 we use the transformation in Equation~\ref{eqn:degenldfiniterho2} to scale the other parameters and plot the resulting light curves in the upper subpanels.
The line colors go from lightest to darkest for increasing values of $\Gamma$.
The bottom subpanels show the fractional residuals, where $F_0$ is the flux from the event with $\Gamma=0.2$ and $\Delta F=F-F_0$. 
We note the larger residuals in the $\rho=3.6$ case than in the $\rho=11.4$ case due to the shape induced by changing $\rho$ on $\tsub{f}{ws}$.}
\label{fig:degenldfiniterho}
\end{figure}

Analogous to Figure \ref{fig:degennoldfiniterho}, we demonstrate the severity of this degeneracy in Figure \ref{fig:degenldfiniterho}.
We use fiducial values of $b_0=0.0$, $\tsub{\mu}{rel}=6.5$ mas yr$^{-1}$, $\tsub{f}{s}\simeq0.13$, and $\Gamma=0.2$ but use $\rho=3.6$ in the top panel and $\rho=11.4$ in the bottom panel.
Rather than scale the other parameters based on $b_0$, we use nine values that uniformly sample $\Gamma$ from 0.2 to 0.4.
The upper subpanels show the resulting light curve, whereas the fractional residuals compared to the $\Gamma=0.2$ light curves are shown in the lower subpanels. 
In both the top and bottom panels, the color of the light curves and their residuals go from lightest to darkest for increasing values of $\Gamma$.
Similar again to Figure \ref{fig:degennoldfiniterho}, the prominent deviations in the top panel are due to the changing values of $\rho$ and $\beta$ impacting the observable $\tsub{f}{ws}$. 
But for the $\rho=11.4$ case, we see the magnitude of these deviations decreases significantly.
We also note that the ``trough'' between the peaks in the $\rho=3.6$ case between $\tau_c=-1$ and 1 has a larger amplitude than in Figure \ref{fig:degennoldfiniterho}.
This is due to effects of higher order that affect the shape of the light curve induced by the limb-darkening profile at larger impact parameters. 
While light curves arising from different chords for a linear limb-darkening profile have similar shapes, they are not identical except when $\Gamma=0$ or $\Gamma=1$ (see Figure 4).
We also note that these scaling require extreme values of the degenerate parameters. 
For instance in the $\rho=11.4$ case, increasing $\Gamma$ from 0.2 to 0.4 requires $\rho\approx80$ and $\tsub{f}{S}\approx10$, the latter of which is unphysical.

We reiterate that so far we have explored these degeneracies as {\it purely mathematical} degeneracies.  
In reality, $b_0$, $t_0$, $\tsub{F}{S}$, $\rho$, and $\Gamma$ are not free to take on an arbitrary range of values, alone and particularly when constrained by the values of the other parameters.
For example, since the operand of ${\cal S}(b)$ can only vary between the values of $b=[0,1]$, ${\cal S}(b)$ can only take on the values between ${\cal S}(0)=1+\Gamma/2$ and ${\cal S}(1)=1-\Gamma$. 
Therefore, even by setting $b_0=0$, it is not possible to reproduce the maximum magnification of $A=1+\frac{2\tsub{F}{S}}{\rho^2}(1-\Gamma/2)$ of such a light curve with another light curve with finite $\Gamma$ by changing $b_0$.
This is because the maximum magnification of a light curve with $b_0=0$ is indeed $A=1+\frac{2\tsub{F}{S}}{\rho^2}(1-\Gamma/2)$, $\Gamma$ has a maximum value of unity, and all light curves with $b_0 > 0$ have smaller maximum flux differences as the trajectory of the lens passes over only more limb-darkened portions of the source star.
However, it {\it would} be possible by increasing $\tsub{f}{S}$ or decreasing $\rho$, although doing so may lead to either nonphysical or physically unlikely values of the other parameters. 

When exploring the mathematical degeneracies in the following Sections 5 and 6, we adopt the following constraints: $0 \le \Gamma \le 1$, $b_0\ge 0$ ($\beta \le 1$), and $\tsub{F}{S} > 0$ (or $f_s>0$).  
However, we allow the blend flux to be negative; specifically we constrain $f_s \le 3$.  
As discussed in Section \ref{sec:mathvsphysical}, somewhat negative blend fluxes are possible, and indeed have been observed in some events. 

We further discuss the extent to which the mathematical degeneracies analyzed analytically in the previous sections and explored numerically in Sections 5 and 6 may manifest themselves observationally in Section 7. 

\subsection{Summary of the Parameters and Observables for EFSEs}
\label{subsec:paramsefses}

In summary, there are seven model parameters that describe EFSE events when a linear limb-darkening profile is assumed: (1) the time at which the lens is minimally separated from the center of the source at the midpoint of the event $t_0$, (2) the source radius crossing time $t_*$, (3) the angular impact parameter of the lens relative to the center of the source in units of the angular radius of the source $\tsub{b}{0}$, which we will generally parameterize as $\beta$ for convenience\footnote{Although we will generally use $\beta$ in this paper, we note that the intrinsic distribution of $b_0$ is uniform, whereas the distribution of $\beta$ is not.  Thus when exploring the severity of the degeneracies discussed in this paper, we will consider uniformly-distributed values of $b_0$.} (4) the baseline flux $\tsub{F}{base}$, (5) the flux of the source $\tsub{F}{S}$, which is related to the blend flux $\tsub{F}{B}$ through $\tsub{F}{S}=\tsub{F}{base}-\tsub{F}{B}$, (6) the dimensionless radius of the star in units of the Einstein ring radius $\rho$, and (7) the dimensionless limb-darkening coefficient for a linear profile $\Gamma$.  We emphasize again that, in the extreme case we are considering here, the light curve is better parameterized by $b_0$ and $t_*$ rather than $u_0$ and $\tsub{t}{E}$\citep{agol2003}.
    
Although the seven model parameters $t_0$, $t_*$, $\beta$, $\tsub{F}{base}$, $\tsub{F}{S}$, $\rho$, $\Gamma$ are all technically free parameters, only five ($t_*$, $\beta$, $\tsub{F}{S}$, $\rho$, $\Gamma$) participate in the degeneracies explored mathematically in Section~\ref{subsec:degenmathfreeld}. 
For well-sampled events, the parameter $t_0$ is not covariant with the other observables, because the derivative of the flux with respect to $t_0$ is anti-symmetric about $t_0$, whereas the derivatives of the flux with respect to the other parameters are symmetric about $t_0$. Thus $t_0$ is (essentially) directly constrained by observations. As discussed above, $\tsub{F}{base}\equiv \tsub{F}{S}+\tsub{F}{B}$ is also essentially a direct (and typically precisely-measured) observable based on observations outside of the microlensing event. 

On the other hand, there are five gross observables for EFSE events for $\rho \rightarrow \infty$, namely: (1) the baseline flux $\tsub{F}{base}$, (2) the peak change in flux $\Delta \tsub{F}{max}$, (3) the FWHM of the event, which is roughly equal to the half-chord crossing time, $\tsub{t}{c}\simeq \tsub{t}{FWHM}/2$, (4) the fractional peak-to-limb flux difference $\tsub{f}{pl}$, and (5) the midpoint $t_0$ of the (time-symmetric) event. 
Since $\tsub{F}{base}$ and $t_0$ are both parameters and direct observables, they do not participate in the degeneracy, and thus we only have to consider three observables: $\Delta \tsub{F}{max}$, $\tsub{f}{pl}$, and $\tsub{t}{c}$. 
When $\rho\gg1$ but finite, we have a fourth observable, which is the duration of the event that exhibits wings/shoulders during the event $\tsub{f}{ws}$.

In summary, for $\rho \gg1$ but finite, fixed limb-darkening, and arbitrary $\Gamma$, there are an equal number of parameters and observables and thus there is no formal mathematical degeneracy.  
For the special cases of $\Gamma=0$ and $\Gamma=1$, there is one more free parameter than observables, and thus there is a one-parameter degeneracy in each case.  
However, the scalings of the parameters required to keep the observables fixed is different in the two cases (Equations \ref{eqn:degennoldfiniterho} and \ref{eqn:degengammaequalone1}).
In the case of free limb darkening, there is again one more parameter than observable, and thus a one-parameter degeneracy.  

\section{Exploring Components of the Degeneracy for Fixed Limb-Darkening}
\label{sec:degenfixedld}

\subsection{Set-up}
In order to investigate the nature of the degeneracies that we have explored mathematically in the previous section, we will qualitatively and quantitatively explore two fiducial events. 
For each, we will present a suite of light curves that highlight the various steps we have used to approach our explanation of the degeneracies.  
We also perform a quantitative investigation of the severity of the degeneracy with similar groupings of relevant parameters incorporated in the degeneracy.

\subsubsection{Fiducial Events}
\label{sec:fiducial}

For all the following considerations, we will be comparing light curves to two ``fiducial'' light curve models.
For the first of these models, we select parameter values such that the physical limits of the degeneracy will fundamentally limit its mathematical extent and that are more akin to recent FFP candidates \citep[e.g.,][]{mroz2020a}.
For the second, we choose parameters that will allow us to explore the degeneracy more thoroughly within its physical bounds. 
We will refer to these fiducial models as `Event 1' and `Event 2' throughout the remainder of this paper.
All Figures pertaining to Event 1 use green accents while Figures pertaining to Event 2 use red accents. 
As discussed above, we note that in some cases the degenerate light curves we find will require parameter values that are not physical.
Such situations can therefore be ruled out, and a {\it physical} degeneracy (as opposed to a {\it mathematical} degeneracy) will not exist.  
Similarly, in other cases, some of the degeneracies will require parameter values that are physically unlikely.  
Such situations can therefore be considered implausible, and thus although both a {\it physical} and {\it mathematical} degeneracy exist, it is unlikely that this degeneracy will ever be manifested in actual microlensing surveys.  
Such situations can be dealt with by applying appropriate prior information.
To allow the degeneracy to demonstrate a subset of these unlikely or impossible regions, we use a prior that $\tsub{f}{S}<3.0$ for all investigations. 
Note that this prior was not applied in Figures \ref{fig:degennoldfiniterho} and \ref{fig:degenldfiniterho}.

The parameter values we adopt for Event 1 are \mbox{$\tsub{f}{S}=1.00$}, $\tsub{f}{B}=\tsub{F}{B}/\tsub{F}{base}=0.00$, $\Gamma=0.4$, $b_0=0.0$, $\tsub{\pi}{rel}=0.125~{\rm mas}$, $M=0.11~M_\oplus$, $\rho=10$, $R_*=10~R_\odot$, $\theta_*=5.8~{\mu}$as, $t_*=7.8$~d, and $\tsub{\mu}{rel}=6.5~{\rm mas~yr^{-1}}$. 
For Event 2, all the parameters are identical except \mbox{$\tsub{f}{S}=0.20$}, $\tsub{f}{B}=0.80$, $M=0.55~M_\oplus$, and  $\rho=4.47$.
These fiducial light curves are shown as the back-most light curve (green/red) in the third from the top group ($\Gamma=0.4$) of Figure \ref{fig:gammaFs_vary_rho}. 
In the following figures, we specify which parameters vary from this fiducial set.
%Of course, in showing the degeneracy, we will vary some subsets of the other parameters as appropriate for the other light curves.

\begin{figure*}[t]
\epsscale{\epsScaleFactorOne}
\plottwo{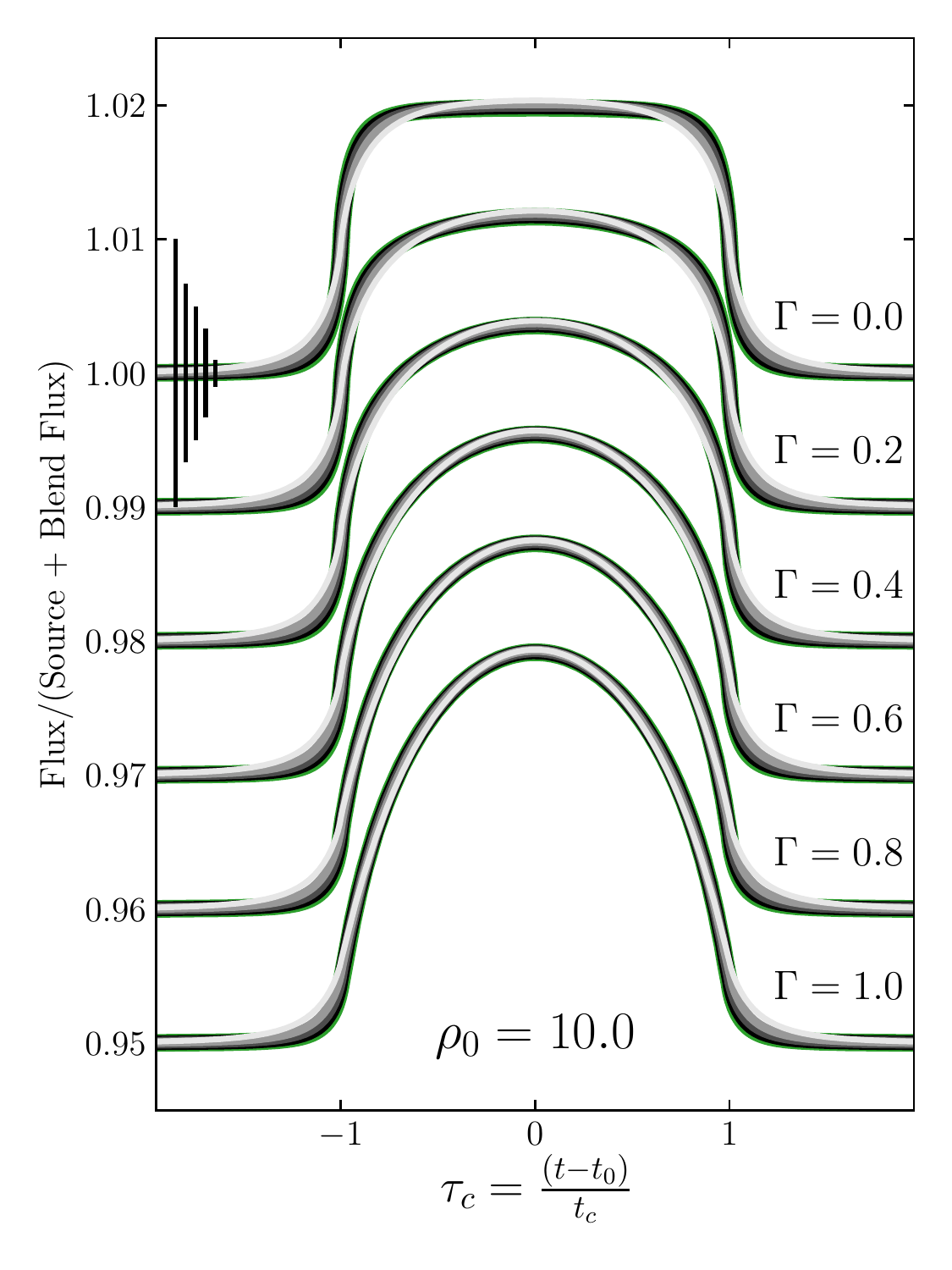}{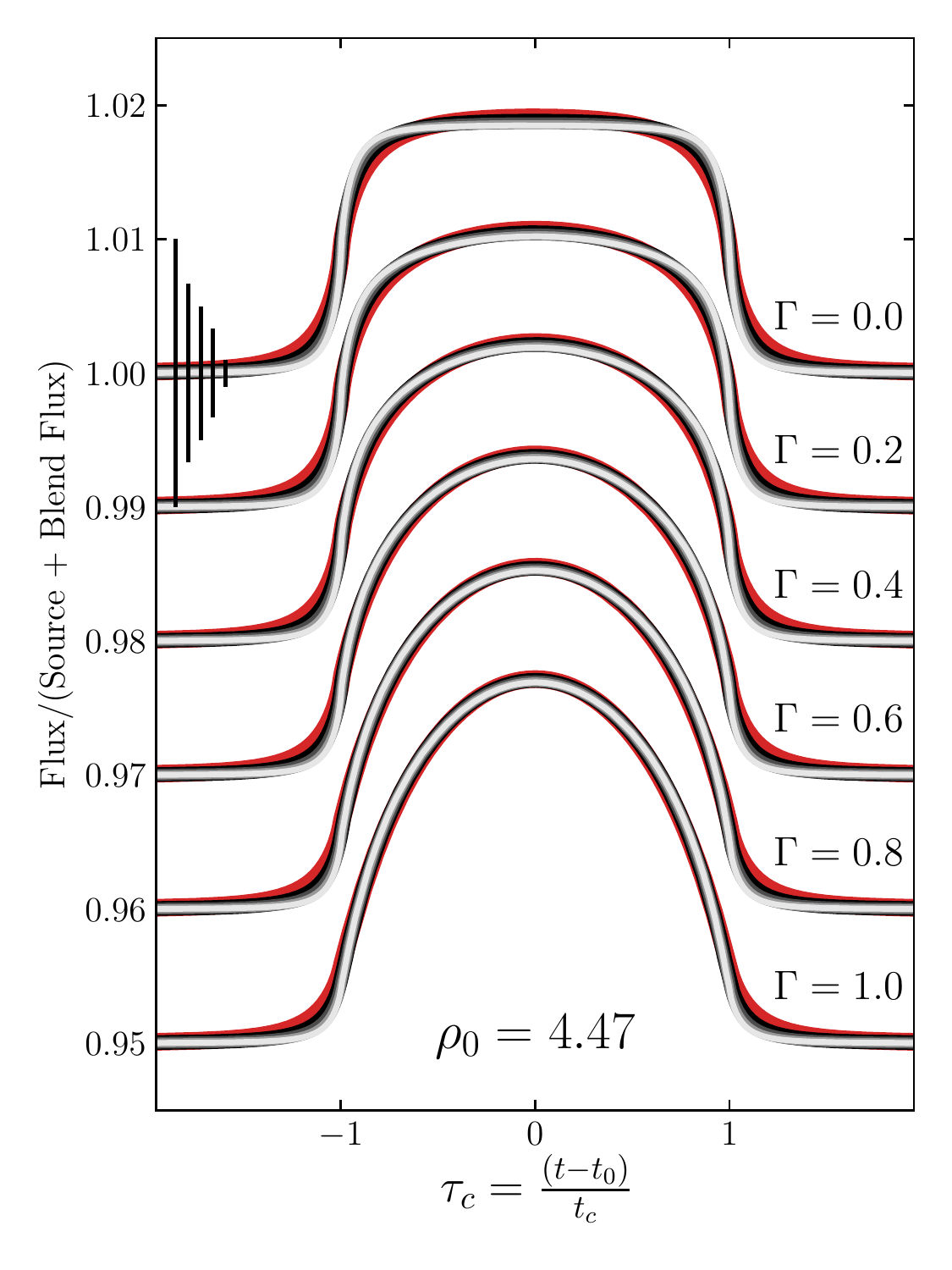}
\caption{Qualitative demonstration of the $\rho$--$\tsub{F}{S}$ degeneracy for the Event 1 (left, green) and Event 2 (right, red).
In both panels, we fix $b_0=0$ and show six sets of light curves with different limb-darkening coefficient $\Gamma$. 
For each $\Gamma=0.0,0.2,0.4,0.6, 0.8, 1.0$, five light curves with source flux $\tsub{f}{S}=1.0, 0.8, 0.6, 0.4, 0.2$ are shown from backmost to frontmost (green/red, black, dark gray, light grey, white) respectively in the left panel for Event 1.
The order of $\tsub{f}{S}$ values are reversed in the right panel for Event 2.
For each light curve, $\rho$ was varied to minimize the $\Delta\chi^2$ from the fiducial light curve.
Fractional uncertainties of 1\%, 0.667\%, 0.5\%, 0.333\%, and 0.1\% (left to right on the $\Gamma=0.0$ light curves) are shown.}
\label{fig:gammaFs_vary_rho}
\end{figure*} 

\begin{figure*}[t]
\epsscale{\epsScaleFactorOne}
\plottwo{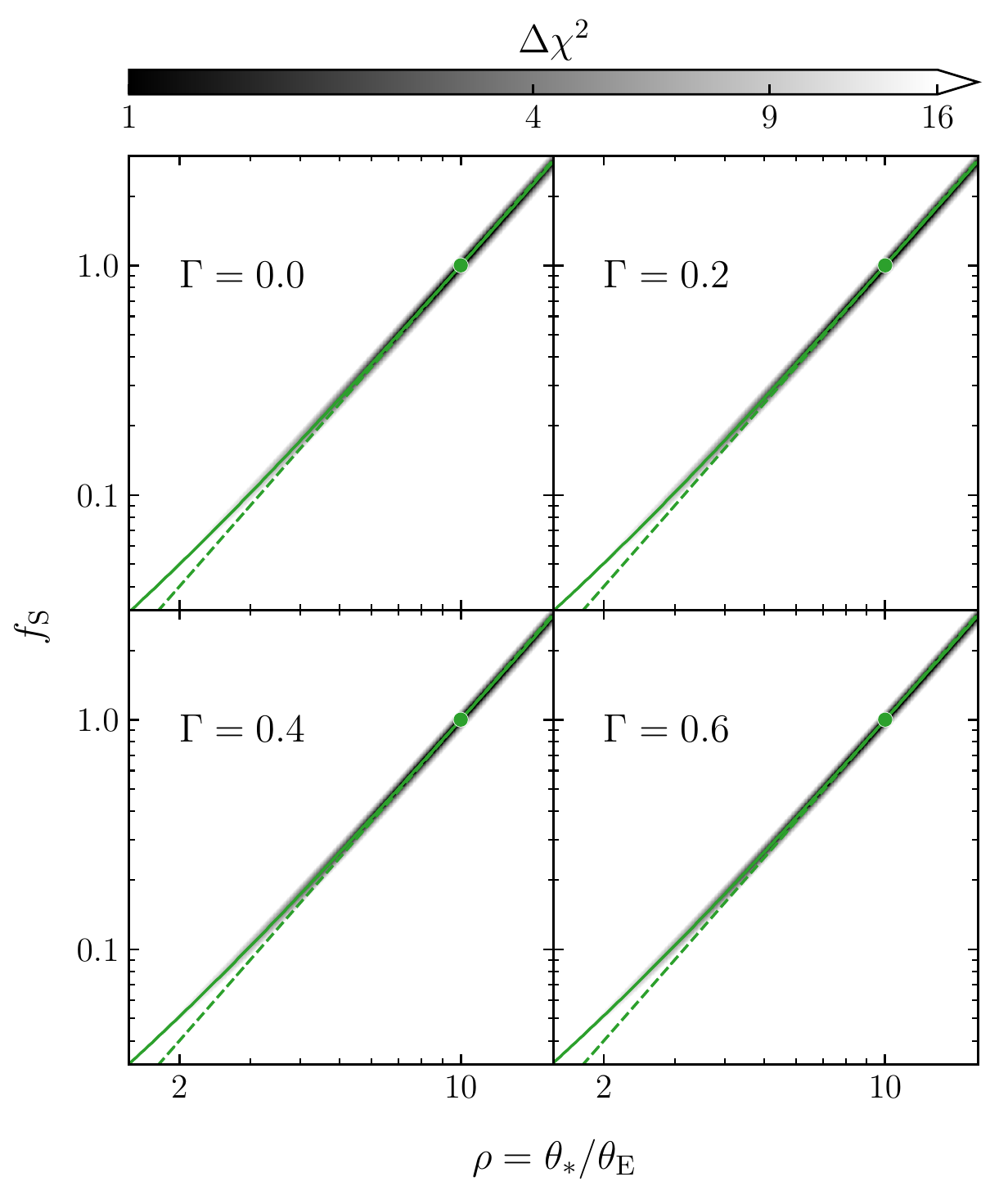}{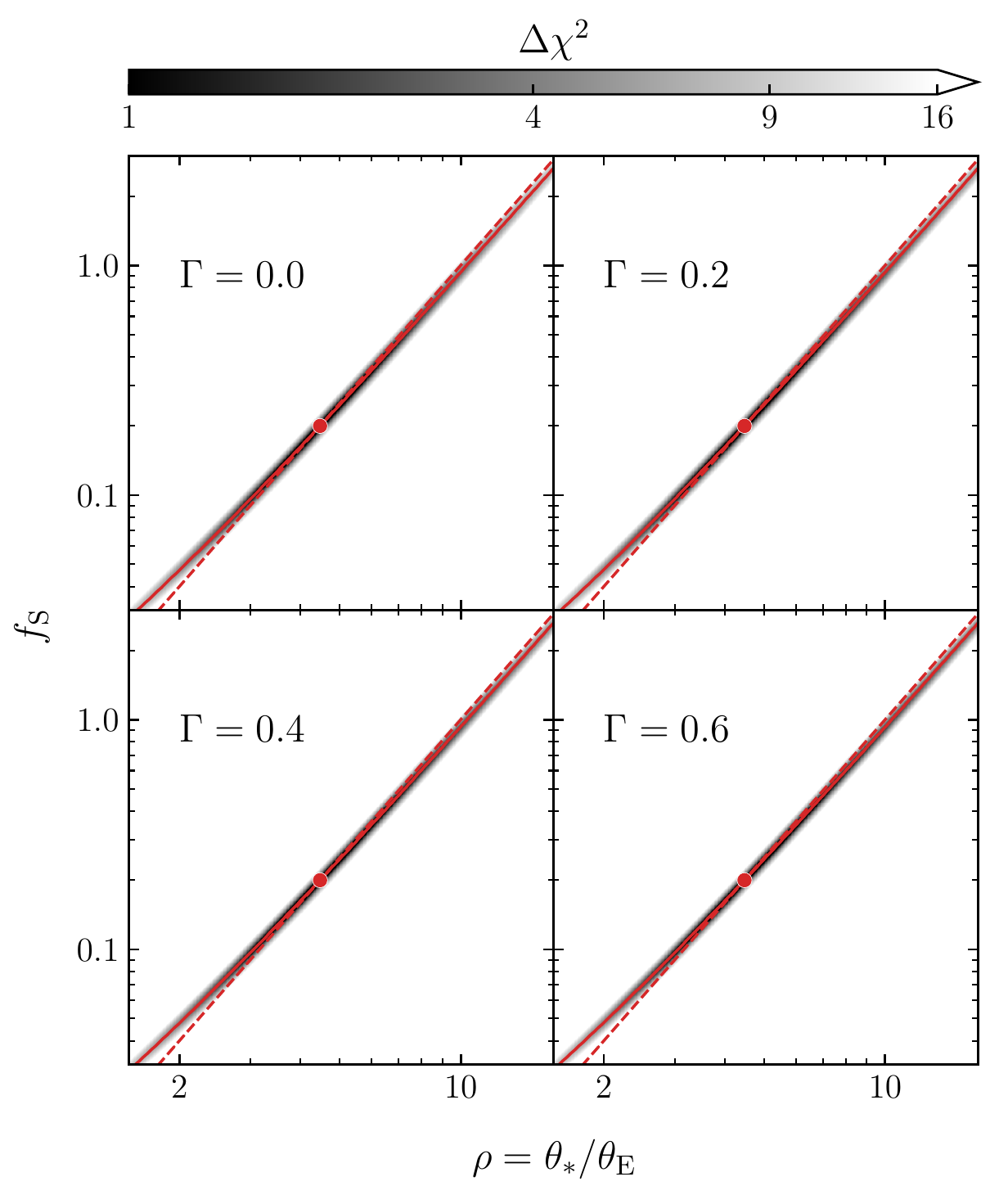}
\plottwo{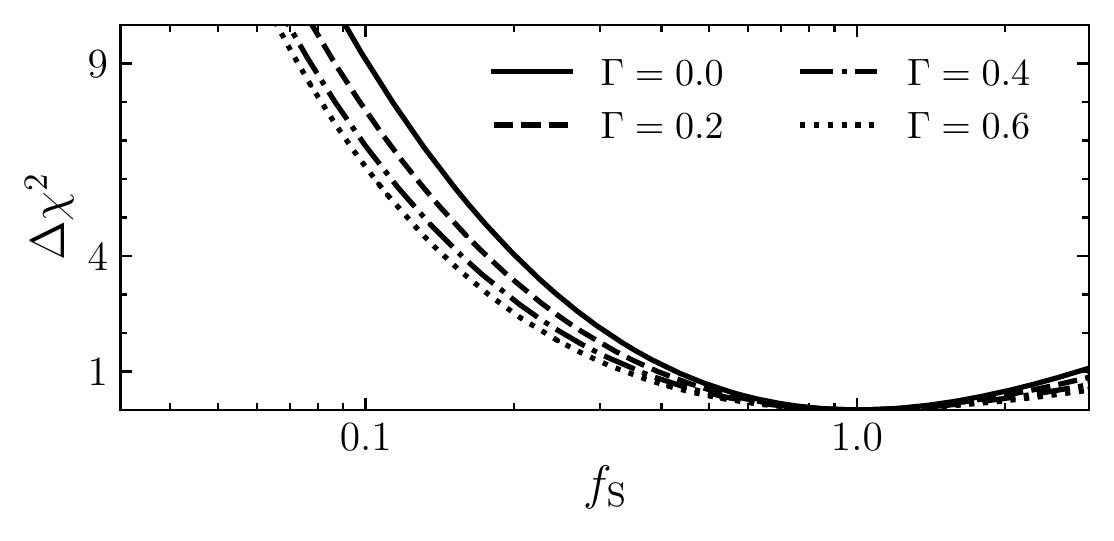}{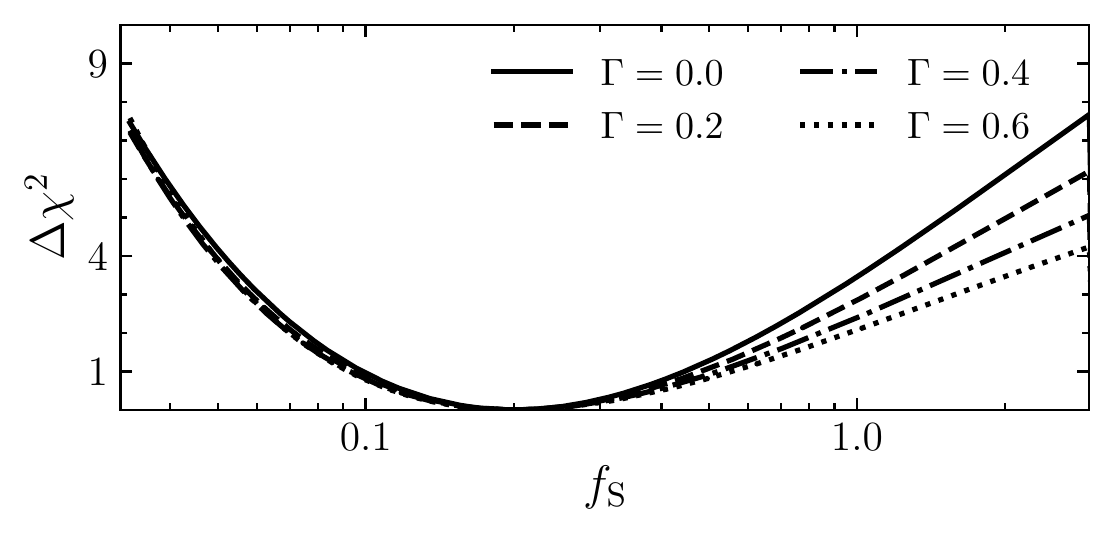}
\caption{
Quantitative demonstration of the $\rho-\tsub{f}{S}$ degeneracy. 
\textit{Upper Panels}: At each $(\rho,\tsub{f}{S})$ pair, all other parameters are fixed at the fiducial values and the  $\Delta\chi^2$ is calculated between the proposed and the fiducial models assuming an $0.5\%$ uncertainty and 15 min observation cadence. 
No $\Delta\chi^2$ minimization is performed in these comparisons.
The limb-darkening parameter $\Gamma$ assumed in each subpanel is labelled.
The fiducial model is shown as the green/red circle. 
The analytic prediction for the degeneracy (Equation \ref{eqn:flux_degen}) is shown as the dashed green/red lines and the calculated minimum $\Delta\chi^2$ is shown as a solid green/red lines.
\textit{Lower Panels}: The calculated minimum $\Delta\chi^2$ projected onto $\tsub{F}{S}$. 
Each line style corresponds to one of the assumed values of $\Gamma$ in the Upper Panel.
}
\label{fig:map.rhofs}
\end{figure*} 

\subsubsection{Explanation of Figure Layouts}
\label{subsec:plot_exp}
\begin{figure*}[t]
\epsscale{\epsScaleFactorOne}
\plottwo{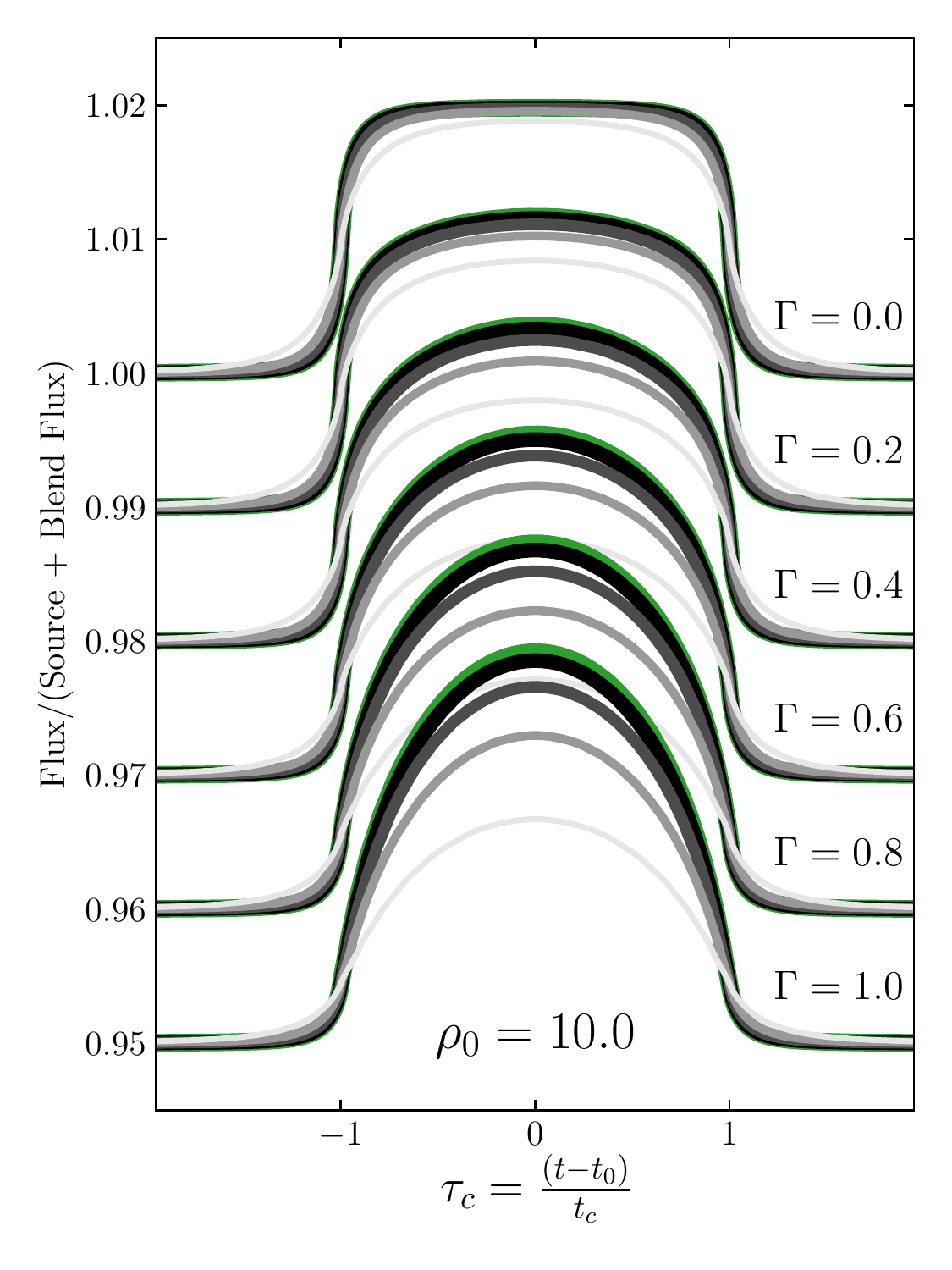}{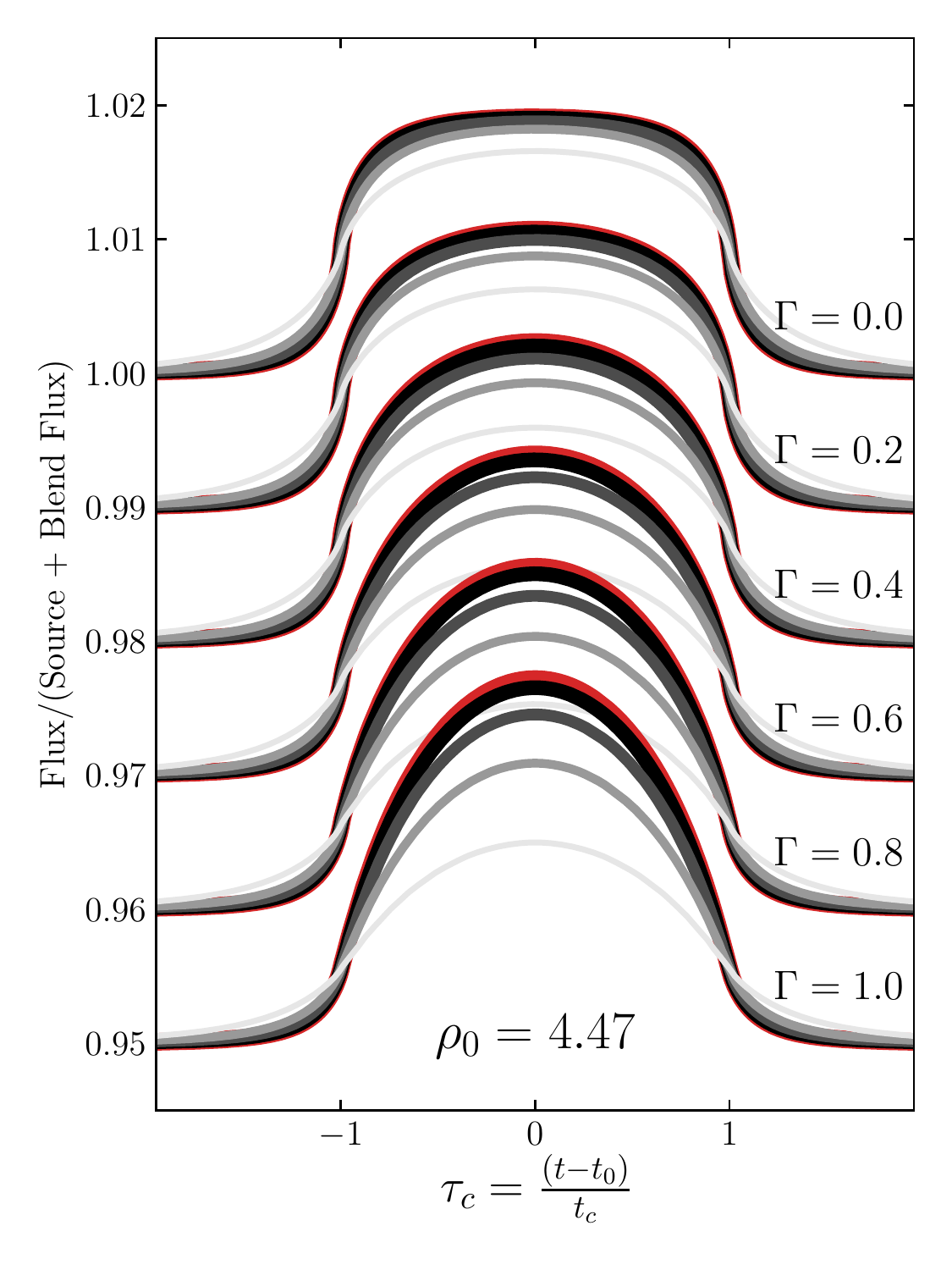}
\caption{Qualitative demonstration of the $t_*-b_0$ degeneracy, only matching the timescale.
For each $\Gamma$, five light curves are shown with a different value for the impact parameter $b_0=0.0, 0.2, 0.4, 0.6, 0.8$ from backmost to frontmost, respectively. For each, $t_*$ was varied match $\tsub{t}{FWHM}$.  
Note the decrease in $\Delta \tsub{F}{max}$ as $b_0$ increases even for $\Gamma = 0$.}
\label{fig:gammaustar_vary_muONLY}
\end{figure*}

In each of the following subsections, two figures depict qualitative and quantitative representations of different aspects of the degeneracy for Events 1 and 2.
The first of these of these two Figures contain two panels that show a qualitative comparison of sets of light curves with fixed limb-darkening.
The majority of light curves within a set lie on top of each other, making distinguishing them difficult. 
The color order is constant, so when a list of values is provided the respective line-coloring is green/red, black, dark gray, light gray, and white.
All of Figures \ref{fig:gammaFs_vary_rho}, \ref{fig:gammaustar_vary_muONLY}, and \ref{fig:gammaustar_vary_murhofs} show six sets of light curves, each with fixed linear limb-darkening parameters of $\Gamma=0.0$, 0.2, 0.4, 0.6, 0.8, and 1.0.
We will relax the assumption of fixed limb-darkening in Section \ref{sec:degenfull}.
The first set of light curves in a Figure also includes five fractional uncertainties in the flux, which have values of 1\%, 0.667\%, 0.5\%, 0.333\%, and 0.1\% from left to right. 
This visual comparison of the light curves demonstrates how similar they are, but leads to little understanding how severe the degeneracy is quantitatively.

\begin{figure*}[t]
\epsscale{\epsScaleFactorOne}
\plottwo{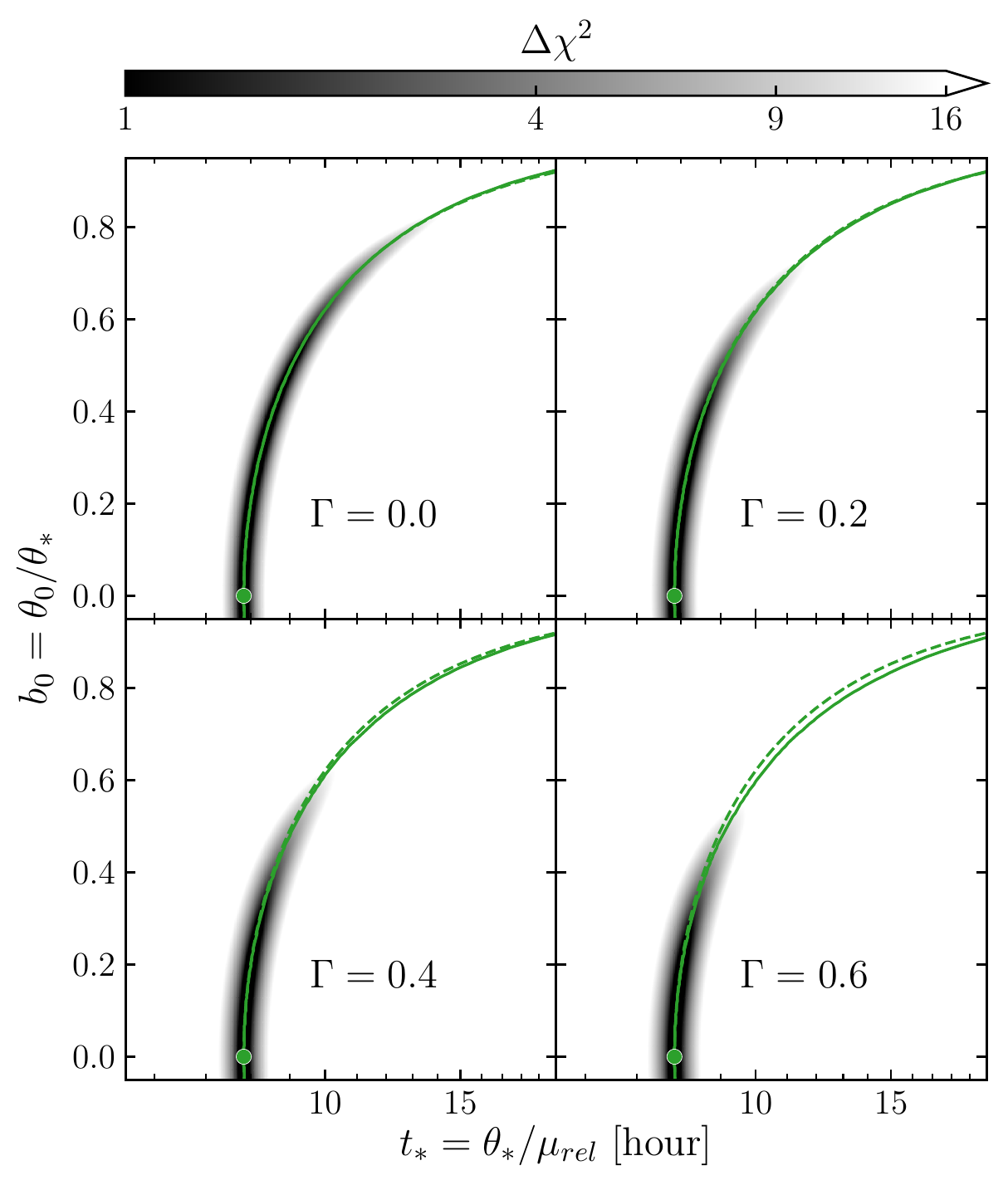}{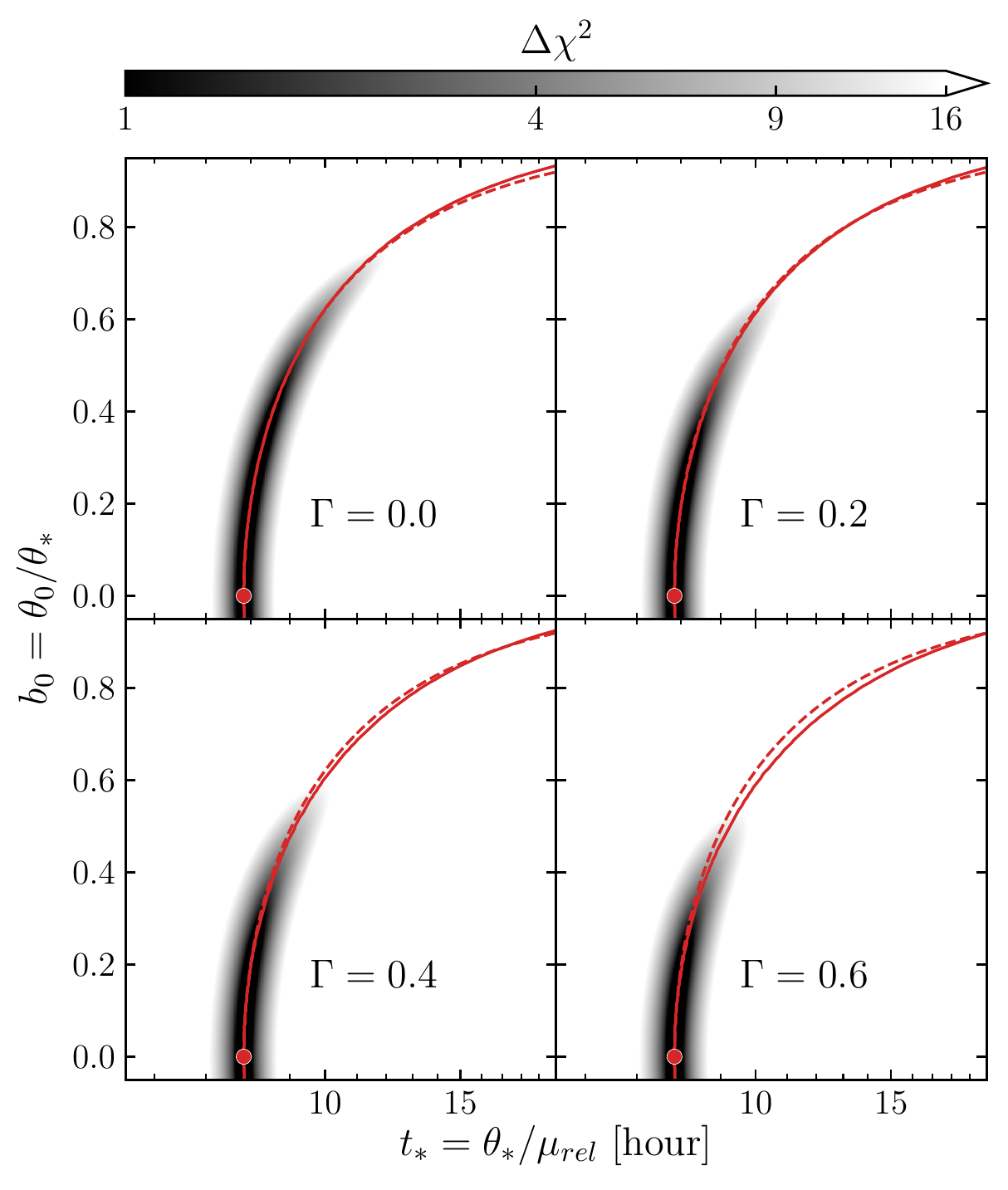}
\plottwo{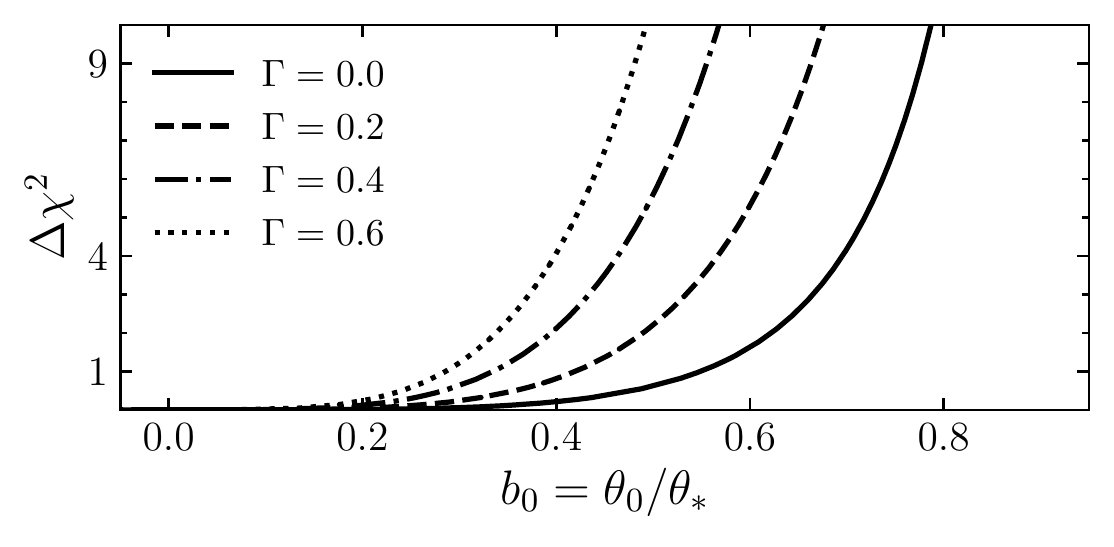}{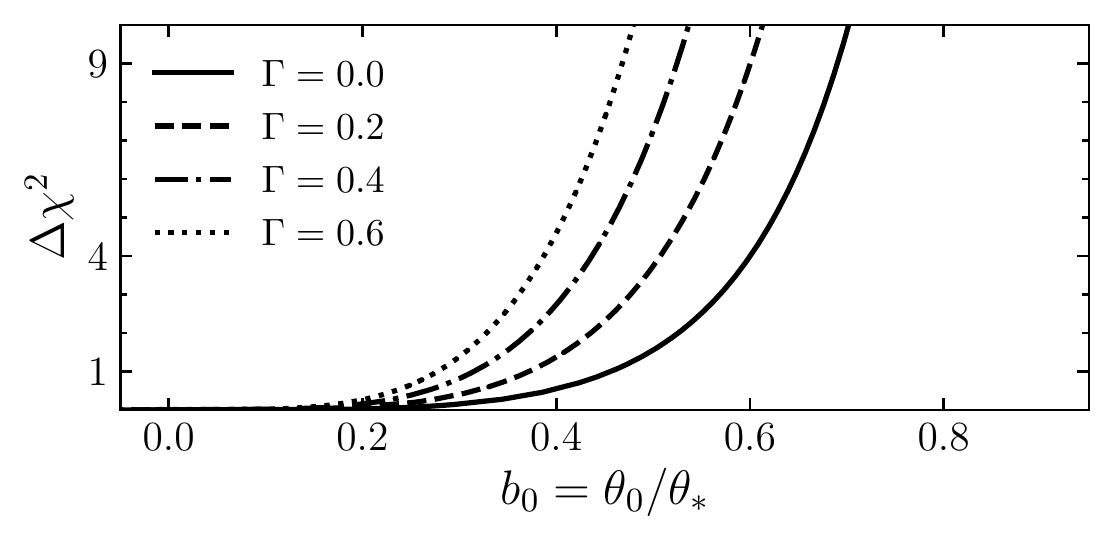}
\caption{Quantitative demonstration of the $t_*-b_0$ degeneracy, analogous to Figure \ref{fig:map.rhofs}. 
\textit{Upper panels}: at each $(t_*,b_0)$ pair, all other parameters are fixed at the fiducial values and the $\Delta\chi^2$ is calculated between the proposed and the fiducial models. 
No minimization is performed in these comparisons.
The analytic prediction for the degeneracy (Equation \ref{eqn:time_degen}) is shown as the dashed green/red lines and the calculated minimum is shown as solid green/red lines.
\textit{Lower panels}: the calculated minimum $\Delta\chi^2$ projected onto $b_0$.} 
\label{fig:map.tstarustarOnly}
\end{figure*}

\begin{figure*}[t]
\epsscale{\epsScaleFactorOne}
\plottwo{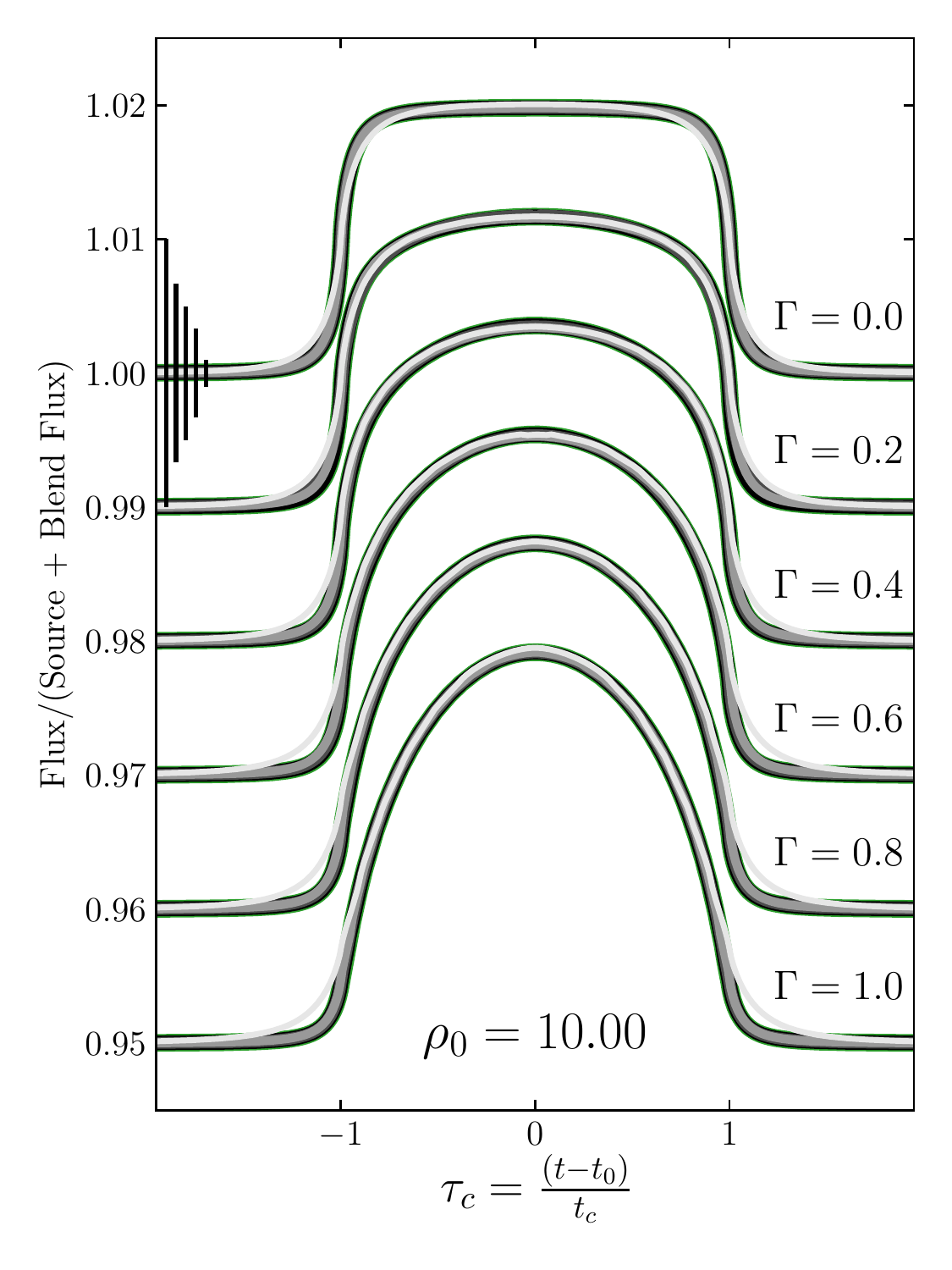}{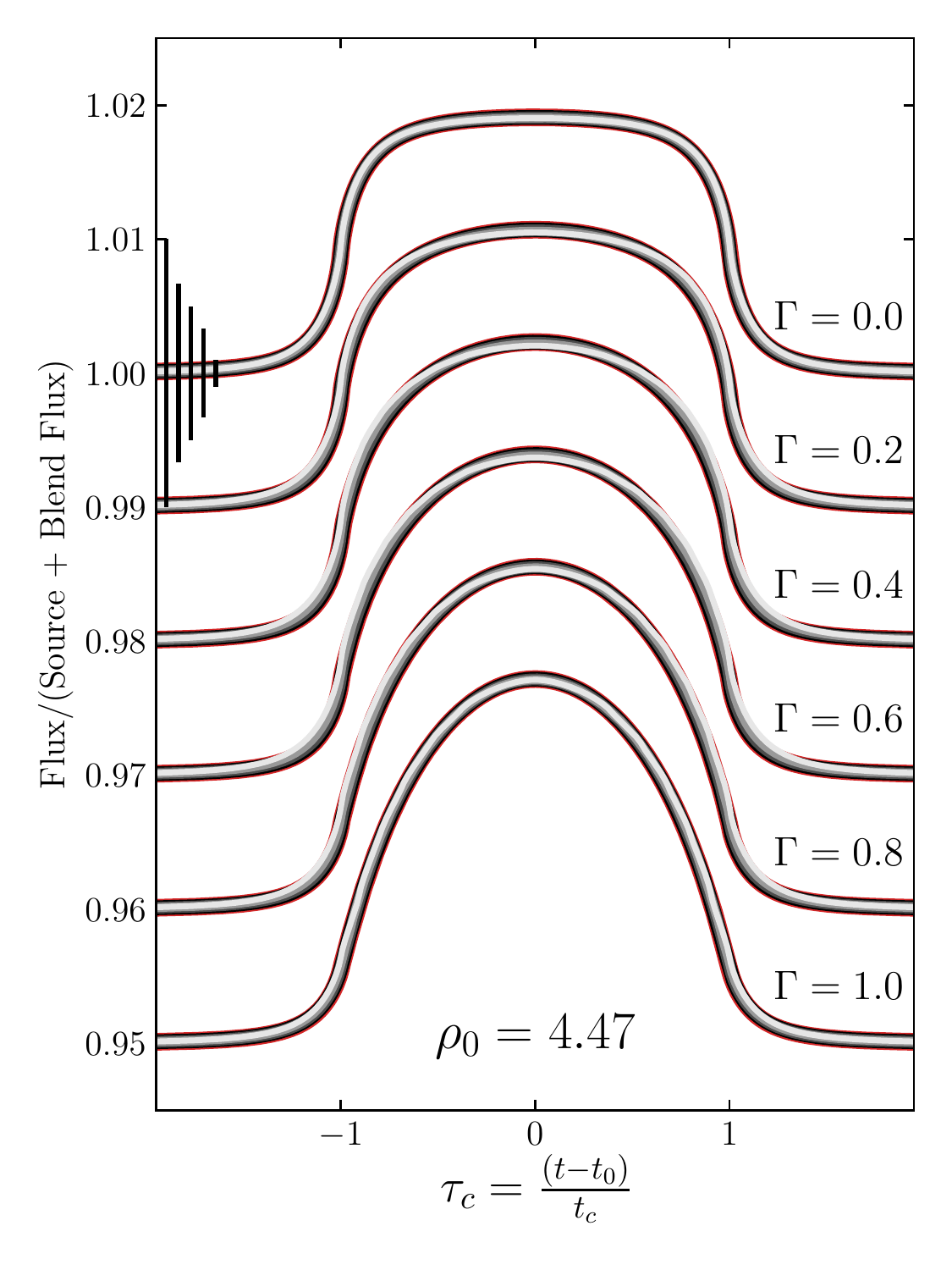}
\caption{
Qualitative demonstration of the $t_*-b_0-\rho-\tsub{F}{S}$ degeneracy.
For each of $\Gamma=0.0,0.2,0.4,0.6$, five light curves with $b_0=0.0, 0.2, 0.4,0.6,0.8$ are shown from backmost to frontmost light curves, respectively.
The timescales $\tsub{t}{FWHM}$ can simply be matched by changing $t_*$.
In this case, both $\tsub{f}{S}$ and $\rho$ were varied to account for the change in flux due to larger impact parameters/decreased surface brightness. 
We include the values of the fit parameters in Tables \ref{tbl:rho10} and \ref{tbl:rho45} in Appendix \ref{apdx:tables}.}
\label{fig:gammaustar_vary_murhofs}
\end{figure*}

\begin{figure*}[t]
\epsscale{\epsScaleFactorOne}
\plottwo{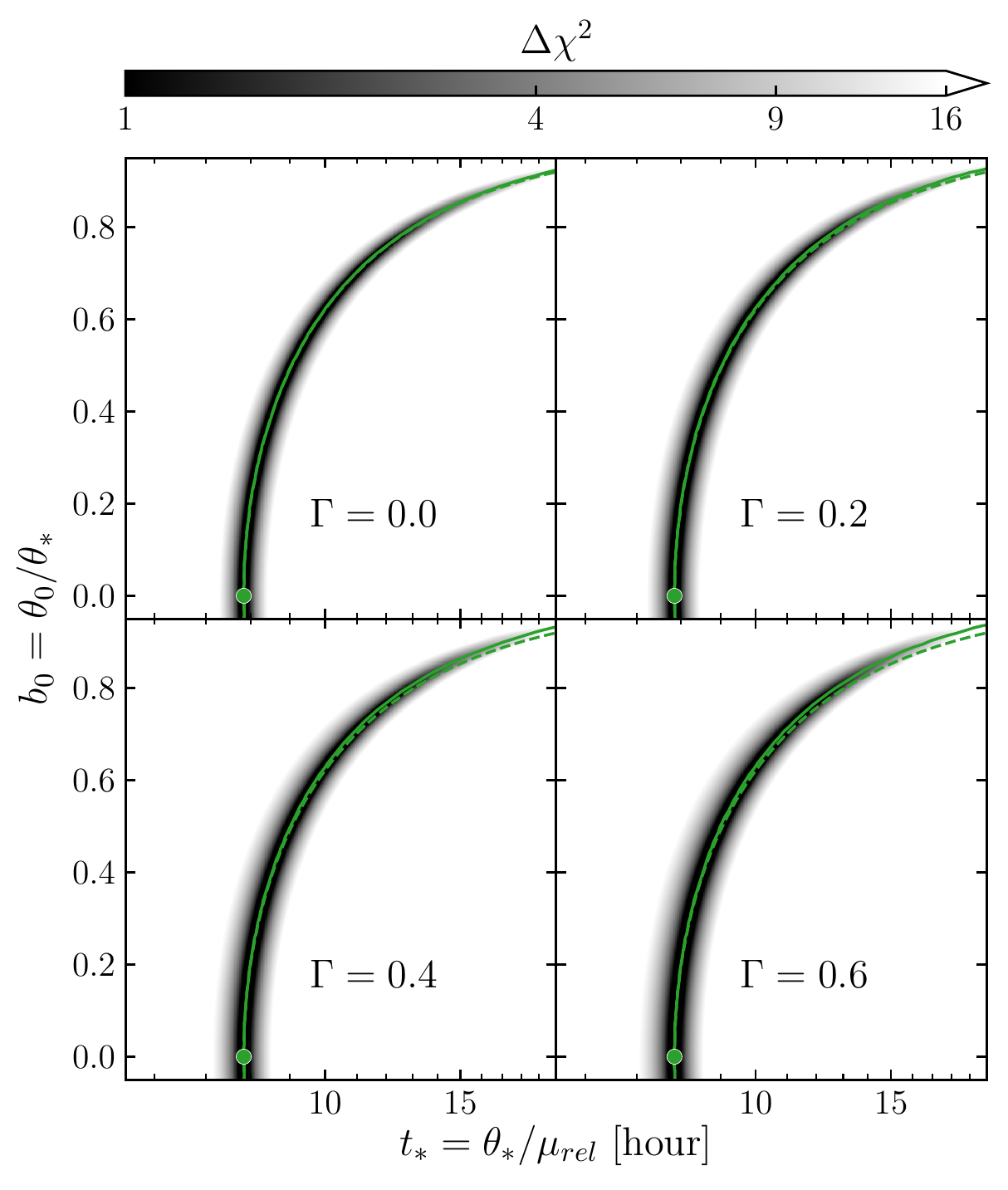}{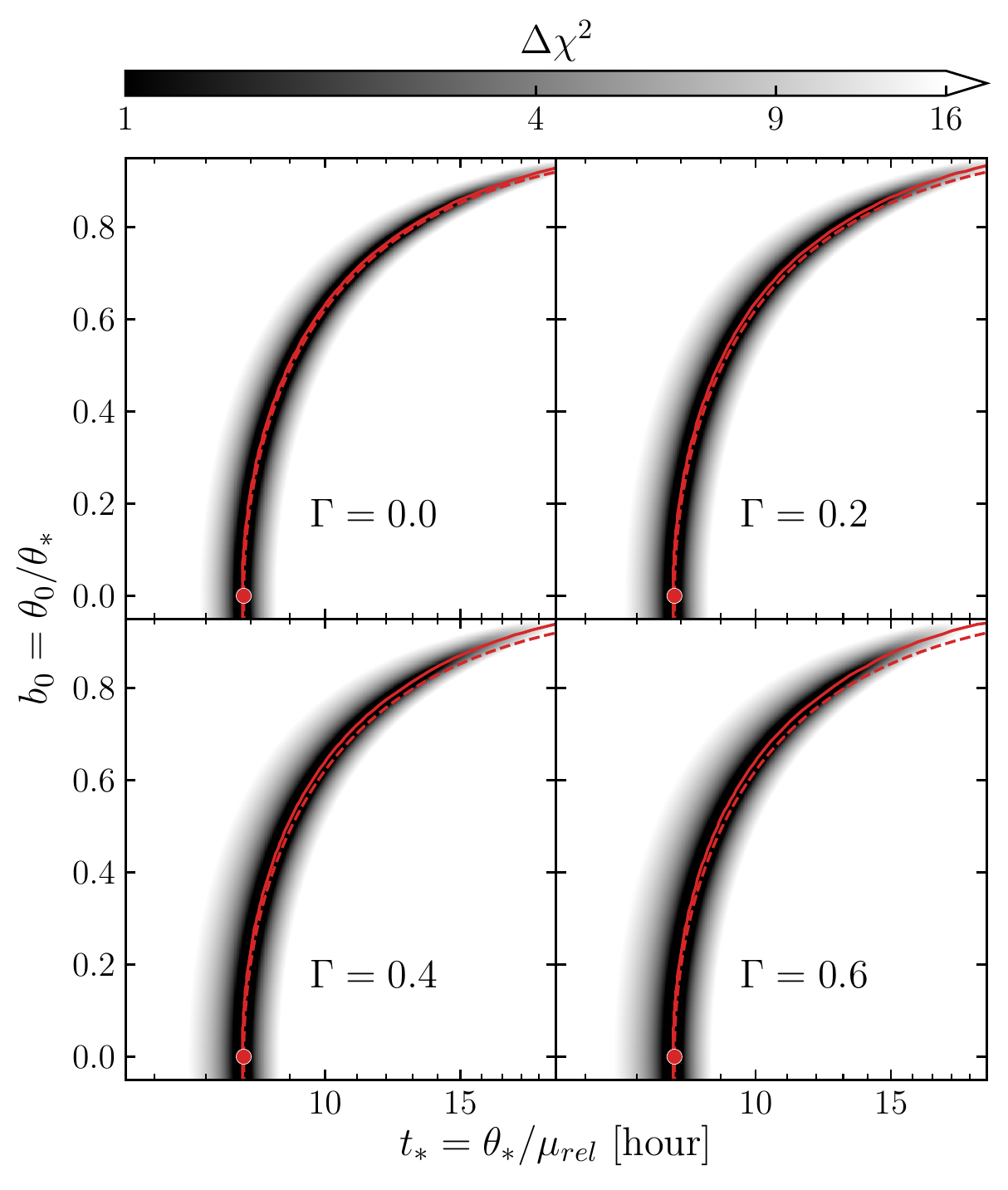}
\plottwo{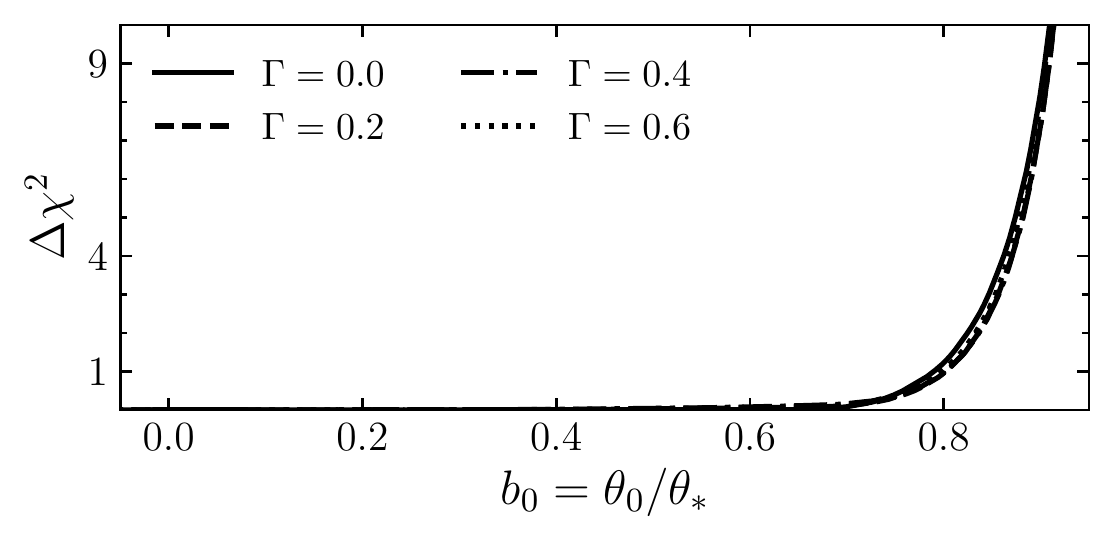}{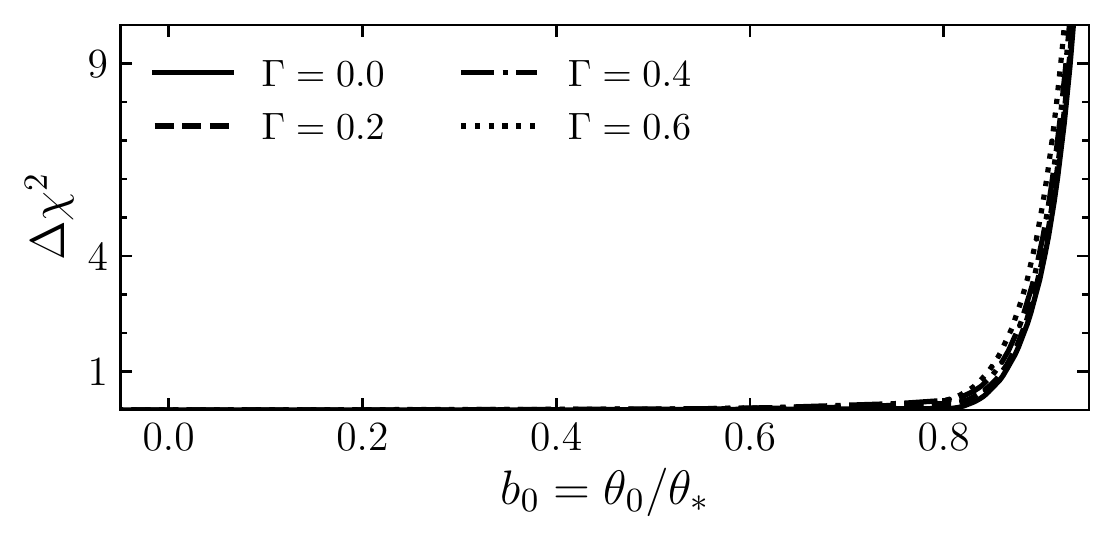}
\caption{Quantitative demonstration of the $t_*-b_0-\rho-\tsub{F}{S}$ degeneracy, similar to Figure \ref{fig:map.tstarustarOnly}.
\textit{Upper panels}: At each $(t_*,b_0)$ pair, we perform a grid search over $\rho$ and a least squares minimization over $\tsub{f}{S}$ to find those values which minimize the $\Delta \chi^2$ between the proposed and fiducial models.
\textit{Lower Panels}: The calculated minimum $\Delta\chi^2$ projected onto $b_0$.
}
\label{fig:map.tstarustarrhofs}
\end{figure*}

We therefore proceed to numerically evaluate the severity of these degeneracies. 
We do so using the $\Delta\chi^2$ statistic, which simply evaluates the difference in $\chi^2$ between a proposed model and the fiducial model above for the simulated data.  
In such a scheme, broad swaths of parameter space may have  $\Delta\chi^2 = \tsub{\chi}{trial}^2-\tsub{\chi}{fiducial}^2$ that is significantly less than $\sim 9$, or $3\sigma$, leading to poor constraints on the subset of degenerate parameters for those events.
The second figure in each subsection shows a map of $\Delta\chi^2$ values for a much finer grid of the relevant parameters for each both fiducial cases. 
These are shown in Figures \ref{fig:map.rhofs}, \ref{fig:map.tstarustarOnly}, and  \ref{fig:map.tstarustarrhofs}, where we include grey-scaled maps of the $\Delta\chi^2$ for a proposed model compared to the fiducial model under varying treatments for a labeled, fixed value of the limb-darkening parameter.
Across all of these Figures, we adopt a uniform $\sigma=0.5\%$ photometric uncertainty and a 15 min observation cadence in order to evaluate $\Delta\chi^2$. 
We do not introduce any noise in the photometric data points, such that each lies perfectly on the model of the event.  
This is appropriate as it represents the severity of the degeneracies when averaged over a large ensemble of light curve realizations. 
We parameterize our calculations in terms of the blending parameter $\tsub{f}{S}$ such that we can ignore units of flux. 
We limit $\tsub{f}{S}\le 3.0$, but note that the negative blend fluxes implied by source fluxes at the upper end of this range are unlikely to be physical.
Finally, we note that since we are adopting a constant fractional uncertainty, $\Delta\chi^2$ is proportional to the root-mean-square (RMS) fractional difference between the trial light curve and the fiducial light curve.

The top panels of Figures \ref{fig:map.rhofs}, \ref{fig:map.tstarustarOnly}, and  \ref{fig:map.tstarustarrhofs} have four subpanels, each corresponding to a fixed value of $\Gamma=0.0,0.2,0.4,$ or 0.6 as labeled in the subpanel.
In these Figures, we sample $t_*=\theta_*/\tsub{\mu}{rel}$ rather than $\tsub{\mu}{rel}$, but as we hold $\theta_*$ constant these are effectively interchangeable.
In each subpanel, we indicate the appropriate fiducial model parameters with a green/red circle.
We plot the trace of the minimum $\Delta\chi^2$ as a solid green/red line.
We also plot the appropriate analytic prediction (Equations \ref{eqn:flux_degen} or \ref{eqn:time_degen}) for the degeneracies as a dashed green/red line.
Note that all axes in these Figures are logarithmically scaled except for $b_0$.

In the bottom panel of Figures \ref{fig:map.rhofs}, \ref{fig:map.tstarustarOnly}, and  \ref{fig:map.tstarustarrhofs} we include a plot of the marginalized minimum $\Delta\chi^2$ as a function of the ordinate $\tsub{f}{S}$ or $b_0$ (the solid green/red line in the Upper Panels).
In these bottom panels, each value of $\Gamma=0.0, 0.2, 0.4$ and $0.6$ use a line style of solid, dashed, dot-dashed, and dotted, respectively. 

\subsection{The $\rho$--$\tsub{F}{S}$ degeneracy}
\label{subsec:rhofs}
We begin by examining the degeneracy that results from constraining $\Delta \tsub{F}{max}$, which depends on both $\rho$ and $\tsub{F}{S}$.
In Figure \ref{fig:gammaFs_vary_rho} we fix the impact parameter $b_0=0$ for each set of light curves.
For each $\Gamma$, five values of $\tsub{f}{S}=1.0, 0.8, 0.6, 0.4$, and 0.2 are plotted from the backmost to the frontmost light curve in the left panel for Event 1.
The order of these $\tsub{f}{S}$ values is reversed in the right panel for Event 2.
For each, we find the value of $\rho$ that minimizes $\Delta\chi^2$ with respect to the fiducial model.  
Note that to keep the timescale constant, $\theta_*$ and $\tsub{\mu}{rel}$ are held constant and we are only varying $\rho$ by changing $\tsub{\theta}{E}$.
Also note that the backmost (green/red) light curve in the $\Gamma=0.4$ sets are the fiducial light curves for Event 1 and Event 2.
In the $\Gamma=0.0$ set, we see a slight rounding during the event as $\tsub{f}{S}$ decreases in the left panel, and the opposite on the right. 
This is caused by the compensating increase (decrease) in $\rho$ through $\tsub{\theta}{E}$, which causes a departure from (to) the EFSE regime. 
However, as $\Gamma$ increases to 0.6 the shape induced by the presence of limb-darkening decreases this discrepancy near the peak of the event.
For all values of $\Gamma$, as $\rho$ decreases the wings and shoulders of the event becomes more prominent as $\tsub{f}{ws}$ increases (Equation \ref{eqn:fws}). 
This is again due to the increase in the size of the Einstein ring, which begins to magnify the source at earlier times during the event. 
In both panels, we see that even just by varying $\rho$ and $\tsub{f}{S}$ that the $\Delta\tsub{F}{max}$ degeneracy is quite severe, with fractional differences on the order of 0.1\%. 
 
Next we perform the quantitative investigation of the $\rho-\tsub{F}{S}$ degeneracy. 
In Figure \ref{fig:map.rhofs} we fix all event parameters to those of the fiducial models and find the $\Delta\chi^2$ for each ($\rho,\tsub{f}{S}$) pair in the grids mapped in the upper panels. 
We perform no minimization in computing the $\Delta\chi^2$ of these grids.
In both of the upper panels, we see a region of ($\rho,\tsub{f}{S}$) pairs that follow the analytic prediction between $\rho$ and $\tsub{f}{S}$ (Equation \ref{eqn:flux_degen}, dashed green/red line). 
We see that the analytic prediction of the minimum $\Delta\chi^2$ nearly follows the numerical values except for a departure at small $\rho$ where the events leave the EFSE regime (e.g., $\rho$ becomes comparable to unity). 
In the upper right panel (Event 2), the range of ($\rho,\tsub{f}{S}$) pairs that have $\Delta\chi^2<9$ increases significantly as Event 2 has values that better explore the physical extent of the degeneracy.
In the bottom panels, we see that for both Events 1 and 2 the numerical minimum traces of $\Delta\chi^2$ projected onto $\tsub{f}{S}$ have large ranges with values less than 9 (3$\sigma$). 
The numerical traces of $\Delta\chi^2$ for Event 2 are lower than those for Event 1, especially for small $\tsub{f}{S}$ where Event 1 would require values of $\rho$ that depart from the EFSE regime. 
However, both Events 1 and 2 have definite minima at the fiducial values of $\tsub{f}{S}$.
Also note that as $\Gamma$ increases, we see a slight increase in the range of $\tsub{f}{S}$ values that have $\Delta\chi^2<9$ as the shape induced in the light curve by the limb-darkening of the source better matches the topology of the peak of the light curve (see Figure \ref{fig:gammaFs_vary_rho}). 

For both Events 1 and 2, we see that by only altering $\rho$ and $\tsub{F}{S}$ the degeneracy is strong but resolvable. 
These two parameters can alter $\Delta\tsub{F}{max}$ of an event, but the detailed shape of the light curve will lead to a potential resolution in this aspect of the degeneracy if the photometric precision is sufficiently high. 

\subsection{The $t_*$-$b_0$ degeneracy}
Next we investigate the the piece of the degeneracy that results from constraining $\tsub{t}{FWHM}$, which is the trade-off between $t_*$ and $b_0$.
However, in this section we are only matching the timescale of the event; in particular we are not modifying the $\Delta\tsub{F}{max}$ of the event through changing either $\rho$ or $\tsub{F}{S}$.
In both panels of Figure \ref{fig:gammaustar_vary_muONLY}, we plot the six sets of light curves for each value of $\Gamma$.
For each $\Gamma$, five values of $b_0=0.0, 0.2, 0.4, 0.6,$ and $0.8$ are plotted from the backmost to the frontmost light curve. 
We then vary $t_*$ to match $\tsub{t}{FWHM}$ without accounting for the change in flux. 
Note that for increasing impact parameters, $\Delta \tsub{F}{max}$ decreases even in the absence of limb-darkening. 
This is due to the fact that as the impact parameter approaches the edge of the source ($\beta \rightarrow 0$ or $b_0\rightarrow1$), significant regions of the lens' magnification pattern will lie outside of the source.
This effect is more prominent for Event 2, as the Einstein ring of the lens falls further off the disk of the source.
Also note that the wings of the events become much more prominent as $b_0$ increases causing $\tsub{f}{ws}$ to increases (Equation \ref{eqn:fws}). 
This is because $\tsub{\mu}{rel}$ for these events must decrease, leading to a longer ramp-up and -down.
This is fundamentally different than the previous case for the discrepancy in wing shape coming from the $\rho-\tsub{F}{S}$ trade-off.
For the second model, this change in morphology is actually less significant as $\tsub{f}{ws}$ is less impacted due to the higher value of $\rho$.
Overall, we see that this aspect of the degeneracy must act in concert with matching $\Delta\tsub{F}{max}$ for it to become severe.

Next we investigation of the $t_*-b_0$ degeneracy more quantitatively.
Analogous to Figure \ref{fig:map.rhofs}, Figure \ref{fig:map.tstarustarOnly} displays the map of $\Delta\chi^2$ values across a grid of $(t_*,b_0)$ pairs of values for the same four values of $\Gamma$.
We fix all event parameters to those of the fiducial models and find the $\Delta\chi^2$ for each $(t_*,b_0)$ pair.
Similarly, no minimization is performed here.
The analytic prediction of the minimum (Equation \ref{eqn:time_degen}) nearly perfectly approximates the calculated minima for these maps. 
When the limb-darkening profile is steeper however, $\Delta\tsub{F}{max}$ decreases more and more as $b_0$ increases to chords of the source with surface brightnesses much lower than at the center of the source.
The impact of not altering $\Delta\tsub{F}{max}$ is evident by the region of small $\Delta\chi^2$ becoming more localized near the true values of $(t_*,b_0)$ of the event for increasing $\Gamma$. 
This is clearly shown in the bottom panels, where the traces of the minimum with $\Delta\chi^2<9$ cover consecutively smaller ranges of $b_0$ for increasing $\Gamma$.
In fact, the larger value of $\rho$ in Event 2 results in the range of $b_0$ values with $\Delta\chi^2<9$ being smaller than for Event 1 due to the region of magnification falling off of the disk of the source at large impact parameters.
In contrast though, the range of $t_*$ values for Event 2 with small $\Delta\chi^2$ is wider about the minima.
This is because the larger fiducial value of $\rho$ in these events lead to a more commensurate value of $\tsub{f}{ws}$.
In the bottom panels, we see that the trace of the minima $\Delta\chi^2$ is essentially flat and much smaller than one for $b_0\in[0,0.2]$ for all values of $\Gamma$ as these chord lengths are all essentially the same and the impact of limb-darkening is small for these impact parameters. 

In all, this aspect of the degeneracy manifests most strongly in a range of impact parameters that are close to zero, where $\Delta\tsub{F}{max}$ is minimally impacted by limb-darkening and chord crossing times vary little. 
That said, this aspect of the degeneracy is not as significant, as $b_0$ is not a physically informative parameter and for much of its range of values the value of $t_*$ stays relatively constant.
If the impact parameter of the event increases too much though, then larger regions of the magnification pattern will lie outside the source, resulting in a deficit in $\Delta\tsub{F}{max}$ that must be compensated for by variation of other parameters.

\subsection{The $t_*$-$b_0$-$\tsub{F}{S}$-$\rho$ degeneracy}

Finally, we consider the $t_*$-$b_0$ degeneracy but instead we hold $\Delta\tsub{F}{max}$ constant by varying both $\tsub{F}{S}$ \textit{and} $\rho$.
We demonstrate the $t_*-b_0-\rho-\tsub{F}{S}$ degeneracy in Figure \ref{fig:gammaustar_vary_murhofs} for the same six values of $\Gamma$.
For each $\Gamma$, five values of $b_0=0.0, 0.2, 0.4, 0.6,$ and $0.8$ are chosen.
For each, we vary $t_*$ to match $\tsub{t}{FWHM}$ and then find the values of $\tsub{f}{S}$ and $\rho$ that minimizes $\Delta\chi^2$ with respect to the fiducial model using least squares minimization. 
In all cases in Event 1, we see extremely close agreement between the light curves except for the $b_0=0.8$  case (the white light curve) where the wings are still more prominent. 
This is due to the prior that $\tsub{f}{S}\le3.0$, requiring the value of $\rho$ to increase beyond the expected analytic value, causing an increase in $\tsub{f}{ws}$.
For Event 2, we see the expected near-perfect degeneracy in the $\Gamma=0.0$ and 1.0 cases, and only slight variations for intermediate values of $\Gamma$.
The values of the parameters for these light curves are included in Appendix \ref{apdx:tables} in Tables \ref{tbl:rho10} and \ref{tbl:rho45} for Events 1 and 2, respectively.

The quantitative investigation of this four-parameter degeneracy demonstrates its severity.
For each fixed pair of $(t_*,b_0)$ values, we vary both $\tsub{f}{S}$ \textit{and} $\rho$ to minimize the $\Delta\chi^2$ between the events. 
To do so, we use a grid search over $\rho$ and a least squares minimization over $\tsub{f}{S}$.
We use a grid search in $\rho$ (rather than least squares minimization as before) because the minimization algorithm found local minima in unphysical locations far from the trace of the minimum $\Delta\chi^2$. 
Our grid search for $\rho$ was logarithmically spaced over $\log(\rho)=[-2,\log(30)]$ and we maintained the prior that $\tsub{f}{S}<3.0$.
Compared to the plots in Figure \ref{fig:map.tstarustarOnly}, we see a significant increase in the area of the $\Delta\chi^2\gtrsim4$ values and a dramatic increase in the range of $b_0$ values out to $\sim0.8$ that have $\Delta\chi^2\lesssim1$.
The extent of this region is slightly larger for Event 2.
Although not visible in the lower panel, the average value of $\Delta\chi^2$ along the minimum for $b_0<0.8$ dropped by roughly an order of magnitude from $\Delta\chi^2{\sim}0.1$ to ${\sim}0.01$, and so the degeneracy is more challenging to break via improved photometric precision. 
By allowing both $\rho$ and $\tsub{f}{S}$ to vary for fixed limb-darkening, the true nefarious nature of this degeneracy becomes increasingly apparent. 

Even when isolating aspects of the degeneracy by varying a subset of parameters, the maximum differences in the light curves can be on the order of 0.1\%. 
The most obvious deviations are in the wings/shoulders of the events with $b_0$=0.8, where the changes in $\rho$ or limb-darkening profile at different impact parameters are strongest.
However, we note that these light curves still do not demonstrate the most severe form of the degeneracy discussed in this paper. 
The most severe form of the degeneracy involves varying $t_*$, $\rho$, $b_0$, $\tsub{F}{S}$, {\it and} $\Gamma$ simultaneously in order to minimize the $\chi^2$ or the root-mean-square of the difference in the light curves with respect to the fiducial light curve. 
We further explore the isolated degeneracies and the ``complete" degeneracy in Section \ref{sec:degenfull}.

\section{Investigation of the Full Degeneracy}
\label{sec:degenfull}

In the previous section, we only considered varying two, three, or four parameters relevant in this degeneracy at a time.
However, the full degeneracy involves five free parameters: $b_0$, $t_*$, $\tsub{F}{S}$, $\rho$, and $\Gamma$, with four observables (ignoring the central time of the event $t_0$, which is not degenerate with the above parameters, and eliminating $\tsub{F}{base}$, as discussed previously). 
Exploring the degeneracy by varying these five parameters simultaneously to best match the light curve (in the $\Delta\chi^2$ sense) of the fiducial parameters involves a multi-parameter minimization, which is more well suited to Markov Chain Monte Carlo techniques.  
We use the Markov Chain Monte Carlo (MCMC) package {\tt emcee}~ \citep{2013PASP..125..306F} to sample all of the parameters for both Event 1 and Event 2 (see Section \ref{sec:fiducial}).
This includes those five listed above as well as $t_0$ and $\tsub{F}{base}'/\tsub{F}{base}$.
We sample $\tsub{F}{base}'$ normalized to the fiducial value of $\tsub{F}{base}$ instead of $\tsub{f}{B}$ as we can sample values logarithmically (similar to how we sample $\tsub{f}{S}$), which allows $\tsub{f}{B}$ to take on negative values to account for negative blending ($\tsub{f}{S}>1$).
We do not introduce any noise or scatter to the data points of the light curves.
We consider three values of the relative photometric precision per measurement: 0.100\%, 0.333\%, 0.667\%. 
For each value of the photometric precision, we run 20 separate chains until all parameters converge with autocorrelation times $>25$\footnote{We note this a formally weak convergence criteria, but it is sufficient for this demonstration.
Most parameters had autocorrelation times$>50$ when the runs were terminated.}.
We start each chain at a range of positions along the degeneracies and discard the first 1000 samples as a burn-in.
We use wide, uninformative (uniform) priors to prevent the chains running into artificial boundaries.
These are listed in Table \ref{tbl:priors}.

\begin{figure*}[ht]
\epsscale{\epsScaleFactorTwo}
\plotone{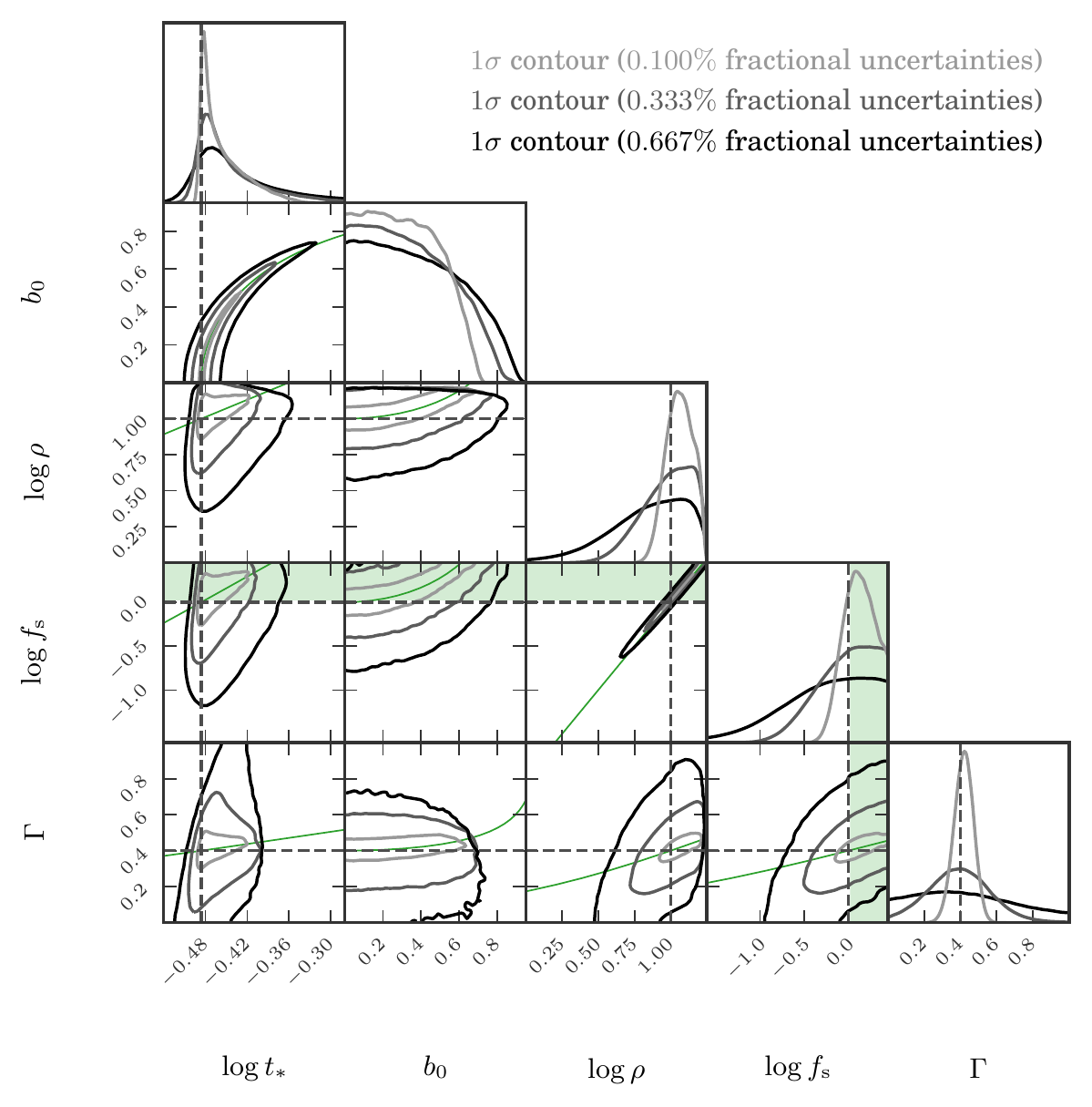}
\caption{Three MCMC posterior samples for Event 1 but with differing photometric uncertainties.
The black contour is the $1\sigma$ boundary for the event with 0.667\% fractional photometric uncertainties, dark gray for 0.333\%, and light gray for 0.1\%.
This is for Fiducial Event 1 detailed in Section \ref{sec:fiducial}, with true values marked by dashed lines. 
We also included $t_0$ and $\tsub{F'}{base}/\tsub{F}{base}$ as free parameters, but excluded them in the plot as they do not contribute to the degeneracies being investigated here.
We include green lines in all the panels the follow the relations from Equation \ref{eqn:degenldfiniterho2}.
Green shaded regions denote values of $\tsub{f}{S}>1.05$ which would require a significant amount of negative blending.}
\label{fig:mcmcrho10}
\end{figure*} 

\begin{table}[b]
  \caption{Priors used in MCMC investigation. $\mathcal{U}(X,Y)$ indicates a uniform distribution with lower bound $X$ and upper bound $Y$.}
  \begin{tabular}{ll}
    Parameter & Prior\\
    \hline
    $t_0~[\textrm{days}]$ & $\mathcal{U}(-4,4)$\\
    $b_0$ & $\mathcal{U}(-3,3)$\\
    $\Gamma$ & $\mathcal{U}(0,1)$\\
    $\log(t_*/\textrm{days})$ & $\mathcal{U}(-3,1)$\\
    $\log(\rho)$ & $\mathcal{U}(-1,2)$\\
    $\log(\tsub{f}{S})$ & $\mathcal{U}(-3,0.477)$\\
    $\log(\tsub{F}{base}'/\tsub{F}{base})$ & $\mathcal{U}(-3,0.477)$\\
    \hline
  \end{tabular}
  \label{tbl:priors}
\end{table}

In these chains, we sample $t_0$, $\Gamma$, and $b_0$ in linear space.
For $b_0$, we sample linearly to allow for negative impact parameters.
This way, there is no artificial boundary imposed that prevents the impact parameter from crossing the equator of the source star. 
We then simply find the absolute value of the impact parameters and then include those links in the chain in the Figures. 

\begin{figure*}[ht]
\epsscale{\epsScaleFactorTwo}
\plotone{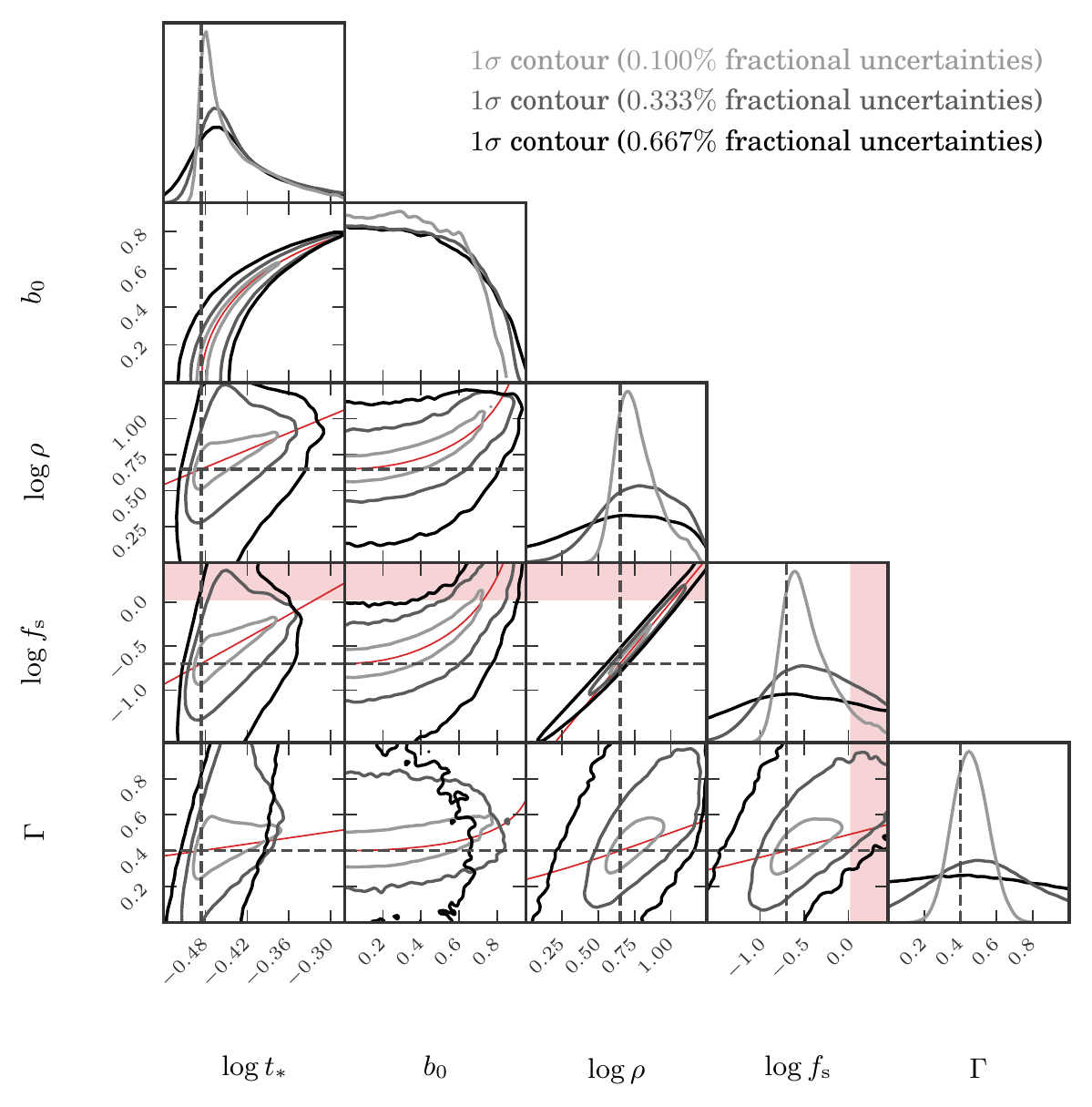}
\caption{Same as Figure \ref{fig:mcmcrho10} but for Event 2. 
As this event has parameters that allow a fuller range of the degeneracy to manifest, the contours demonstrating the degeneracies increase dramatically.
We note that in some cases, the gray contours extend beyond the black contours as these are independent MCMC runs.}
\label{fig:mcmcrho45}
\end{figure*} 

We show the posterior distributions of these MCMC runs in Figures \ref{fig:mcmcrho10} and \ref{fig:mcmcrho45}.
We also include the predicted degeneracy from Equation \ref{eqn:degenldfiniterho2} as red/green lines in each subpanel.
These lines agree with the sampling fairly well near the event parameter values, but as parameters stray from their true values, the contours truncate the degenerate relationship between many of these parameters. 
Especially notable are that the contours from the MCMC run with the lowest assumed photometric uncertainty which tightly trace the $\rho-\tsub{F}{S}$ and $b_0-t_*$ degeneracies. 
These tight correlations expand as the photometric precision increases to 0.667\%, evident by the neatly nested 1-$\sigma$ contours in the appropriate subpanels of Figure \ref{fig:mcmcrho10} and \ref{fig:mcmcrho45}.
As the photometric precision increases from 0.01\% to 0.667\%, we see an increase in the area of the projected posterior distributions of $\rho$ and $\tsub{f}{S}$, although they are still relatively centered on their fiducial values. 
We also shade the regions with $\tsub{f}{S}>1.05$ green/red to indicate regions in parameter space that would require a significant amount of negative blending.
This is much more restrictive for Event 1 (Figure \ref{fig:mcmcrho10}) than for Event 2 (Figure \ref{fig:mcmcrho45}), as the fiducial value for Event 1 is $\tsub{f}{S}=1.0$. 

The contours for $b_0$ and $t_*$ are also neatly nested, but the posterior for $b_0$ is wide with a steep cliff around $b_0\approx0.6-0.8$. 
This is due to that fact that events with $b_0\lesssim\sqrt{2}$ all have very similar cord lengths and thus similar event durations.
Despite this, $t_*$ \textit{appears} to be well-recovered from its posterior distribution.

\begin{figure}
\epsscale{\epsScaleFactorOne}
\plotone{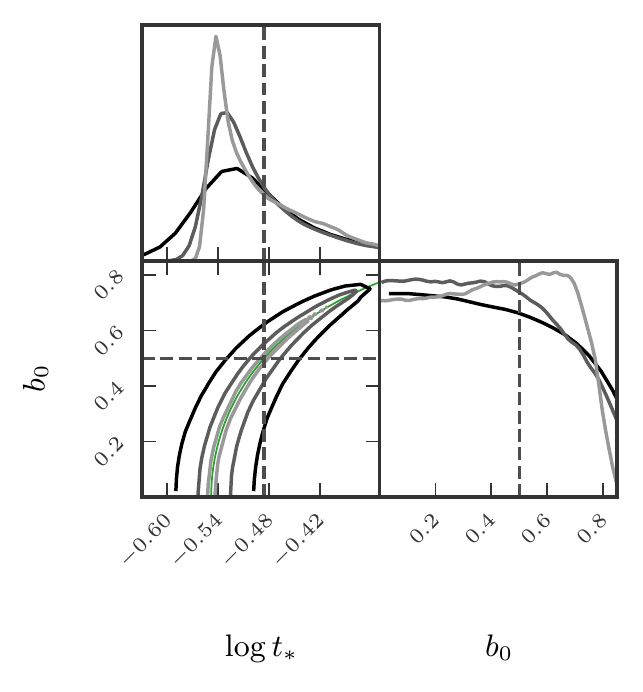}
\caption{Same as Figure \ref{fig:mcmcrho10} (including line colors), but the event has a fiducial value of $b_0=0.5$ rather than zero and $\tsub{f}{S}=0.75$ rather than one.
All other parameters are equal.
Here, we can see that the posterior for $t_*$ systematically underestimates its value. 
This can be potentially mitigated by using a more informative prior on $t_*$.}
\label{fig:mcmcb005}
\end{figure}

We also perform the same MCMC analysis on a third event akin to Event 1, except the only differences are that $b_0=0.5$ and $\tsub{f}{S}=0.75$.
All other values and priors are identical, and the difference in $\tsub{f}{S}$ is negligible. 
The posteriors for just $t_*$ and $b_0$ are shown in Figure~\ref{fig:mcmcb005}.
Perhaps the most notable difference is that the posterior of $t_*$ systematically underestimates its value. 
The shape of the posterior can be understood as a sharp rise resulting from a firm lower bound on $t_{\ast}$, and a long tail to infinite $t_{\ast}$ resulting from the chord length approaching zero.
However, as the chord length for impact parameters $b_0\lesssim\sqrt{2}$ are similar and in fact more probable when uniformly distributed, smaller values of the impact parameter are preferred.
This explanation is corroborated by the fact that there is a (weak) positive correlation between $\log{t_*}$ and $\log{\rho}$ in Figures \ref{fig:mcmcrho10} and \ref{fig:mcmcrho45}.  
As $\rho$ increases, the change in the morphology of the light curve decreases, and thus $t_*$ becomes more degenerate.  

We identify correlations of varying strengths between nearly all the parameters shown in Figures \ref{fig:mcmcrho10} and \ref{fig:mcmcrho45}. 
The weakest are between $b_0$ and the other parameters, and essentially only manifest for large impact parameters approaching the limb of the source star. 
This not surprising, as the magnification in the EFSE regime is approximately independent of $b_0$, except when $b_0 \simeq 1$ \citep{agol2003}. 
For Event 1, and as noted before, $t_*$ has a slight positive correlations with both $\rho$ and $\tsub{f}{S}$, however the $\tsub{f}{S}$ correlation occurs mostly for ranges where significant negative blending is required.  
The correlations with $\rho$ are restricted (e.g., with $\Gamma$) as they would require values of $\tsub{f}{S}$ that are outside of the prior distribution. 
We also note that the posterior of $\tsub{f}{S}$ would systematically over predict its value, despite requiring a likely unphysical value. 

We see many of the same patterns for Event 2 in Figure \ref{fig:mcmcrho45}. 
However, many of the correlations identified for Event 1 are much more apparent as the fiducial values for this event are in regions of parameter space that exacerbate the degeneracy. 
Specifically, the correlations between $t_*$ and $b_0$ and the other parameters are all more realized for Event 2. 
The correlation between $\rho$ and $\Gamma$ is also more apparent. 
And as the value of $\tsub{f}{S}$ is lower, the physical regions of parameter space for these degeneracies are much larger. 

Overall, between Events 1 and 2 we observe the correlations derived and expected in earlier sections. 
All five parameters show correlations with each other with varying severity that depends on the true values of these parameters in the events. 
There is good agreement between our analytical predictions for the form of the degeneracy, and the shape of the posterior for each pair of parameters.
This shows that our analytic approximations do lend insight into the nature of this degeneracy.
Next, we discuss in more detail the severity of this degeneracy with focus on the aspects that are purely mathematical and those that could arise physically. 

\section{Mathematical versus Physical Degeneracy}
\label{sec:mathvsphysical}

We have shown there is a strong \textit{mathematical} degeneracy for EFSE microlensing events. 
However, as we noted earlier, there are values of many of these parameters that while mathematically possible are physically unlikely or even impossible.

One case to consider is extremely large negative blending with $\tsub{f}{S}>1$. 
As described previously, while negative blending may at first sight appear unphysical, it does occasionally manifest itself in observed microlensing events.  
Negative blending typically occurs in events whose source stars reside in a local deficit in the mottled, semi-resolved background of fainter stars that is omnipresent in crowded microlensing fields. 
However, even when present, negative blending generally does not lead to $\tsub{f}{S}\gg1$.
In our investigations, we allow for some cases of extreme negative blending by placing a prior of $\tsub{f}{S}<3$. 
This allows us to further explore the degeneracy, as while values of $\tsub{f}{S}\gtrsim1.05$ are mathematically capable of satisfying the $\rho$--$\tsub{F}{S}$ degeneracy, they are physically unlikely to become realized in actual observations.
In real world cases, priors on blending can be placed based on the characteristics of individual events, in order to constrain the region of mathematically allowable parameter space to one that is physical. 
For example, negative blending for bright clump giant sources is extremely unlikely, as these sources are much brighter than the partially-resolved stellar background \citep{mroz2020a}. 

To this effect, almost all EFSE free-floating planet events reported to date have source stars that are giants or sub-giants in the bulge \citep{mroz2018,mroz2019,mroz2020a,mroz2020b}. 
In these cases, it is a good assumption that the source flux is essentially the baseline flux (e.g., \citealt{mroz2020b}).
However, unlike ground-based surveys, some fraction of EFSE events detected through the \textit{Nancy Grace Roman Space Telescope} (\romanst) Galactic Bulge Time Domain Survey will have main sequence sources for which such an assumption may not be valid.  
In these cases, each situation must be considered individually.
Companions to the source or lens, or potentially undetected stars which are blended with the source-star's point spread function, could lead to FFP candidates in which the blend flux could be significant \citep{johnson2020}.
As such, the physically plausible range of the mathematical degeneracy described here between $\tsub{F}{S}$ and $\rho$ may be considerably larger than for the cases of brighter source stars. 

\begin{figure}[t]
\plotone{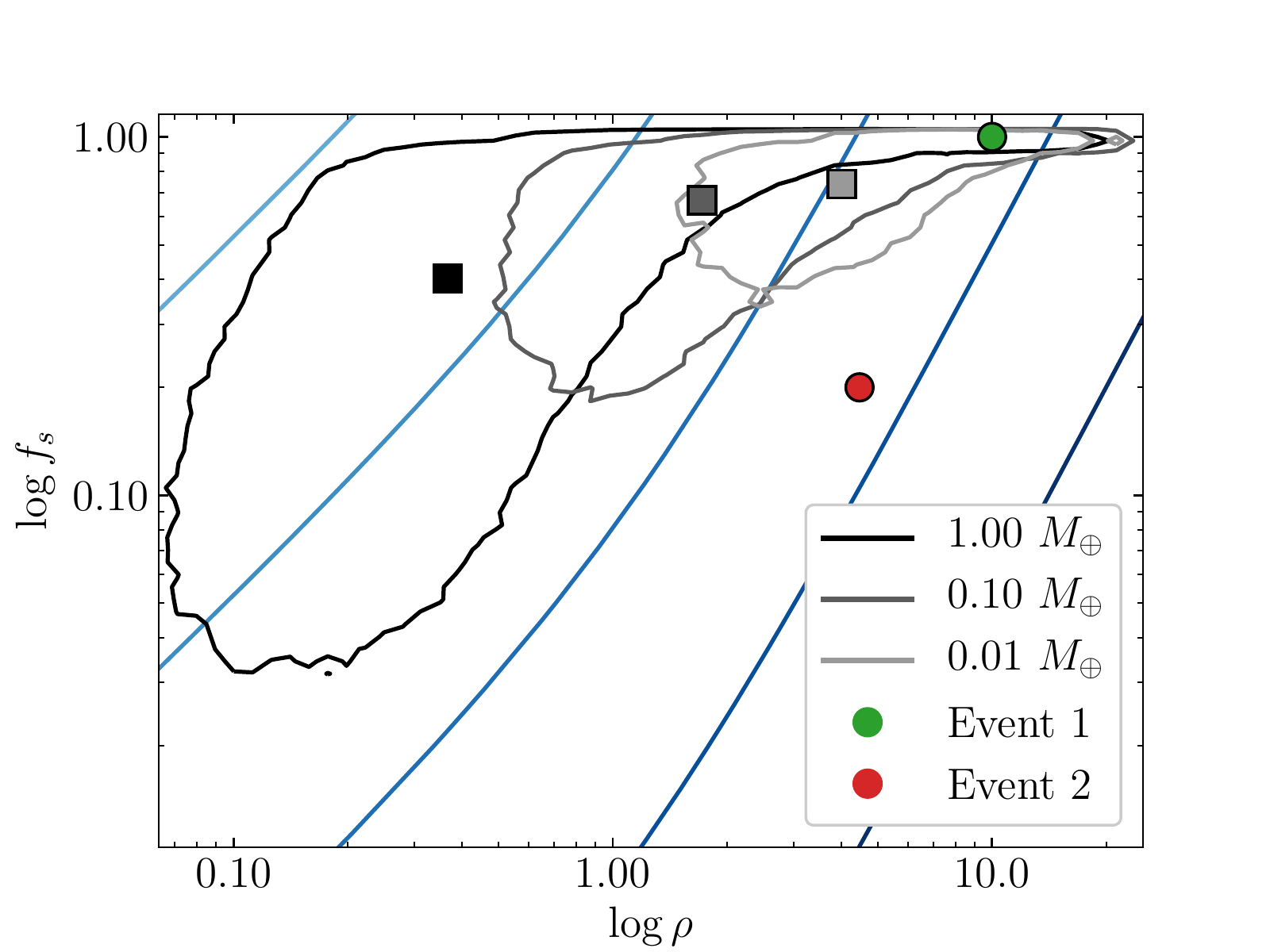}
\caption{The $\rho-\tsub{f}{S}$ distribution of free-floating planet (FFP) events detected by \textit{Roman } \citep{johnson2020}.
The open contours enclose 95\% of events with lens masses of 1.00 $M_\oplus$ (black), 0.10 $M_\oplus$ (gray), and 0.01 $M_\oplus$ (light gray) and a color-matched square indicates the marginalized medians of these distributions. 
The blue lines in the background are contours of constant $\Delta\tsub{F}{max}/\tsub{F}{base}$ in steps of factors of 10, where the lightest (leftmost) is for $\Delta\tsub{F}{max}/\tsub{F}{base}=10$ and the darkest (rightmost) is for $\Delta\tsub{F}{max}/\tsub{F}{base}=0.001$.
The green/red circle indicate the positions of Events 1 and 2.
Here, Event 1 could be among detected the detected events in all the discrete masses considered.}
\label{fig:rhofs_dist}
\end{figure}

We demonstrate this using results from the simulations presented in \citet{johnson2020}.
In Figure \ref{fig:rhofs_dist} we show the $\rho-\tsub{f}{s}$ distributions for \textit{Roman} detected FFP events with only its primary \textit{W146} filter.
Here, detected means that events have a $\Delta\chi^2\geq300$ compared to a flat baseline and a consecutive number of data points at least 3$\sigma$ above $\tsub{F}{base}$ $n_{3\sigma}\geq6$ \citep[see ][]{johnson2020}. 
These results are for three discrete masses of 1.00, 0.10, and 0.01 $M_\oplus$ shown using black, gray, and light gray contours, respectively.
Each contour contains 95\% of the detected events for each discrete mass.
We include the marginalized medians for $\rho$ and $\tsub{f}{S}$ with squares that match the color of their respective contour.
We plot the positions of Event 1 and 2 as green and red circles. 
The blue lines are contours of constant $\Delta\tsub{F}{max}/\tsub{F}{base}$, which is the predicted maximum fractional deviation from the baseline for an event. 
From the leftmost, lightest blue curve with $\Delta\tsub{F}{max}/\tsub{F}{base}=10$, we include contours with steps of factors of ten to the rightmost, darkest blue line with $\Delta\tsub{F}{max}/\tsub{F}{base}=0.001$.

Here it is apparent that the average $\rho$ for detected events increases with decreasing lens mass. 
As discussed in \citet{johnson2020}, the detection of lower mass lenses is actually facilitated by EFSE as for large $\rho$ the chord crossing timescale can be much longer than the expected microlensing timescale. 
If any FFP lenses with masses of 0.10 or 0.01 $M_\oplus$ are detected by \romanst, they will likely have $\rho>1$ and would be close to the EFSE regime.
However, unless these objects have have a high occurrence rate there could potentially be very few detections of such lenses (see Table 2 of \citet{johnson2020}).

As lens mass decreases, so does the potential peak magnification of the event leading to the fact that lower mass lenses will be detected in events with smaller amounts of blending (larger $\tsub{f}{S}$). 
Although a smaller value of $\rho$ for these extremely low-mass lenses would require a main sequence source, main sequence sources are likely to be significantly blended in the bulge, in contrast to giant/sub-giant sources, which generally dominate the flux in the source PSF. 
If these lenses cause only few percent deviations for the source flux, this could easily pass undetected if the amount of blend flux is significant. 
This is one of the reasons that Event 1 could be among the detectable events in each of these mass bins, but Event 2 is not even though they should have the same fractional maximum flux deviation.

However, even in the case of faint sources, the brightness and blending distribution of source stars is not arbitrary and can be estimated.
Priors on $\tsub{F}{S}$ and $\tsub{F}{B}$ can be placed, thereby restricting the plausible range of the mathematical degeneracy (see, e.g., Figures 10 and 11 of \citep{johnson2020} and the surrounding discussion).
Such priors will be even more informative with a measurement of the color of the baseline flux.

Similarly, we note that it is possible to place physical priors that limit the range of the mathematical $t_*$-$b_0$ degeneracy. 
The reasonably well-known proper motion distributions of Galactic disk and bulge stars limit the possible range of plausible values of $\tsub{\mu}{rel}$ for microlensing events.
For example, if we assume a range of $\tsub{\mu}{rel}=[3,10]$ mas yr$^{-1}$, and a giant source star in the bulge with $\theta_*$=5 $\mu$as, the resulting range of $t_*\approx[5,17]$. 
A more detailed prior could be imposed by using a Galactic model that incorporates kinematics under the assumption that free-floating planets share the same kinematics as stars.

As pointed out by \cite{mroz2020a}, even if the source flux cannot be well constrained, a measurement of the source color based on observations in multiple filters during an EFSE event can allow for a tight constraint on $\tsub{\theta}{E}$.  
At first glance, this is surprising: for `typical' microlensing events, one measures $\tsub{F}{S}$ in a given filter via multiple observations of the time-variable magnification of the source.  The source color can be derived in a model-independent way by linear regression between observations taken in multiple filters during the microlensing event \citep{gould2010}.  
With an estimate of extinction to the source based on, e.g., the color of the red clump \citep{yoo2004,dominik99}, the unextinguished source flux and color can be determined.  
These, together with empirical color-surface brightness relations (e.g., \citealt{kervella2004}), allow for an estimate $\theta_*$.
Combined with a measurement of $\rho$ from finite source effects in the light curve, it is possible to measure $\tsub{\theta}{E}$.  
As we have shown in the case of EFSE events, it is not possible to measure $\tsub{F}{S}$ and $\rho$ separately to high precision with only single band photometry, as they are strongly degenerate.  
Nevertheless, as identified by \cite{mroz2020a} and shown mathematically here in Equation~\ref{eqn:dfexpannold}, a measurement of the excess flux $\Delta F$ during an EFSE event allows one to measure the combination $2\tsub{F}{S}/\rho^2$. Noting that $\tsub{F}{S}=\tilde S \pi \theta_*^2$, where $\tilde S$ is the average surface brightness of the source, we have that $\Delta F \propto 2\pi \tsub{\theta}{E}^2 \tilde S$.  
The average surface brightness $\tilde S$ can be estimated from multi-color observations during the event \citep{gould2010}, and thus it is possible to constrain $\tsub{\theta}{E}$, despite the fact that neither the source flux or angular radius are well-constrained \citep{mroz2020a}.

\section{Discussion and Conclusion
\label{sec:discussconclude}}

We have uncovered and explored a multi-parameter degeneracy for microlensing events that exhibit extreme finite source effects (EFSEs).
This degeneracy arises fundamentally from the fact that (in the absence of limb-darkening) EFSEs with $\rho\gg 1$ have a peak magnification that is constant and depends only on the flux of the source $\tsub{F}{S}$ divided by the dimensionless source size $\rho$ squared. 
Furthermore, the duration of an EFSE event is decoupled from its peak magnification and depends only on the chord crossing time $t_c = 2t_*\sqrt{1-b_0^2}$. 
For finite $\rho$, both $\rho$ and $b_0$ give shape to the light curve through wings and shoulders during the event, which have a fractional duration of $\tsub{f}{ws}$.
Thus there are four model parameters ($\tsub{F}{S}, \rho, t_*, b_0$) to describe three observables ($\Delta\tsub{F}{max},\tsub{t}{FWHM},\tsub{f}{ws}$). 
In the presence of limb-darkening, the peak magnification of the event becomes covariant with the duration of the event, but the degeneracy remains. 
In particular, changing the impact parameter $b_0$ changes the maximum flux and shape of the light curve, but this can be completely compensated for by changing the limb-darkening, the ratio of angular source radius to the angular Einstein ring radius, and the source flux. 

We have largely explored these as purely mathematical degeneracies without detailed consideration as to what degree they will manifest in physically realistic situations, and how they may be ameliorated by changing the survey parameters.  

We encourage detailed consideration of the plausible severity of these degeneracies in individual events, as has been done in the seven likely free-floating (or wide-separation) planet candidates to date \citep{mroz2018,mroz2019,mroz2020a,mroz2020b,kim2021,ryu2021}.  
Regarding the second condition (ameliorating the degeneracy by changing the survey parameters), the fundamental difficultly arises from the fact that the morphologies of the degenerate light curves appear very similar for different values of the parameters. 
However, the morphologies are not identical and therefore can be broken with precise and dense photometry. 
Unfortunately, `precise' in this context means relative photometry of $\ll 1\%$, as these deviations will also be of very low amplitude \mbox{($\la$ few percent)}.  
Furthermore, these events are short \mbox{($\sim$ several hours)}, and thus obtaining both dense and precise photometry is difficult.  
Overall, obtaining the required precision and cadence to break these degeneracies will be challenging for the relatively faint bulge sources that will be monitored by, e.g., \romanst.  
Fortunately, many EFSE events will be due to giant source stars \citep{mroz2018,mroz2019,mroz2020a,mroz2020b}, which are on the bright end of the magnitude distribution of sources monitored by \romanst. 
A possible solution to avoid large sources entirely would be to conduct a microlensing survey towards the Magellanic Clouds or M31, which have source distances much larger than the bulge but would be sampling lenses belonging to the Galactic Halo \citep[][Slaybaugh et al. 2022, in preparation]{montero-camacho2019,sajadian2021}.

Observations in multiple filters while the source is magnified will allow for a constraint on $\tsub{\theta}{E}$ \citep{mroz2020a}.
The baseline plan for the \romanst Galactic Exoplanet Survey \citep{penny2019,johnson2020} is to observe in an alternative filter once every 12 hours, although this has not been finalized. 
The typical $\sim 10$~hr duration of EFSEs means that at most one color observation will be made during an EFSE event, which is likely insufficient to break this degeneracy. 
We concur with \citet{mroz2020b} that the cadence of \romanst supplemental filters should be increased, if at all possible, but note that the achievable cadence may ultimately be limited by engineering constraints on the lifetime rotations of {\it Roman's} filter wheel.

Currently, most large $\rho$ events are consistent with having source stars that are giants in the bulge. 
It is {\it a priori} more likely for such sources to dominate the baseline flux of these events, although at least a subset of clump stars will still be blended. 
For low enough lens masses, even main sequence stars could be sources for EFSEs. 
\romanst can detect these events \citep{johnson2020}, but it would likely be even more challenging to break the degeneracies using the methods stated above.  
Detailed simulations of the event rate of EFSEs, along with detailed fitting to these simulated events, will be necessary to determine to what extent these degeneracies will hinder the characterization of microlensing events attributable to very low-mass lenses.

\acknowledgements
We thank Przemek Mr\'{o}z and Radek Poleski for their extremely helpful suggestions to improve this manuscript over several iterations.
SAJ dedicates his contribution to this work to Ty Colin Harrison, who filled countless lives with light and joy. 

SAJ, MTP, and BSG were supported by NASA grant  NNG16PJ32C. BSG was supported by the Thomas Jefferson Chair for Discovery and Space Exploration at the Ohio State University.  
MTP acknowledges support from Louisiana Board of Regents
Support Fund (RCS Award Contract Number: LEQSF(2020-23)-RD-A-10).

\appendix
\section{Tables for Figure 10}
\label{apdx:tables}
In this Appendix, we include the parameters for the light curves in Figure \ref{fig:gammaustar_vary_murhofs} for Event 1 (Table \ref{tbl:rho10}) and Event 2 (Table \ref{tbl:rho45}).
For six values of $\Gamma$ (Column 1) and for five values of $b_0=0.0,0.2,0.4,0.6,0.8$ (Column 2) we found values of $t_*$, $\rho$, and $\tsub{f}{S}$ that minimize $\Delta\chi^2$.
Column 3 is simply the value of $\beta$ for these values of $b_0$ (Equation \ref{eqn:tchord}).
From these values of $b_0$, we calculate the values of $\eta_1$ and $\eta_2$ for scaling the other parameters in Equations \ref{eqn:degennoldfiniterho} and \ref{eqn:degengammaequalone1}, respectively and in Columns 4 and 5.
The values of $\rho$, $\tsub{f}{S}$, and $t_*$ that we found minimize $\Delta\chi^2$ are included in Columns 6, 9, and 12 and have a subscript ``n'' for the parameter. 
The two columns following each of Columns 6, 9, and 12 are the predicted values from scaling relations in Equations \ref{eqn:degennoldfiniterho} and \ref{eqn:degengammaequalone1}, with subscripts ``1'' and ``2'' respectively. 
Note that the first row for each $\Gamma$ is the fiducial values for Event 1 and 2.
For small values of $b_0$, the predictions of Equation \ref{eqn:degennoldfiniterho} agree better with the numerical values, and for large $b_0$ the numerical values agree better with predictions from Equation \ref{eqn:degengammaequalone1} (see Section \ref{subsec:degenmathfixedld}). 
However, the values diverge one the scaling would require values of $\tsub{f}{S}>3$ and are restricted by our prior. 
This forces a lower value of $\rho$ than expected to increase the magnification (Equation \ref{eqn:fse_peak}).
\begin{deluxetable}{l|rrrrrrrrrrrrr}[h]
\tablecolumns{18}
\label{tbl:rho10}
\tablecaption{The parameters for the Event 1 light curves in the left panel of Figure 9. 
}
\tablehead{\colhead{${\Gamma}$}&\colhead{$b_{0}$}&\colhead{$\beta$}&\colhead{$\eta_1$}&\colhead{$\eta_2$}&\colhead{$\tsub{\rho}{n}$}&\colhead{$\rho_1$}&\colhead{$\rho_2$}&\colhead{$\tsub{f}{S,n}$}&\colhead{$\tsub{f}{S,1}$}&\colhead{$\tsub{f}{S,2}$}&\colhead{$t_{*,\text{n}}$}&\colhead{$t_{*,1}$}&\colhead{$t_{*,2}$}}
\startdata
0.0 & 0.0 & 1.000 & 1.00 & 1.00 & 10.00 & 10.00 & 10.00  & 1.00 & 1.00 & 1.00 & 0.33 & 0.33 & 0.33 \\
 & 0.2 & 0.980 & 1.09 & 1.11 & 10.41 & 10.42 & 10.42  & 1.08 & 1.09 & 1.11 & 0.33 & 0.33 & 0.33 \\
 & 0.4 & 0.917 & 1.42 & 1.55 & 11.89 & 11.90 & 11.90  & 1.41 & 1.42 & 1.55 & 0.36 & 0.36 & 0.36 \\
 & 0.6 & 0.800 & 2.44 & 3.05 & 15.58 & 15.62 & 15.62  & 2.43 & 2.44 & 3.05 & 0.41 & 0.41 & 0.41 \\
 & 0.8 & 0.600 & 7.72 & 12.86 & 17.08 & 27.78 & 27.78  & 3.00 & 7.72 & 12.86 & 0.54 & 0.54 & 0.54 \\

\hline
0.2 & 0.0 & 1.000 & 1.00 & 1.00 & 10.00 & 10.00 & 10.00  & 1.00 & 1.00 & 1.00 & 0.33 & 0.33 & 0.33 \\
 & 0.2 & 0.980 & 1.09 & 1.11 & 10.34 & 10.42 & 10.42  & 1.07 & 1.09 & 1.11 & 0.33 & 0.33 & 0.33 \\
 & 0.4 & 0.917 & 1.42 & 1.55 & 11.60 & 11.90 & 11.90  & 1.37 & 1.42 & 1.55 & 0.36 & 0.36 & 0.36 \\
 & 0.6 & 0.800 & 2.44 & 3.05 & 14.60 & 15.63 & 15.63  & 2.24 & 2.44 & 3.05 & 0.41 & 0.41 & 0.41 \\
 & 0.8 & 0.600 & 7.72 & 12.86 & 16.23 & 27.78 & 27.78  & 3.00 & 7.72 & 12.86 & 0.54 & 0.54 & 0.54 \\

\hline
0.4 & 0.0 & 1.000 & 1.00 & 1.00 & 10.00 & 10.00 & 10.00  & 1.00 & 1.00 & 1.00 & 0.33 & 0.33 & 0.33 \\
 & 0.2 & 0.980 & 1.09 & 1.11 & 10.31 & 10.42 & 10.42  & 1.07 & 1.09 & 1.11 & 0.33 & 0.33 & 0.33 \\
 & 0.4 & 0.917 & 1.42 & 1.55 & 11.39 & 11.90 & 11.90  & 1.35 & 1.42 & 1.55 & 0.36 & 0.36 & 0.36 \\
 & 0.6 & 0.800 & 2.44 & 3.05 & 13.92 & 15.63 & 15.63  & 2.14 & 2.44 & 3.05 & 0.41 & 0.41 & 0.41 \\
 & 0.8 & 0.600 & 7.72 & 12.86 & 15.41 & 27.78 & 27.78  & 3.00 & 7.72 & 12.86 & 0.53 & 0.54 & 0.54 \\

\hline
0.6 & 0.0 & 1.000 & 1.00 & 1.00 & 10.00 & 10.00 & 10.00  & 1.00 & 1.00 & 1.00 & 0.33 & 0.33 & 0.33 \\
 & 0.2 & 0.980 & 1.09 & 1.11 & 10.29 & 10.42 & 10.42  & 1.07 & 1.09 & 1.11 & 0.33 & 0.33 & 0.33 \\
 & 0.4 & 0.917 & 1.42 & 1.55 & 11.47 & 11.90 & 11.90  & 1.39 & 1.42 & 1.55 & 0.36 & 0.36 & 0.36 \\
 & 0.6 & 0.800 & 2.44 & 3.05 & 13.89 & 15.63 & 15.63  & 2.23 & 2.44 & 3.05 & 0.40 & 0.41 & 0.41 \\
 & 0.8 & 0.600 & 7.72 & 12.86 & 14.61 & 27.78 & 27.78  & 3.00 & 7.72 & 12.86 & 0.53 & 0.54 & 0.54 \\

\hline
0.8 & 0.0 & 1.000 & 1.00 & 1.00 & 10.00 & 10.00 & 10.00  & 1.00 & 1.00 & 1.00 & 0.33 & 0.33 & 0.33 \\
 & 0.2 & 0.980 & 1.09 & 1.11 & 10.37 & 10.42 & 10.42  & 1.09 & 1.09 & 1.11 & 0.33 & 0.33 & 0.33 \\
 & 0.4 & 0.917 & 1.42 & 1.55 & 11.65 & 11.90 & 11.90  & 1.46 & 1.42 & 1.55 & 0.36 & 0.36 & 0.36 \\
 & 0.6 & 0.800 & 2.44 & 3.05 & 14.86 & 15.62 & 15.62  & 2.65 & 2.44 & 3.05 & 0.41 & 0.41 & 0.41 \\
 & 0.8 & 0.600 & 7.72 & 12.86 & 13.87 & 27.78 & 27.78  & 3.00 & 7.72 & 12.86 & 0.53 & 0.54 & 0.54 \\

\hline
1.0 & 0.0 & 1.000 & 1.00 & 1.00 & 10.00 & 10.00 & 10.00  & 1.00 & 1.00 & 1.00 & 0.33 & 0.33 & 0.33 \\
 & 0.2 & 0.980 & 1.09 & 1.11 & 10.44 & 10.42 & 10.42  & 1.11 & 1.09 & 1.11 & 0.33 & 0.33 & 0.33 \\
 & 0.4 & 0.917 & 1.42 & 1.55 & 12.03 & 11.90 & 11.90  & 1.57 & 1.42 & 1.55 & 0.36 & 0.36 & 0.36 \\
 & 0.6 & 0.800 & 2.44 & 3.05 & 15.54 & 15.62 & 15.62  & 3.00 & 2.44 & 3.05 & 0.41 & 0.41 & 0.41 \\
 & 0.8 & 0.600 & 7.72 & 12.86 & 13.19 & 27.78 & 27.78  & 3.00 & 7.72 & 12.86 & 0.53 & 0.54 & 0.54 \\

\enddata
\end{deluxetable}

\begin{deluxetable}{l|rrrrrrrrrrrrr}[h]
\tablecolumns{18}
\label{tbl:rho45}
\tablecaption{The parameters for the Event 2 light curves in the right panel of Figure 9. 
}
\tablehead{\colhead{${\Gamma}$}&\colhead{$b_{0}$}&\colhead{$\beta$}&\colhead{$\eta_1$}&\colhead{$\eta_2$}&\colhead{$\tsub{\rho}{n}$}&\colhead{$\rho_1$}&\colhead{$\rho_2$}&\colhead{$\tsub{f}{S,n}$}&\colhead{$\tsub{f}{S,1}$}&\colhead{$\tsub{f}{S,2}$}&\colhead{$t_{*,\text{n}}$}&\colhead{$t_{*,1}$}&\colhead{$t_{*,2}$}}
\startdata
0.0 & 0.0 & 1.000 & 1.00 & 1.00 & 4.47 & 4.47 & 4.47  & 0.20 & 0.20 & 0.20 & 0.33 & 0.33 & 0.33 \\
 & 0.2 & 0.980 & 1.09 & 1.11 & 4.66 & 4.66 & 4.66  & 0.22 & 0.22 & 0.22 & 0.33 & 0.33 & 0.33 \\
 & 0.4 & 0.917 & 1.42 & 1.55 & 5.31 & 5.32 & 5.32  & 0.28 & 0.28 & 0.31 & 0.36 & 0.36 & 0.36 \\
 & 0.6 & 0.800 & 2.44 & 3.05 & 6.96 & 6.99 & 6.99  & 0.48 & 0.49 & 0.61 & 0.41 & 0.41 & 0.41 \\
 & 0.8 & 0.600 & 7.72 & 12.86 & 12.14 & 12.42 & 12.42  & 1.46 & 1.54 & 2.57 & 0.54 & 0.54 & 0.54 \\

\hline
0.2 & 0.0 & 1.000 & 1.00 & 1.00 & 4.47 & 4.47 & 4.47  & 0.20 & 0.20 & 0.20 & 0.33 & 0.33 & 0.33 \\
 & 0.2 & 0.980 & 1.09 & 1.11 & 4.65 & 4.66 & 4.66  & 0.22 & 0.22 & 0.22 & 0.33 & 0.33 & 0.33 \\
 & 0.4 & 0.917 & 1.42 & 1.55 & 5.24 & 5.32 & 5.32  & 0.28 & 0.28 & 0.31 & 0.36 & 0.36 & 0.36 \\
 & 0.6 & 0.800 & 2.44 & 3.05 & 6.76 & 6.99 & 6.99  & 0.48 & 0.49 & 0.61 & 0.40 & 0.41 & 0.41 \\
 & 0.8 & 0.600 & 7.72 & 12.86 & 11.32 & 12.42 & 12.42  & 1.41 & 1.54 & 2.57 & 0.53 & 0.54 & 0.54 \\

\hline
0.4 & 0.0 & 1.000 & 1.00 & 1.00 & 4.47 & 4.47 & 4.47  & 0.20 & 0.20 & 0.20 & 0.33 & 0.33 & 0.33 \\
 & 0.2 & 0.980 & 1.09 & 1.11 & 4.64 & 4.66 & 4.66  & 0.22 & 0.22 & 0.22 & 0.33 & 0.33 & 0.33 \\
 & 0.4 & 0.917 & 1.42 & 1.55 & 5.23 & 5.32 & 5.32  & 0.28 & 0.28 & 0.31 & 0.35 & 0.36 & 0.36 \\
 & 0.6 & 0.800 & 2.44 & 3.05 & 6.71 & 6.99 & 6.99  & 0.49 & 0.49 & 0.61 & 0.40 & 0.41 & 0.41 \\
 & 0.8 & 0.600 & 7.72 & 12.86 & 11.26 & 12.42 & 12.42  & 1.54 & 1.54 & 2.57 & 0.53 & 0.54 & 0.54 \\

\hline
0.6 & 0.0 & 1.000 & 1.00 & 1.00 & 4.47 & 4.47 & 4.47  & 0.20 & 0.20 & 0.20 & 0.33 & 0.33 & 0.33 \\
 & 0.2 & 0.980 & 1.09 & 1.11 & 4.64 & 4.66 & 4.66  & 0.22 & 0.22 & 0.22 & 0.33 & 0.33 & 0.33 \\
 & 0.4 & 0.917 & 1.42 & 1.55 & 5.26 & 5.32 & 5.32  & 0.29 & 0.28 & 0.31 & 0.35 & 0.36 & 0.36 \\
 & 0.6 & 0.800 & 2.44 & 3.05 & 6.77 & 6.99 & 6.99  & 0.52 & 0.49 & 0.61 & 0.40 & 0.41 & 0.41 \\
 & 0.8 & 0.600 & 7.72 & 12.86 & 11.37 & 12.42 & 12.42  & 1.73 & 1.54 & 2.57 & 0.53 & 0.54 & 0.54 \\

\hline
0.8 & 0.0 & 1.000 & 1.00 & 1.00 & 4.47 & 4.47 & 4.47  & 0.20 & 0.20 & 0.20 & 0.33 & 0.33 & 0.33 \\
 & 0.2 & 0.980 & 1.09 & 1.11 & 4.66 & 4.66 & 4.66  & 0.22 & 0.22 & 0.22 & 0.33 & 0.33 & 0.33 \\
 & 0.4 & 0.917 & 1.42 & 1.55 & 5.32 & 5.32 & 5.32  & 0.30 & 0.28 & 0.31 & 0.35 & 0.36 & 0.36 \\
 & 0.6 & 0.800 & 2.44 & 3.05 & 6.97 & 6.99 & 6.99  & 0.57 & 0.49 & 0.61 & 0.40 & 0.41 & 0.41 \\
 & 0.8 & 0.600 & 7.72 & 12.86 & 12.27 & 12.42 & 12.42  & 2.19 & 1.54 & 2.57 & 0.53 & 0.54 & 0.54 \\

\hline
1.0 & 0.0 & 1.000 & 1.00 & 1.00 & 4.47 & 4.47 & 4.47  & 0.20 & 0.20 & 0.20 & 0.33 & 0.33 & 0.33 \\
 & 0.2 & 0.980 & 1.09 & 1.11 & 4.61 & 4.66 & 4.66  & 0.22 & 0.22 & 0.22 & 0.33 & 0.33 & 0.33 \\
 & 0.4 & 0.917 & 1.42 & 1.55 & 5.41 & 5.32 & 5.32  & 0.31 & 0.28 & 0.31 & 0.36 & 0.36 & 0.36 \\
 & 0.6 & 0.800 & 2.44 & 3.05 & 7.26 & 6.99 & 6.99  & 0.64 & 0.49 & 0.61 & 0.40 & 0.41 & 0.41 \\
 & 0.8 & 0.600 & 7.72 & 12.86 & 13.29 & 12.42 & 12.42  & 2.80 & 1.54 & 2.57 & 0.53 & 0.54 & 0.54 \\

\enddata
\end{deluxetable}

\end{document}